\documentclass[fleqn,usenatbib]{mnras}

\usepackage[T1]{fontenc}

\DeclareRobustCommand{\VAN}[3]{#2}
\let\VANthebibliography\thebibliography
\def\thebibliography{\DeclareRobustCommand{\VAN}[3]{##3}\VANthebibliography}

\usepackage{graphicx}	% Including figure files
\usepackage{amsmath}	% Advanced maths commands
\usepackage{amssymb}	% Extra maths symbols
\usepackage{lipsum}
\usepackage{xcolor}

\usepackage{multirow}
\usepackage{nicematrix}

\newcommand{\Msun}{\hbox{M$_\odot$}}

\usepackage{newtxtext,newtxmath}

\title[Galaxy Sizes in Two Rest-frame Wavelengths
]{
Two rest-frame wavelength measurements of galaxy sizes at $z<1$: the evolutionary effects of emerging bulges and quenched newcomers 
}

\author[A. George et al.]{Angelo George,$^{1}$\thanks{E-mail: \href{mailto:angelo.george@smu.ca}{angelo.george@smu.ca}}
Ivana Damjanov,$^{1}$\thanks{Canada Research Chair} 
Marcin Sawicki,$^{1}$\thanks{Canada Research Chair}
St\'ephane Arnouts,$^{2}$
Guillaume Desprez,$^{1}$
\newauthor
Stephen Gwyn,$^{3}$
Vincent Picouet,$^{2}$
Simon Birrer$^{4, 5}$
and John Silverman$^{6, 7, 8}$
\\
$^{1}$Institute for Computational Astrophysics and Department of Astronomy \& Physics, Saint Mary's University, 923 Robie Street, Halifax, NS B3H 3C3, Canada\\
$^{2}$Aix-Marseille University, CNRS, CNES, LAM, Marseille, France\\
$^{3}$NRC Herzberg Astronomy and Astrophysics, 5071 West Saanich Road, Victoria, BC V9E 2E7, Canada\\
$^{4}$Department of Physics, Kavli Institute for Particle Astrophysics and Cosmology, Stanford University, Stanford, CA 94305, USA\\
$^{5}$SLAC National Accelerator Laboratory, Menlo Park, CA 94025, USA \\
$^{6}$Kavli Institute for the Physics and Mathematics of the Universe, The University of Tokyo, Kashiwa, 277-8583, Japan \\
$^{7}$Center for Data-Driven Discovery, Kavli IPMU (WPI), UTIAS, The University of Tokyo, Kashiwa, Chiba 277-8583, Japan \\
$^{8}$Department of Astronomy, School of Science, The University of Tokyo, 7-3-1 Hongo, Bunkyo, Tokyo 113-0033 Japan \\
}

\date{Accepted XXX. Received YYY; in original form ZZZ}

\pubyear{2024}

\begin{document}
\label{firstpage}
\pagerange{\pageref{firstpage}--\pageref{lastpage}}
\maketitle

% Abstract of the paper
\begin{abstract}

We analyze the size evolution of $16000$ star-forming galaxies (SFGs) and $5000$ quiescent galaxies (QGs) with mass $M_*>10^{9.5}$\Msun\ at $0.1<z<0.9$ from the COSMOS field using deep CLAUDS+HSC imaging in two rest-frame wavelengths, $3000$~\AA\ (UV light) and $5000$~\AA\ (visible light). With half-light radius ($R_e$) as proxy for size, SFGs at characteristic mass $M_0 = 5\times10^{10}$\Msun\ grow by $20\%$ ($30\%$) in UV (visible) light since $z\sim1$ and the strength of their size evolution increases with stellar mass. After accounting for mass growth due to star formation,  we estimate that SFGs grow by $75\%$ 
in all stellar mass bins and in both rest-frame wavelengths. Redder SFGs are more massive, smaller and more concentrated than bluer SFGs and the fraction of  red SFGs increases with time. These results point to the emergence of bulges as the dominant mechanism for the average size growth of SFGs. We find two threshold values for the stellar mass density within central $1$~kpc ($\Sigma_1$): all SFGs with $\log\Sigma_1\gtrsim9$ are red and only QGs have  $\log\Sigma_1\gtrsim9.7$. The size of $M_*=M_0$ QGs grows by $50\%$ ($110\%$) in the UV (visible) light. Up to $\sim20\%$ of this increase in size of massive QGs is due to newcomers (recently quenched galaxies). However, newcomers cannot explain the observed pace in the size growth of QGs; that trend has to be dominated by processes affecting individual galaxies, such as minor mergers and accretion.

\end{abstract}


%%%%%%%%%%%%%%%%%%%%%%%%%%%%%%%%%%%%%%%%%%%%%%%%%%
%%%%%%%%%%%%%%%%% BODY OF PAPER %%%%%%%%%%%%%%%%%%

%%%%%%%%%%%%%%%%%%%%%%%%%%%%%%%%%%%%%%%%%%%%%%%%%%%%%%%%%%%%%%%%%%
%%%%%%%%%%%%%%%%%%%%%%%%%%%%%%%%%%%%%%%%%%%%%%%%%%%%%%%%%%%%%%%%%%
%%%%%%%%%%%%%%%%%%%%%%%%%%%%%%%%%%%%%%%%%%%%%%%%%%%%%%%%%%%%%%%%%%

\section{Introduction}
\label{sec:intro}

The assembly history of galaxies in the aging universe depends on their evolutionary stage at the cosmic time (redshift) of observations.
Galaxy transformations over time include changes both in the average properties of their stellar population (e.g., age and metallicity) and in galaxy structure (i.e., size and shape). 
By observing evolutionary trends in morphological properties of galaxies in large (spectro-)photometric surveys (e.g., SDSS, \citealt{yorkSloanDigitalSky2000, abazajianSeventhDataRelease2009, aiharaEighthDataRelease2011}; HSC-SSP \citealt{aiharaHyperSuprimeCamSSP2018, aiharaSecondDataRelease2019}; DESI, \citealt{deyOverviewDESILegacy2019}), we can probe the assembly histories of different galaxy populations.

\par 
A suite of studies conducted over the last two decades show that for both SFGs and QGs galaxy size depends on stellar mass. \citep[e.g.,][and many others]{shenSizeDistributionGalaxies2003, guoStructuralPropertiesCentral2009, williamsEvolvingRelationsSize2010, newmanCanMinorMerging2012,  moslehEvolutionMassSizeRelation2012,  vanderwel3DHSTCANDELSEvolution2014, langeGalaxyMassAssembly2015, faisstConstraintsQuenchingMassive2017, royEvolutionGalaxySizestellar2018,   mowlaCOSMOSDASHEvolutionGalaxy2019, matharuHSTWFC3Grism2019, matharuHSTWFC3Grism2020, kawinwanichakijHyperSuprimeCamSubaru2021, baroneLEGACSAMIGalaxy2022, mercierScalingRelations2512022, damjanovSizeSpectroscopicEvolution2023}. Both galaxy populations show an expected positive trend in size with stellar mass,  in agreement with theoretical studies 
\citep[e.g.,][]{priceTestingRecoveryIntrinsic2017, genelSizeEvolutionStarforming2018, rositoMasssizePlaneEAGLE2019, degraaffObservedStructuralParameters2022}.  

\par
However, the parameters of the relation between galaxy size and stellar mass (also called the size-mass relation; SMR) 
differ between SFGs and QGs. The SFGs show a weak relation between galaxy size and stellar mass with $R_e\varpropto M_*^{\sim0.2}$, where $R_e$ is the effective radius (radius that includes $50\%$ of galaxy light) in kpc and $M_*$ is the stellar mass in $\Msun$ \citep[e.g.,][]{shenSizeDistributionGalaxies2003,  williamsEvolvingRelationsSize2010,   vanderwel3DHSTCANDELSEvolution2014, mowlaCOSMOSDASHEvolutionGalaxy2019, kawinwanichakijHyperSuprimeCamSubaru2021, baroneLEGACSAMIGalaxy2022}.  
On the other hand, QGs exhibit a more complex SMR. 
Below the pivot mass ($\sim 3\times10^{10}$\Msun),  
QGs show a weak trend with $R_e \varpropto M_*^{-0.1}$, resembling the equivalent relation for SFGs \citep[e.g.,][]{morishitaGrismLensamplifiedSurvey2017, kawinwanichakijHyperSuprimeCamSubaru2021}. In sharp contrast, the exponent of the power-law SMR for QGs above the pivot mass increases significantly \citep[$R_e\varpropto M_*^{\sim0.7}$; e.g.,][]{shenSizeDistributionGalaxies2003, williamsEvolvingRelationsSize2010, langeGalaxyMassAssembly2015, huangRelationsSizesGalaxies2017, mowlaCOSMOSDASHEvolutionGalaxy2019, moslehGalaxySizesPerspective2020, kawinwanichakijHyperSuprimeCamSubaru2021, nedkovaExtendingEvolutionStellar2021, damjanovSizeSpectroscopicEvolution2023}.

\par 
\citet{mowlaMassdependentSlopeGalaxy2019} and \citet{kawinwanichakijHyperSuprimeCamSubaru2021} show that the pivot mass of the SMR of QGs matches the pivot point of the stellar mass-halo mass (SMHM) relation, a relation between galaxies and their host dark matter halos \citep[e.g.,][]{keresHowGalaxiesGet2005, keresGalaxiesSimulatedLCDM2009, bowerBreakingHierarchyGalaxy2006, crotonManyLivesActive2006, behrooziComprehensiveAnalysisUncertainties2010, huangRelationsSizesGalaxies2017, golden-marxImpactEnvironmentLatetime2019, golden-marxObservedEvolutionStellar2022, gabrielpillaiGalaxyFormationSanta2022, zaritskyPhotometricMassEstimation2023}. The SMHM relation describes how efficiently baryons accreted into the halo are cooled and converted into stars. The  pivot point of the SMHM relation is the stellar mass that corresponds to the maximum efficiency of this conversion \citep[e.g.,][]{behrooziComprehensiveAnalysisUncertainties2010, leauthaudNewConstraintsEvolution2012, mosterGalacticStarFormation2013, rodriguez-pueblaConstrainingGalaxyhaloConnection2017}. According to \citet{mowlaMassdependentSlopeGalaxy2019} and \citet{kawinwanichakijHyperSuprimeCamSubaru2021}, because the pivot points of SMR and SMHM relation are similar, the pivot point of the SMR indicates the stellar mass at which \textit{in-situ} galaxy growth via star formation is replaced by the \textit{ex-situ} driven growth via mergers and accretion.  

\par 
In addition, the pace at which galaxy sizes evolve also differ between massive ($M_*\gtrsim 3\times10^{10}$\Msun) SFGs and QGs. 
On average, the evolution in galaxy sizes (i.e., in $R_e$) with cosmic time is stronger for QGs than for SFGs \citep[e.g.,][]{vanderwel3DHSTCANDELSEvolution2014, kawinwanichakijHyperSuprimeCamSubaru2021}. This difference suggests that the observed evolution in galaxy morphology can probe physical processes that affect galaxy growth before and after the star formation in them ceases.

\par 
The strength of SFG size evolution varies  with the observed redshift interval. The observational studies that cover a wider redshift range ($0<z<3$) show a strong evolution of sizes \citep[e.g.,][]{vanderwel3DHSTCANDELSEvolution2014, mowlaCOSMOSDASHEvolutionGalaxy2019}. However, several studies show weak to no evolution since $z\sim1$ \citep[e.g.,][]{lillyHubbleSpaceTelescope1998,ravindranathEvolutionDiskGalaxies2004,bardenGEMSSurfaceBrightness2005, kawinwanichakijHyperSuprimeCamSubaru2021} whereas studies that focus only at higher redshifts ($z>2$) show very strong evolution in SFG sizes with redshift \citep[e.g.;][]{oeschStructureMorphologies782010}.% All these studies indicate 
The range of results for different redshift intervals indicates that the size evolution of SFGs may be slowing down with cosmic time.

\par

There are several reasons why the size growth of SGFs could slow down with decreasing redshift. 
One obvious reason is the availability of cold gas. As the average density of gas decreases with time, there is less gas (star-forming material) in galaxy environment that can be accreted. The lack of star-forming material leads to reduced levels of cosmic star formation at redshift $z\lesssim 1$ \citep{madauCosmicStarFormationHistory2014}. If the size growth of SFG is driven predominantly by the \textit{in-situ} star formation, the reduced amount of available gaseous material will affect the observed pace of their size growth. 

\par
Another reason for slower size growth of SFGs at $z<1$ is the emergence of centrally concentrated bulges in galaxies \citep{martigMorphologicalQuenchingStar2009, sachdevaGrowthBulgesDisk2017, hashemizadehDeepExtragalacticVisible2022}. Older galaxies tend to have a significant bulge component in their center compared to younger galaxies. In massive galaxies at $z<1$, these bulges contain older and less massive red stars that dominate the light in visible and near infra-red wavelength regimes. In contrast, the light in outer disks is dominated by the contribution from young stars that emit strongly in UV 
\citep[e.g.,][]{bredaStellarAgeGradients2020}. Therefore, the overall size of a galaxy appears smaller and more centrally concentrated in the rest-frame $5000$~\AA\ where the light from old stars dominates. In studies that target this redder rest-frame wavelength, the observed slower pace of size evolution with decreasing redshift can indicate the emergence of concentrated red bulges. 

\par 
Galaxies undergo morphological transformations when they become quiescent.  
A quenched galaxy typically has smaller size than a star-forming one. If an SFG is a disk-dominated system and the quenching process does not affect the stellar distribution within it, the stellar disk fades away when the star-formation ceases \citep[e.g.,][]{christleinCanEarlyTypeGalaxies2004, carolloZENSIVSimilar2016, matharuHSTWFC3Grism2020}. 
As galaxy disks are more extended than their more centrally concentrated bulges/spheroids, the half-light radius of a galaxy decreases when the disk fades. 
These transformations can be probed further by analyzing galaxy light profile components (e.g., bulges and disks) separately, investigating mass complete spatially resolved spectroscopic datasets, and measuring galaxy dynamical properties \citep[e.g.,][]{hudsonColoursBulgesDiscs2010, belliMOSFIRESpectroscopyQuiescent2017, belliMOSFIRESpectroscopyQuiescent2019, newmanResolvingQuiescentGalaxies2018}.

\par
QGs, on the other hand, evolve and grow in size through various physical processes, 
including major mergers, minor mergers, accretion and adiabatic expansion \citep[e.g.,][]{ naabFormationEarlyTypeGalaxies2007, naabMinorMergersSize2009, fanDramaticSizeEvolution2008, fanCosmicEvolutionSize2010,   vandokkumGrowthMassiveGalaxies2010, trujilloDissectingSizeEvolution2011, cimattiFastEvolvingSize2012, lopez-sanjuanDominantRoleMergers2012,   huertas-companyEvolutionMasssizeRelation2013, zahidCoevolutionMassiveQuiescent2019, damjanovSizeSpectroscopicEvolution2023}. Additionally, newly quenched galaxies being added to the quiescent population (newcomers) affect the overall size distribution and the average size growth of QGs \citep[progenitor bias; e.g.,][]{vandokkumMorphologicalEvolutionAges2001, sagliaFundamentalPlaneEDisCS2010, carolloNewlyQuenchedGalaxies2013,  genelSizeEvolutionStarforming2018, damjanovSizeSpectroscopicEvolution2023}. 
The estimated level of influence that the newcomers have on the average size evolution of the quiescent population range from minimal \citep[e.g.,][]{zanellaRoleQuenchingTime2016, damjanovSizeSpectroscopicEvolution2023} to substantial \citep[e.g.,][]{carolloNewlyQuenchedGalaxies2013, fagioliMinorMergersProgenitor2016}. 
However, apart from the progenitor bias that affect the bulk of the QG population, a majority of studies tracing morphological evolution of QGs indicate that minor mergers are the major driver of size growth for this galaxy population. \citep{newmanCanMinorMerging2012, beifioriRedshiftEvolutionDynamical2014,  ownsworthMinorMajorMergers2014, belliStellarPopulationsSpectroscopy2015, buitragoCosmicAssemblyStellar2017, zahidCoevolutionMassiveQuiescent2019, hamadoucheCombinedVANDELSLEGAC2022, damjanovSizeSpectroscopicEvolution2023}. These galaxy encounters affect the light in different rest-frame wavelengths by depositing younger and/or metal-poor stellar material at galaxy outskirts \citep[e.g.,][]{suessColorGradientsQuiescent2020}.

So far, the large scale studies have been focused on a single rest-frame wavelength (typically $5000$~\AA). Galaxy light profiles at this wavelength are dominated by the light from older stars that contribute most of the stellar mass of a galaxy \citep[e.g.,][]{vanderwel3DHSTCANDELSEvolution2014, mowlaCOSMOSDASHEvolutionGalaxy2019, kawinwanichakijHyperSuprimeCamSubaru2021}. 
By studying galaxy morphology in ultraviolet (UV) light, we can probe the distribution of recently formed stars in galaxies because the shorter wavelengths ($2000<\lambda<4000$~\AA) are dominated by the light from young stars \citep[e.g.,][]{bruzualStellarPopulationSynthesis2003}. In addition, simultaneous analysis of galaxy morphology at longer wavelengths ($\lambda>4000$~\AA)  enables us to probe the difference in spatial distribution of young and old stars. 

\par

\begin{figure*}
    \centering
    
    \includegraphics[width=2\columnwidth]{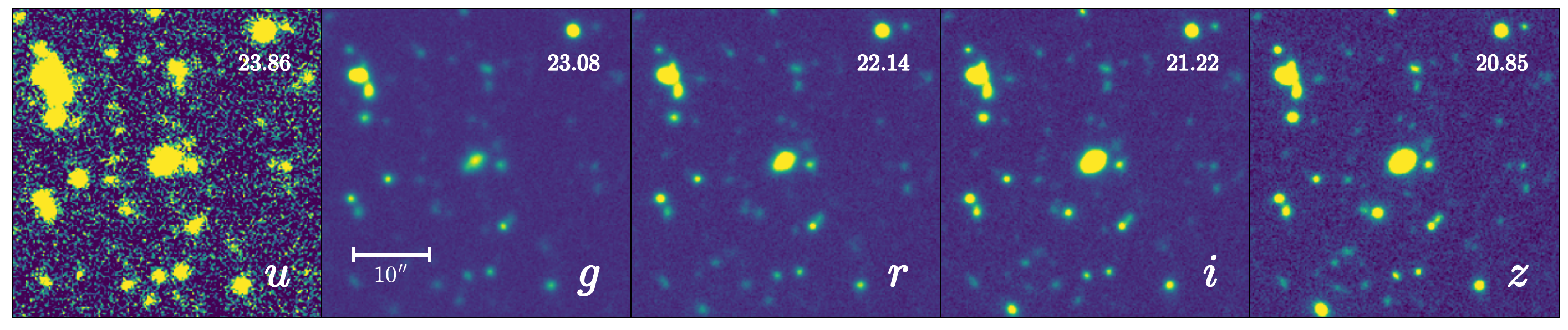}
    
    \caption[Image of a galaxy in CLAUDS+HSC bands]{CLAUDS+HSC images of an SFG (centre) with mass $\log M =10.53$ at $z=0.67$ ($\alpha$: 150.17897, $\delta$: 2.03213) in \textit{u}, \textit{g}, \textit{r}, \textit{i} and \textit{z} bands. Image cutouts are of dimensions $\sim 33.8^{\prime\prime} \times 33.8^{\prime\prime}$ and have the same intensity stretch. A $10^{\prime\prime}$ ($72.3$ kpc) scale is shown for reference in the middle panel.  }
    \label{f:clauds+hsc_data_3066167}
\end{figure*}

\par 
CFHT CLAUDS \citep{sawickiCFHTLargeArea2019} and Subaru HSC-SSP \citep{aiharaHyperSuprimeCamSSP2018, aiharaSecondDataRelease2019} surveys together provide the ideal dataset for studies of galaxy morphology in rest-frame UV and visible lights at $z<1$.
The CLAUDS+HSC survey includes deep images of the sky over an overlapping area of $18.6$ deg$^2$ in $6$ bands ($Ugrizy$) that range from UV to near infrared (IR). In this pilot study, we use a subset of this dataset (the central $\sim1.6$ deg$^2$ region of the COSMOS field) to analyse the SMR and size evolution of $\sim16000$ SFGs and $\sim5200$ QGs at $0.1<z<0.9$ in two rest-frame wavelengths: $3000$~\AA\ and $5000$~\AA. 
We will expand the study to the entire survey volume in the followup analysis.

\par 
The plan of this paper is as follows. Section \ref{sec:Data} provides an overview of CLAUDS+HSC survey, the data and auxiliary data products we use, sample selection and classification of galaxies into SFGs and QGs. In Section \ref{sec:Methodology}, we present details of the methodology for 2D galaxy light profile fitting (Section \ref{sec:Fitting}). In  Section \ref{sec:Methodology} we also provide the results of the simulations we perform to estimate the robustness of the fitting pipeline and the systematic uncertainties present in our morphological measurements (Section \ref{sec:Simulations}). In addition, we describe how we estimate the rest-frame S\'ersic parameters in Section \ref{sec:2-wavelength-measurement}. We validate our size measurements by comparing them with those from the literature (Section \ref{sec:Literature-comparison}). In Section \ref{sec:results}, we describe the analytic fits to the SMR and present the results of SMR fitting and size evolution for SFGs and QGs in two rest-frame wavelengths. We discuss the implications of our findings in Section \ref{sec:discussion}. We summarize the major conclusions based on our analysis in Section \ref{sec:conclusions}.

\par 
Throughout this study, we use $M$ to denote stellar mass in solar units ($M_\star/$M$_\odot$) unless otherwise stated. Additionally, we use the AB magnitude system and adopt a standard cosmology with $\Omega_M=0.3$, $\Omega_\Lambda=0.7$ and $h=0.7$ throughout this paper.

%%%%%%%%%%%%%%%%%%%%%%%%%%%%%%%%%%%%%%%%%%%%%%%%%%%%%%%%%%%%%%%%%%
%%%%%%%%%%%%%%%%%%%%%%%%%%%%%%%%%%%%%%%%%%%%%%%%%%%%%%%%%%%%%%%%%%
%%%%%%%%%%%%%%%%%%%%%%%%%%%%%%%%%%%%%%%%%%%%%%%%%%%%%%%%%%%%%%%%%%

\section{Data and Sample Selection}
\label{sec:Data}

\subsection{CLAUDS+HSC Survey}
\label{sec:clauds}

We use the ultra deep multi-band ($u+griz$) imaging data from the central $1.6$ deg$^2$ region of the COSMOS field from CFHT Large Area $U$-band Deep Survey\footnote{\url{https://www.clauds.net/}} \citep[CLAUDS;][]{sawickiCFHTLargeArea2019} and the Hyper Suprime-Cam Subaru Strategic Program\footnote{\url{https://hsc.mtk.nao.ac.jp/ssp/}} \citep[HSC-SSP;][]{aiharaHyperSuprimeCamSSP2018, aiharaSecondDataRelease2019}. In total, the combined deep CLAUDS+HSC dataset covers an overlapping area of $18.6$ deg$^2$.  
In \textit{u} and \textit{i} bands, the median depth is $27.1$~mag (measured at $5\sigma$ level in $2''$ apertures).  
The depth of the dataset is optimal for the analysis of massive galaxy morphology out to their faint outskirts.
The CLAUDS+HSC bands also have similar sensitivities, except for \textit{y}-band where the limiting magnitude is $25.6$~mag.
Figure \ref{f:clauds+hsc_data_3066167} shows an example galaxy  from our dataset in \textit{u}, \textit{g}, \textit{r}, \textit{i} and \textit{z} bands.

\par
 
Recently, \citet{kawinwanichakijHyperSuprimeCamSubaru2021} conducted an extensive morphological study using HSC-SSP data only (i.e., without the \textit{u}-band images that are available only for the Deep and UltraDeep layers of the HSC-SSP). 
The \textit{u}-band data are important for constraining photometric redshifts and internal galaxy properties. For example, \textit{u}-band photometry is essential for bracketing the Balmer and $4000$~\AA \, breaks in order to estimate photometric redshifts (photo-$z$'s) accurately for galaxies at intermediate redshifts \citep[$0<z\lesssim0.7$,][]{sawickiCFHTLargeArea2019}. At these intermediate redshifts, \textit{u}-band data also play a vital role in constraining star formation rates (SFRs) of galaxies. 
With the addition of \textit{u}-band data from CLAUDS to the Deep and UltraDeep layer of HSC-SSP, we are able to minimize the contamination when separating SFGs and QGs at $z<1$ and to analyze structural evolution of the two populations in a unique way. 

\par
Although our final goal is to analyze the entire $18.6$ deg$^2$ of the survey region, we limit this pilot study to the COSMOS/UltraVISTA field, covering $\sim1.6$ deg$^2$ on the sky. We target this smaller area because it is a widely studied  survey region in the sky. Choosing a widely studied field is beneficial as we can compare the results of our work with those of previous and parallel studies. This approach ensures quality control of certain steps implemented in this project and enables us to draw scientific conclusions regarding the physical processes driving the observed trends with more certainty. In addition, the COSMOS field is rich in ancillary data, which we utilize in this study for mass and redshift estimates. In the next stages, we will expand the study to the entire CLAUDS+HSC data to improve statistics and analyze the impact of environment on size evolution of galaxies.

\begin{figure*}
    \centering
	\includegraphics[width=1.8\columnwidth]{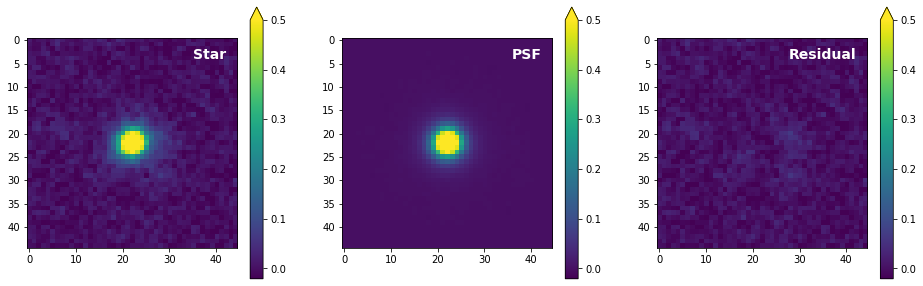}
    \caption[u-PSF]{A sample PSF generated for \textit{u-}band using PSFEx. The left panel shows the image of a star in \textit{u-}band. The middle panel shows the PSF image generated for the same location in the sky. The last panel shows the residual image obtained by subtracting the PSF from the stellar image. }
    \label{f:u-psf}
\end{figure*}

\subsection{Point Spread Function}
\label{sec:PSF}

\par Atmospheric turbulence (atmospheric seeing) dominates the blurring of astronomical images in ground-based observations. 
A point spread function (PSF) describes how a point source in the sky (a star) looks like in the image due to this blurring. Hence, to study the intrinsic shapes of galaxies, we must remove the effect of PSF from galaxy images.
 
\par
The HSC Pipeline uses the PSFEx algorithm\footnote{\url{http://ascl.net/1301.001}} \citep{bertinAutomatedMorphometrySExtractor2011} to characterize the PSF of the HSC images. The PSFex adaptation in the HSC Pipeline models PSF from the images of unsaturated stars in an iterative way to remove contamination from neighbouring objects \citep{boschHyperSuprimeCamSoftware2018}.
PSF models at any position within the footprint of the survey are available through the PSF picker\footnote{\url{https://hsc-release.mtk.nao.ac.jp/psf}} utility in the Public Data Release by HSC-SSP.  

\par
Although PSFs for every sky coordinate covered by the HSC-SSP are available via the PSF picker, fetching a PSF model for every galaxy would be computationally expensive. Hence, we divide every HSC image patch\footnote{Each image patch consists of $4200\times4200$
pixels where each image pixel covers $\sim 0.168^\prime \times 0.168^\prime$ of the sky. The pixel and patch sizes are the same for both CLAUDS and HSC images.} into $36$ sub-regions of dimensions $117.6^{\prime\prime} \times 117.6^{\prime\prime}$. For all galaxies within a sub-region, we use the PSF image at the center of this region. 

\par
The PSF images for CLAUDS \textit{u-}band data are not available externally and thus we model PSFs for these images using PSFEx. Although the software selects the bright unsaturated stars using the catalogue generated by SExtractor, we refine the automated star selection by identifying a region from peak in the surface brightness versus magnitude relation where the unsaturated stars lie. PSEFx then uses these stars to model PSFs for the 36 sub-regions in each image patch.

\par 
Figure \ref{f:u-psf} shows a sample PSF in \textit{u-}band we generate using PSFEx. The third panel shows the residual after PSF model (second panel) is removed from an image of a star in \textit{u-}band (first panel). The residual image shows no visible structure, confirming that the \textit{u-}band PSFs we generated are reliable.

\par 
The PSF changes with the band of observations and location in the sky. Among the HSC bands, the \textit{i}-band has the best seeing with full width at half maximum (FWHM) of the PSF ranges between $0.5^{\prime\prime}$ and $0.7^{\prime\prime}$. The FWHM in \textit{u}-band is also in the same range. 
The median FWHM in other bands is $\sim0.7^{\prime\prime}$.

\subsection{Other Data Products} 
\label{sec:Data-products} 

We make use of the COSMOS2020 \citep{weaverCOSMOS2020PanchromaticView2022}, the latest photometric catalog from Cosmic Evolution Survey (COSMOS, \citet{scovilleCosmicEvolutionSurvey2007a}).
This catalog is based on new imaging and incorporates CFHT CLAUDS \textit{u}-band data, HSC-SSP PDR2 \textit{grizy} data, VISTA VIRCAM YJHKs-band data of UltraVISTA DR4 \citep{mccrackenUltraVISTANewUltradeep2012} and Spitzer/ IRAC channel $1$, $2$, $3$, and $4$ data of the Cosmic Dawn Survey \citep{collaborationEuclidPreparationXVII2022}. The catalogue covers an area of $2$ deg$^2$ and includes multiwavelength photometric measurements and derived properties for $1.7$ million sources.
From the catalogue, we use the photometric redshifts, stellar masses and star formation rates  derived using LePhare \citep{ilbertAccuratePhotometricRedshifts2006} and rest-frame UVJ colours from \textsc{Eazy} code \citep{brammerEAZYFastPublic2008}.

\begin{figure}
    \centering
	\includegraphics[width=\columnwidth]{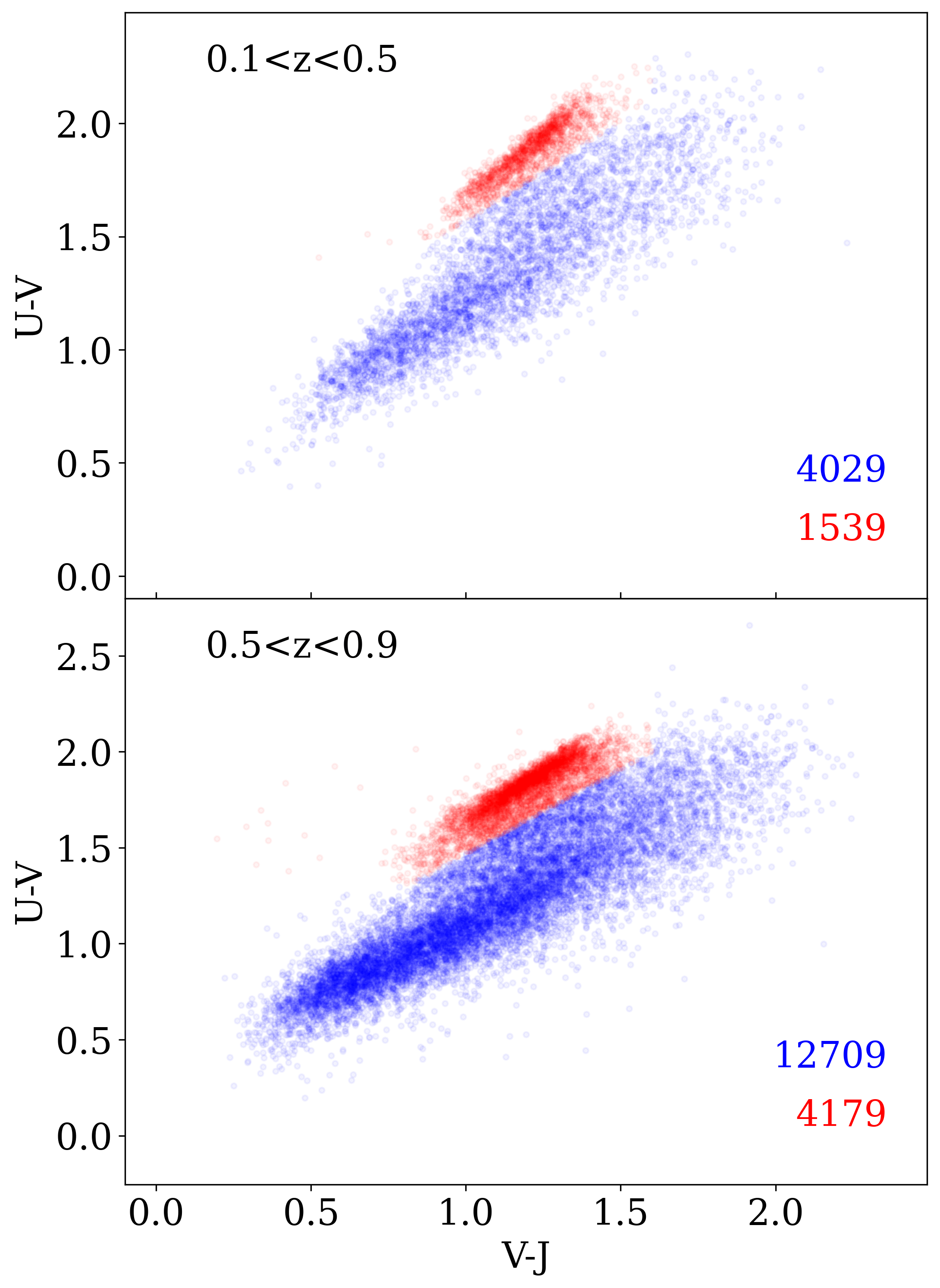}
    \caption[UVJ Diagram]{Classification of galaxies into SFGs (blue) and QGs (red) in the UVJ diagram using Equation \ref{eq:uvj-selection}. The upper and lower panels show the classification in two redshift bins. Rest-frame UVJ colours are taken from \citet{weaverCOSMOS2020PanchromaticView2022}.}
    \label{f:uvj_sel}
\end{figure}

\begin{figure*}
    \centering
	\includegraphics[width=2\columnwidth]{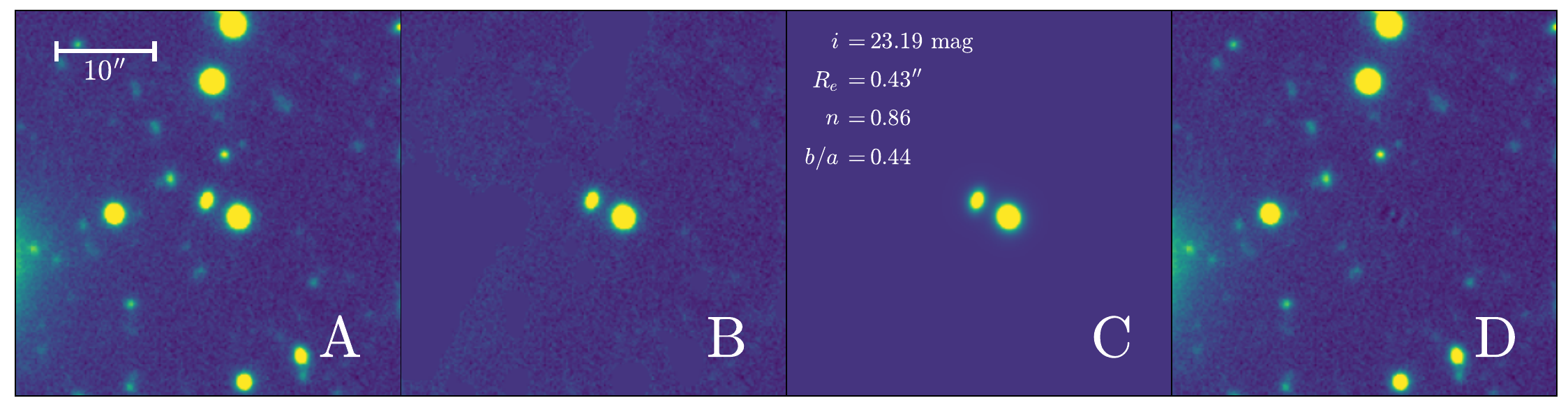} 
    \caption[Images from profile extraction of galaxies in COSMOS using \textsc{Galfit}]{An example of galaxy profile modelling using \textsc{Galfit}. Panels from left to right show the input cutout image of an SFG ($z=0.81$; $\log M = 9.56$) in \textit{i}-band (A), masked image (B), best-fit models (C) and residual image (D). The fitting of the target galaxy (center) is performed simultaneously with a neighbouring bright galaxy on its bottom right. The best-fit parameters of the target galaxy is given in panel D. We also show a $10^{\prime\prime}$ ($77.6$ kpc) scale in panel A for reference. }
    \label{f:hsc-output-images}
\end{figure*}

\subsection{Sample Selection and Galaxy Classification}
\label{sec:sample-selection} 

We limit the redshift range of our work to $0.1<z<0.9$ because we cannot constrain the stellar mass of galaxies well beyond $z\sim0.9$ by using CLAUDS+HSC data alone.  This problem arises due to the limitation in the available wavelength coverage to bracket several spectral features like the $4000$~\AA\  break. 
Although this pilot study utilises COSMOS2020 catalogue with wider wavelength coverage (Section \ref{sec:Data-products}), our future studies will cover the complete CLAUDS+HSC deep survey region and thus we will be relying on the data from $Ugrizy$ bands alone. 

\par
Furthermore, we select only massive galaxies for this study. The mass completeness limits varies with redshift as well as with the level of galaxy star formation activity. The SFGs at lower redshifts have lower stellar mass completeness limit than the QGs at higher redshifts. To have a uniform mass completeness limit for the entire dataset, we limit our study to galaxies more massive than $\log M > 9.5$ (Chen et al. 2024, in prep).

\par
We classify selected  galaxies into SFGs and QGs using UVJ colour-colour diagrams \citep[e.g.,][]{wuytsWhatWeLearn2007, williamsDetectionQuiescentGalaxies2009, brammerNumberDensityMass2011, muzzinEvolutionStellarMass2013,carnallInferringStarFormation2018} 
by applying a selection criterion from \citet{whitakerNEWFIRMMediumbandSurvey2011}, 
\begin{equation}
\begin{split}
    &(U-V) >  0.88\times(V-J) + 0.69 \ [\mathrm{for}\ z<0.5], \\
    &(U-V) >  0.88\times(V-J) + 0.59 \ [\mathrm{for}\ z>0.5],  
    \label{eq:uvj-selection}   
\end{split}
\end{equation} 
where $U-V$ and $U-J$ are rest-frame colours.
Figure \ref{f:uvj_sel} shows the UVJ classification of galaxies in two redshift regimes. The sample we analyze in this study contains in total $\sim16000$ SFGs (blue points) and $\sim5200$ QGs  (red points).

%%%%%%%%%%%%%%%%%%%%%%%%%%%%%%%%%%%%%%%%%%%%%%%%%%%%%%%%%%%%%%%%%%
%%%%%%%%%%%%%%%%%%%%%%%%%%%%%%%%%%%%%%%%%%%%%%%%%%%%%%%%%%%%%%%%%%
%%%%%%%%%%%%%%%%%%%%%%%%%%%%%%%%%%%%%%%%%%%%%%%%%%%%%%%%%%%%%%%%%%

\section{Methodology}
\label{sec:Methodology}

\begin{figure*}
    \centering
	\includegraphics[width=2\columnwidth]{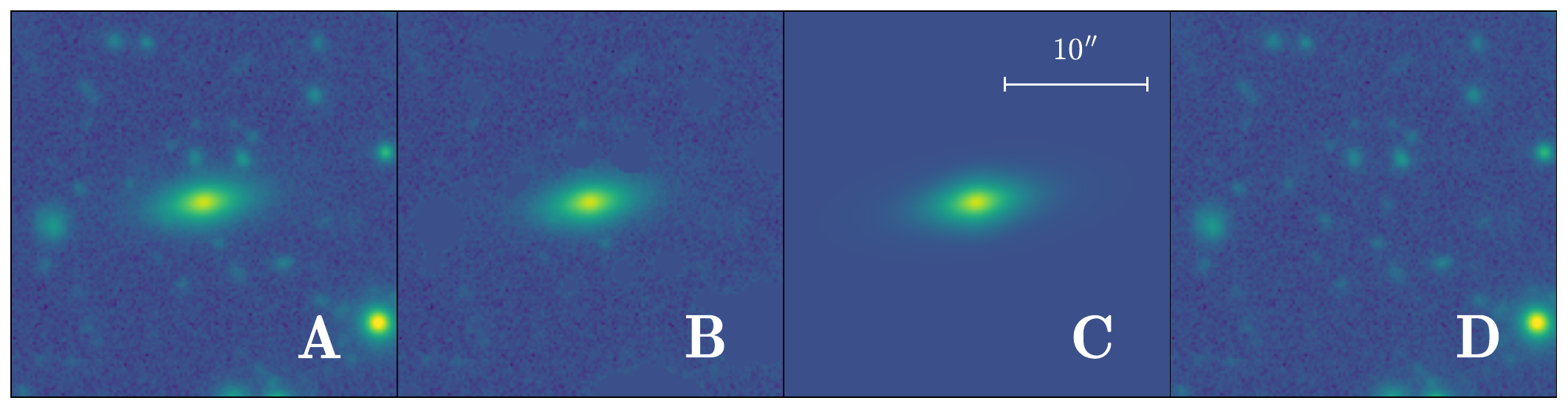}
    \caption[Fitting Pipeline]{An example for profile modelling of simulated galaxies using \textsc{Galfit} in \textit{i}-band. The panels are same as in Figure \ref{f:hsc-output-images} but for a simulated galaxy instead for a real one. 
    }
    \label{f:sim-i-galaxy}
\end{figure*}

\subsection{Galaxy Profile Fitting using \textsc{Galfit}}
\label{sec:Fitting}

We model galaxy light profiles with a single S\'ersic profile, 

\begin{equation} 
    \log\left(\frac{\Sigma(R)}{\Sigma_e}\right)=-b_n \left[ \left(\frac{R}{R_e}\right)^{1/n} - 1 \right],
    \label{eq:sersic}
\end{equation}
where $R_e$ is the effective radius, $\Sigma_e$ is the surface brightness at $R_e$ and $n$ is the S\'ersic index, which defines the concentration of the profile. The coefficient $b_n$ is not a free parameter; it is a function of $n$ and it ensures that the region within $R_e$ encloses half of the total luminosity of the galaxy. 
Although galaxies usually have several internal components such as disks and bulges, we can fit the overall light profile of a galaxy using a single S\'ersic function to obtain its global features \citep[size, light concentration, axis ratio, etc.;][]{vanderwelStructuralParametersGalaxies2012}. 

\par In this work we use \textsc{Galfit}, a two-dimensional ($2$D) parametric light profile fitting software tool to fit galaxy profiles \citep{pengDetailedStructuralDecomposition2002, pengDetailedDecompositionGalaxy2010}. The software extends the one dimensional S\'ersic profile from Equation \ref{eq:sersic} to $2$D space for direct fitting of galaxy images.
Based on the simulations we perform (and describe in Section \ref{sec:Simulations}), we introduce magnitude cut in each band (\textit{u}$<26$; \textit{g}$<25.5$; \textit{r}, \textit{i}, \textit{z}$<25$) that corresponds to $\sim75\%$ success rate for our fitting pipeline. We do not apply any limit based on other galaxy structural parameters such as size, S\'eric index or axis ratio, because they do not affect the robustness of our fitting pipeline (see Appendix \ref{appendix:Simulations}). Additionally, we do not fit \textit{y}-band data because the simulations show that our pipeline does not produce accurate measurements of the structural parameters in that band even for fairly bright galaxies (e.g., $24<y<25$; see Section \ref{robustness}). 

\par
At the start of the fitting procedure, we make an image cutout centred around the target galaxy. We ensure that the size of the cutout is sufficiently large for good fitting by adopting a minimum cutout size of $25\times R_{eS}$, where $R_{eS}$ is the SExtractor effective radius from the COSMOS2020 catalogue (this radius does not take into account the effects of seeing). The cutout size does not exceed $84^{\prime\prime}$.  
Our simulations  (Section~\ref{sec:cutout}) show that the fitting of galaxy profiles using the cutout size of $25\times R_{eS}$ performs significantly better than when using cutout sizes $<20\times R_{eS}$ and that the performance level plateaus around $20-25\times R_{eS}$.  
Because the HSC PSF images have dimensions of 43 pixels per side, we ensure that the image cutout is at least 45 pixels per side, which is also the dimension of the convolution box for PSF. 

\par
We flag neighbouring galaxies for simultaneous fitting if their projected distance from the target galaxy is $<2\times (R_{eS}$(target)$ + R_{eS}$(neighbour)$)$, and the magnitude difference is $m_\text{neighbour}-m_\text{target}<0.5$.    
If the target galaxy is fainter than the $23$ mag, we further restrict the magnitude range for neighbour selection to  ${m_\text{neighbour}-m_\text{target}<0.1}$.  

\par
We then mask out all other fainter galaxies and bright objects in the cutout image. We use a watershed segmentation technique implemented in the \textsc{Photoutils} package for \textsc{Python} for this purpose. The watershed algorithm treats pixel values in the image as an inverted local topography where the centroids of bright objects are at the local minima of this topography. The algorithm generates a segmentation map for the light sources in the image cutout. The algorithm then deblends all overlapping regions and separates the objects in into regions (`watersheds'). Finally, it creates a file containing pixel coordinates belonging to the `watersheds' of all objects to be masked out. 

\par
The next step is to estimate the initial parameters for galaxy fitting. We use magnitudes from COSMOS2020 catalogue as the input magnitude value for \textsc{Galfit} in the respective filters. We roughly estimate the initial values of the other S\'ersic parameters (size, S\'ersic index, axis ratio) using \textsc{PetroFit}\footnote{\url{https://github.com/PetroFit/petrofit/releases/tag/v0.4.1}}, a galaxy light profile fitting \textsc{Python} package, and provide a random angle for the position angle. We estimate the mean sky background using the areas outside the watershed segments and provide it as the initial parameter. However, if a galaxy is fainter than $24.5$ mag, we fix the sky brightness at this mean sky value while running \textsc{Galfit}. In addition, the software requires $\sigma$-images, that contain the pixel level uncertainties in the data, and the PSF images.
\par
A \textsc{\textsc{Galfit}} S\'eric model is a fuction of several parameters: centroid pixel coordinates ($x_0$, $y_0$), magnitude ($m_{tot}$), $R_e$, $n$, axis ratio ($b/a$), and position angle ($\theta_{PA}$).
We fit galaxy profiles using \textsc{\textsc{Galfit}} in two steps, a procedure similar to the one used by \citet{matharuHSTWFC3Grism2019}.  In the first step, we keep all parameters free to vary and obtain refined values of galaxy initial position parameters: centroid pixel coordinates ($x_0$, $y_0$), $b/a$ and $\theta_{PA}$. We provide the best-fit parameters from the first step as the input for the second step, where we force three position parameters to remain constant (within uncertainties) and keep the remaining three structural parameters ($m_{tot}$, $R_e$ and $n$) free. We also constrain the parameter space for both steps to ensure that the software reaches global minimum. These constraints are $10\leq m_{tot}\leq30$, $0.1\leq R_e\leq50$ pix, $0.1\leq n\leq20$ and $0.02\leq b/a\leq 1$. In addition, we do not allow the \textsc{Galfit}-derived magnitude to differ from the magnitude obtained using SExtractor by more than $2$ mags.

\par
Figure \ref{f:hsc-output-images} illustrates the steps in fitting of a galaxy light profile. Panel A shows the image cutout, panel B shows the masking using watershed algorithm and panel C the best fit S\'ersic models of the target and a neighbouring galaxy. Panel D provides the residual image  obtained by subtracting panel C from panel A. From the residual image, it is clear that the fitting pipeline finds S\'ersic profile that best matches the observed galaxy (i.e., Panel D shows no features in the region of fitted galaxy). Since we use only a single S\'ersic profile for a galaxy, in some cases structural features (such as spiral arms) are visible in the residual image (see the residuals for the neighbouring galaxy in panel D).

\subsection{Simulations}
\label{sec:Simulations}

\begin{figure*}
    \centering
	\includegraphics[width=2\columnwidth]{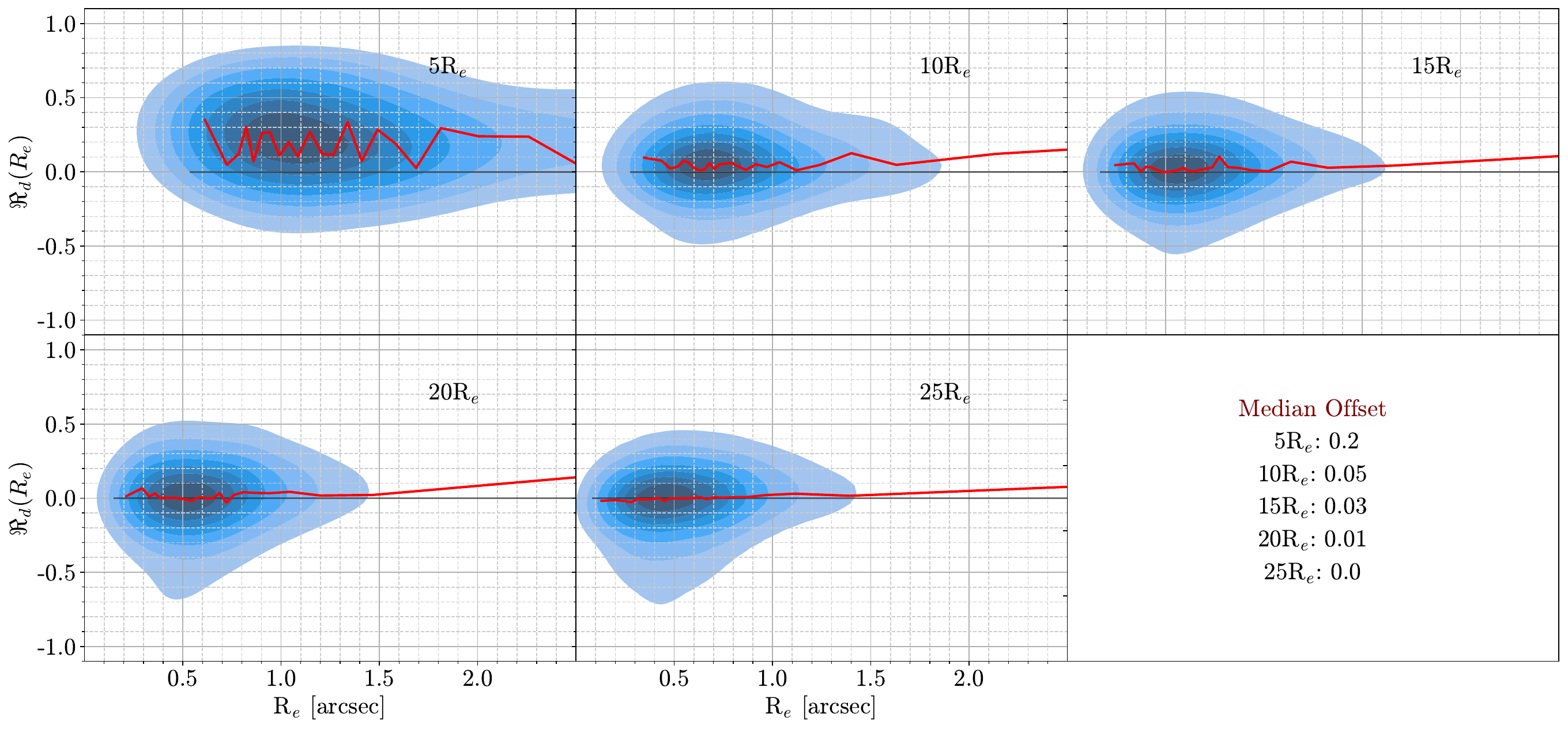}
    \caption[Relative difference i]{The relative difference in galaxy sizes in \textit{i}-band from simulations with different cutout sizes. The contours in each panel show the distribution of the relative difference in size: $\Re_d = $ (input - output)/input (Equation \ref{eq:rd}). The red curve in each panel shows the median relative difference in sizes in bins of input galaxy sizes. We perform five sets of simulations, changing the cutout size in relation to the effective radii of simulated galaxies: $5R_e$, $10R_e$, $15R_e$, $20R_e$ and $25R_e$. 
    In the last panel, we give the median value of the offset in $\Re_d(R_e)$ for each cutout size. 
    
    }
    \label{f:sim-rd-i-csize}
\end{figure*}

\subsubsection{Pipeline}
We perform simulations to test the robustness of our galaxy profile fitting procedure (Section \ref{sec:Fitting}) and to estimate the systematic uncertainties present in the measurements. In these simulations, we generate artificial galaxies and plant them in real images. To ensure that the simulated galaxies reflect the properties of real galaxies, we use the structural parameters of galaxies in the COSMOS field from the Zurich Structure and Morphology Catalogue\footnote{\url{https://irsa.ipac.caltech.edu/data/COSMOS/gator_docs/cosmos_morph_zurich_colDescriptions.html}} \citep[ZSMC;][]{sargentEvolutionNumberDensity2007, scarlataCOSMOSMorphologicalClassification2007}. ZSMC contains Hubble Space Telescope (HST)/ACS F814W-based structural properties of galaxies \citep{sargentEvolutionNumberDensity2007}, derived from a single S\'ersic profile fitting using GIM2D \citep{simardGIM2DIRAFPackage1998}. We first estimate a multivariate probability distribution function (PDF) of GIM2D-derived size ($R_e$), S\'ersic index ($n$) and axis ratio ($q$) along with \textit{u}, \textit{g}, \textit{r}, \textit{i}, \textit{z} and \textit{y} magnitudes from the CLAUDS+HSC-SSP catalogue \citep{desprezCombiningCLAUDSHSCSSP2023} using multi-dimensional kernel density estimation. As there are only a few galaxies with $R_e>1.5''$ in ZSMC, we include additional artificial galaxies with these larger sizes. We then randomly draw parameters from this PDF to simulate galaxy 2D profiles. While doing so, we make sure that we do not draw a parameter set from the multivariate PDF where any of the parameters has a non-physical value (for example, $R_e < 0$ pixels). 

\par
In the next step, we model each artificial galaxy with its randomly chosen parameter set from the PDF using \textsc{Galfit}.  We also convolve these artificial galaxies with PSF at the position where we plant them in the image.  We then plant these galaxies in real images with randomly chosen position angles ($\theta_{PA}$). 
Random values of $\theta_{PA}$ ensure that galaxies are placed in the images with orientations similar to those of real galaxies. We also make sure that the centroid of the simulated galaxy does not fall within the effective radius of another real galaxy in the image. 

\par 
We fit each planted galaxy using the same procedure we use with the real data (Section \ref{sec:Fitting}). Figure \ref{f:sim-i-galaxy} shows an example of a simulated galaxy and its best-fitting model obtained by the pipeline. We perform the planting and fitting procedures using one artificial galaxy at a time to avoid overcrowding of images. One of the caveats of our simulation is that artificial galaxies have single S\'ersic profiles, which is not true for most of the real galaxies. Nevertheless, this estimate is sufficient for the present study as we fit real galaxy profiles with a single-profile model.

\subsubsection{Cutout image size \label{sec:cutout}}
Using simulations we first explore the impact of the image cutout size on the fitting results. We perform several sets of simulations with varying cutout sizes based on the size of the target galaxy: $5R_e$, $10R_e$, $15R_e$, $20R_e$ and $25R_e$. 

\par 
We define relative difference for a given parameter $x$ as 
\begin{equation}
    \Re_d (x) = \frac{x \,\text{(input)}- x\,\text{(output)}} {x\,\text{(output)}}.
\label{eq:rd}
\end{equation}
We then analyse the distribution of $\Re_d (R_e)$ as a function of galaxy size in bins of various cutout sizes (Figure \ref{f:sim-rd-i-csize}). Clearly, the smallest cutout size ($5R_e$; first panel) performs the worst irrespective of the intrinsic size of the target galaxy. In this case, our pipeline yields sizes that are, based on the median relative difference, $\sim20\%$ smaller than input values (red line in the first panel of Figure~\ref{f:sim-rd-i-csize}). In addition the scatter in relative differences is significantly higher than for any other cutout sizes (blue contours). This error in size measurements is mainly due to two reasons: (1) galaxy profiles are extended beyond $5R_e$; (2) there are a few pixels in the cutout to effectively estimate the sky background. 

\par 
The offset in the size measurements decreases with increasing cutout sizes. 
With the $25R_e$ cutout (last populated panel of Figure \ref{f:sim-rd-i-csize}), the median relative offset (red line) shows that the best-fit $R_e$ is equivalent to the real (input) $R_e$ irrespective of the input galaxy size. 
In addition, the scatter in relative galaxy size differences is slightly higher for $20R_e$ than for $25R_e$ cutout size. Since this scatter represents the systematic uncertainty in galaxy size measurements, we consider a cutout size of $\sim25R_e$ as the best among the five cutout sizes chosen for the simulation and adopt that cutout size for our fitting procedure. 

\par
However, unlike simulated galaxies, we do not know the intrinsic sizes of galaxies in real data \textit{apriori}. Therefore, for real galaxies, we use size estimations from SExtractor to determine the required sizes of image cutouts.

\subsubsection{Robustness}
\label{robustness}

\begin{figure}
    \centering
	\includegraphics[width=1\columnwidth]{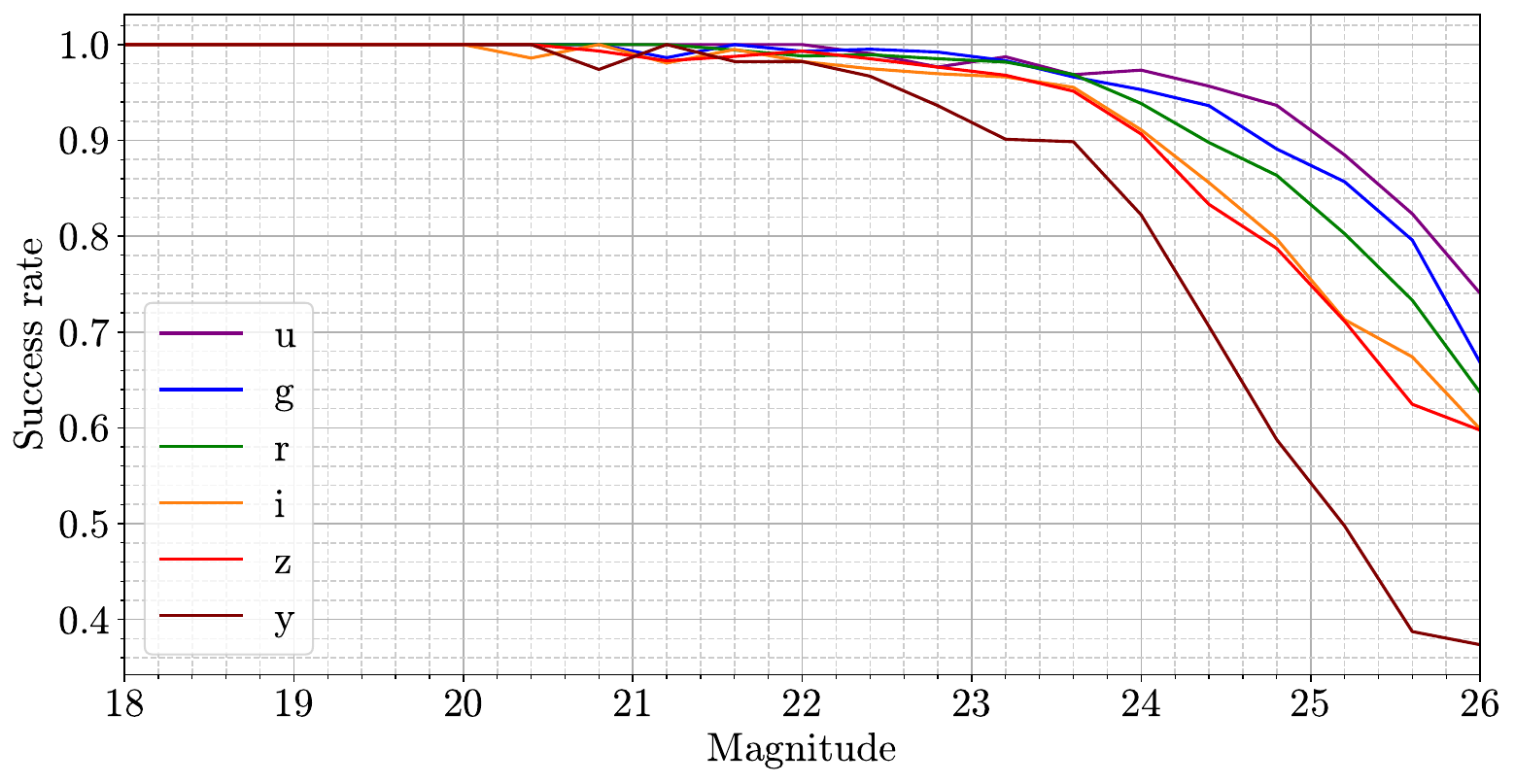}
    \caption[Fit success rate from simulations in CLAUDS+HSC bands]{Simulated success rate (Equation \ref{eq:sim-sr}) as a function of magnitudes in CLAUDS+HSC bands. The \textit{u}-band data performs the best while the \textit{y}-band data performs the poorest (see text for details). Hence, we do not use \textit{y}-band data in this work.}
    \label{f:hsc-sim-sratio}
\end{figure}

\begin{figure*}
    \centering
	\includegraphics[width=2\columnwidth]{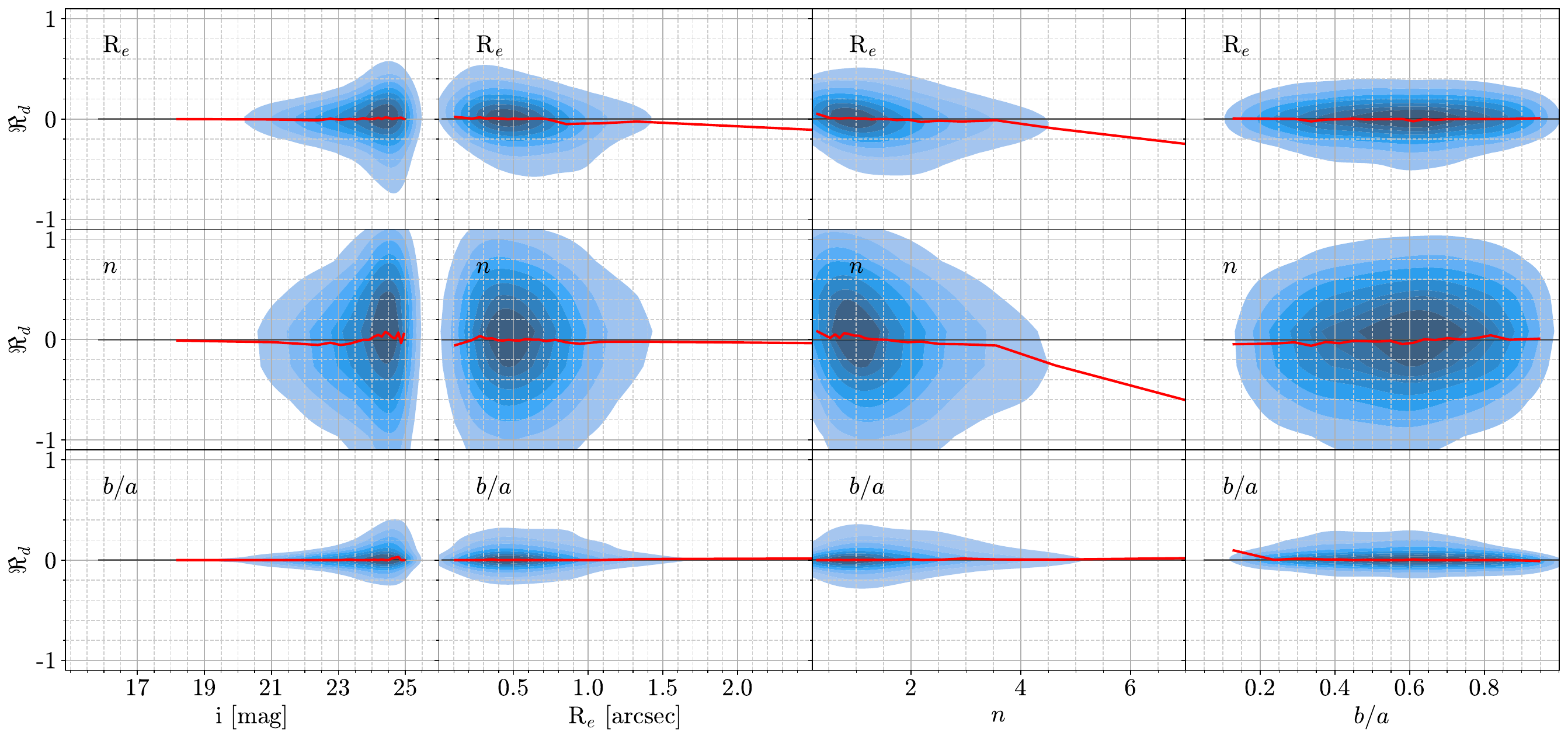}
    \caption[Relative difference i]{Relative difference in S\'ersic parameters in \textit{i}-band from simulations. The contours in each panel show the distribution of the relative difference in parameters: the top, middle and bottom rows show the relative differences in size, S\'ersic index and axis ratios respectively. From left to right, the columns show the relative differences in parameters as functions of S\'ersic parameters: magnitude, output size, output S\'ersic index and output axis ratio. Note that we have used the relative difference here as $\Re_d = $ (input - output)/input (Equation \ref{eq:rd}). In each panel, the red curve represents the median of the relative difference. 
    }
    \label{f:sim-rd-i-o}
\end{figure*}

\par
To investigate the robustness of the pipeline, we define the simulated success rate as

\begin{equation}
    \text{Success rate} = 1 - \frac{\text{Number of failures}}{\text{Total number of simulated galaxies}}.
    \label{eq:sim-sr}
\end{equation}
We consider a fit as failed if any of the parameters lies at the limits of the allowed parameter space. We show this ratio in all the six CLAUDS+HSC bands in Figure \ref{f:hsc-sim-sratio}. 

 \par
 We find that the success rate is a function of magnitude in the fitted band with limiting magnitude ranging from $\sim25$ to $26$ mag for a success rate of $0.75$ in \textit{u}, \textit{g}, \textit{r}, \textit{i} and \textit{z} bands. 
However, we do not find any significant dependence of the success rate on any other S\'ersic parameter (see Appendix \ref{appendix:Simulations}). Hence, we limit our analysis based on the magnitude of galaxies in five CLAUDS+HSC bands. 

\par
We also note that success rate in the \textit{y}-band is the lowest among the CLAUDS+HSC bands. It is because the \textit{y}-band is shallow and has poor seeing compared to other HSC bands. Hence, the fits based on the \textit{y}-band profiles are more likely to fail for even relatively bright galaxies. Therefore, we do not use galaxy morphology in the \textit{y}-band for the analysis of galaxy morphological evolution in this work. This does not impact our ability to probe the light profiles of galaxies in two rest-frame wavelengths. For galaxies in our highest redshift bin ($0.75<z<0.9$, Section~\ref{sec:results}), we use \textit{z}-band images to model light profiles at the redder rest-frame wavelength ($5000$ \AA).

\begin{figure*}
    \centering
	\includegraphics[scale=0.3]{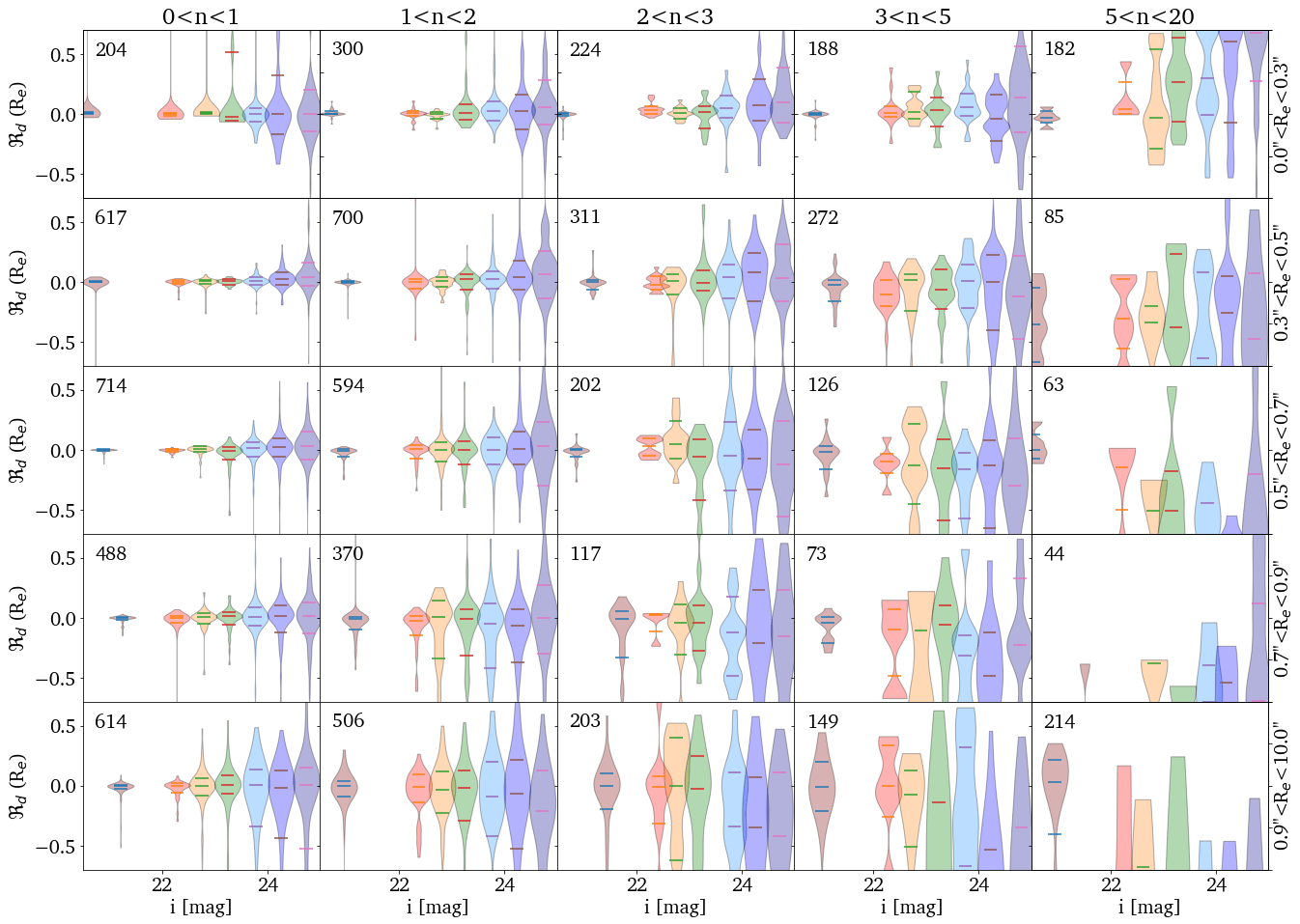}%pdf}
    \caption[Simulation violin plots]{Violin plots showing the distribution of the relative difference in galaxy sizes $\Re_d(R_e)$ in bins of \textit{i}-band magnitude ($x$-axis), its measured size (rows) and measured S\'ersic index (columns). The shape of the violin plots represents the distribution of $\Re_d(R_e)$ in each $3$-D parameter bin, and its $16^\text{th}$, $50^\text{th}$ and $84^\text{th}$ percentiles are shown as horizontal lines. The distributions are approximately Gaussian with their medians around $0$. However, the distributions differ from Gaussian and their median values are significantly different from $0$ for galaxies with higher S\'ersic indices or larger sizes that represent the tail of the distribution of galaxy structural properties. 
    }
    \label{f:violin}
\end{figure*}

\subsubsection{Uncertainty Estimation}
\par 
An important reason for running these simulations is to estimate the systematic uncertainties in our measurements. To do this, 
we analyze the distribution of $\Re_d (x)$ as a function of magnitude, measured size and measured S\'ersic index. The first column of Figure \ref{f:sim-rd-i-o} shows $\Re_d (x)$ as a function of \textit{i}-band magnitude. The median relative difference (red curve) in size, S\'ersic index or axis ratio does not exhibit any significant systematic offset as a function of magnitude. 

\par
If the measured size of a galaxy is large ($R_e>0.8^{\prime\prime}$; the first row in the second column of Figure \ref{f:sim-rd-i-o}), the measurement tends to be slightly larger than the input size. This offset increases with the measured size of the galaxy (offset is above $10\%$ if $R_e>1.5^{\prime\prime}$). We also tend to overestimate the S\'ersic index of a galaxy if its measured size is greater than $1.5^{\prime\prime}$ (second panel in the middle row). These offsets are not significant, and galaxies with large sizes are extremely rare. 
The median relative differences in $R_e$ and $n$ also show offsets if output S\'ersic index is greater than $4$ (red curves in the first and second row of the third column in Figure~\ref{f:sim-rd-i-o}). These  high S\'ersic index galaxies could have complex formation history and are least likely to be described by a single S\'ersic index. Furthermore, the ability of our algorithm to estimate structural parameters do not depend on the $b/a$ of galaxies (third column). Therefore, we do not introduce any offset to measured parameters in this study.  

\par 
In summary, the simulations show that we are able to recover median values for structural parameters in all bands. We do not find any significant systematic offset between the measured $R_e$, $n$ or $b/a$ and their input values irrespective of the brightness, size, S\'ersic index and axis ratio for most of our model galaxies ($R_e\lesssim 1^{\prime\prime}$ and/or $n\lesssim 4$, Figure \ref{f:sim-rd-i-o}). The median difference between input and output parameters (red curves) in the simulations is always close to $0$. We also find that the size measurements are more robust than the measurements of S\'ersic index, as reflected in the scatters (blue contours) of relative differences of $R_e$ and $n$. The scatter in the distribution of $\Re_d(R_e)$ is significantly smaller than the scatter in $\Re_d(n)$. Finally, we are able to recover the size of the simulated galaxies even if they are very small ($R_e<0.2''$). 

\par

We incorporate the $\Re_d$ scatter ( blue contours in Figure \ref{f:sim-rd-i-o}) into the measurement uncertainty as systematic uncertainty. We do not estimate the systematic uncertainty in a given parameter as a function of each S\'ersic parameter separately due to covariance between them. We instead analyze the scatter in the relative differences in multi-dimensional bins of S\'ersic parameters. As we do not find any dependence of the $\Re_d$ scatter on the measured axis ratio, we do not include $b/a$ in our uncertainty analysis. The violin plots in Figure \ref{f:violin} show the distribution of $\Re_d (R_e)$ in 3-D bins of $m$, $R_e$ and $n$.
Except for galaxies with high $R_e$ and $n$ ($R_e>0.7^{\prime\prime}$ and $n>3$; bottom right panels), the distribution is approximately Gaussian. 
Hence, we can use the standard deviation (std) of the distributions in Figure~\ref{f:violin} as a quantitative measure of the systematic uncertainties. We consider the $x\times \text{std}(\Re_d(x))$ to be the systematic uncertainty in the parameter $x$, $\sigma_{sys}(x)$. We then estimate the total uncertainty in a measured parameter to be $\sigma^2 = \sigma_{sys}^2+\sigma_{ran}^2$, where $\sigma_{ran}$ is the random uncertainty reported by \textsc{Galfit} for a given galaxy. We note that in most cases the $\sigma_{sys}$ is significantly larger than $\sigma_{ran}$.

\subsection{Size Measurements in Two Rest-frame Wavelengths}
\label{sec:2-wavelength-measurement}
The goal of this work is to perform morphological analysis of galaxies in two rest-frame wavelengths. We choose rest-frame $3000$~\AA\ (UV light) and $5000$~\AA\ (visible light) because they bracket the $4000$~\AA\ break in galaxy spectrum \citep[][]{bruzualStellarPopulationSynthesis2003, conroyModelingPanchromaticSpectral2013}. Because UV light is dominated by the light from young ($<1$~Gyr) massive stars, this light traces the regions of recent star formation activity in galaxies. In contrast, light from the old ($>1$~Gyr) low mass stars contributes significantly to the visible wavelength region of galaxy spectrum. Since low-mass stars represent the bulk of galaxy stellar mass, the visible light traces closely the stellar mass distribution within galaxies.

\par 
Earlier studies generally focus on the rest-frame $5000$~\AA\ that traces the stellar mass distribution in galaxies. 
Furthermore, though many of the studies have multi-wavelength dataset, they often fit galaxy profiles in a single band alone. They first fit a subset of galaxies in multiple bands and estimate colour gradients (i.e., how galaxy size decreases with increasing wavelength). They then use this colour gradient estimation and correct the size measurements of all galaxies in a selected band to obtain size measurement in any desired rest-frame wavelength \citep[e.g.,][]{vanderwel3DHSTCANDELSEvolution2014, kawinwanichakijHyperSuprimeCamSubaru2021}. 

\par 
In contrast, we fit the light profiles of all galaxies in five bands and use this multi-band structural information to estimate the S\'ersic parameters in two rest-frame wavelength regimes. We first estimate the characteristic redshifts at which effective wavelengths of CLAUDS+HSC bands cover these rest-frame wavelengths. As an example, the rest-frame $5000$~\AA\, corresponds to the effective (observed) wavelengths of \textit{r} and \textit{i} bands for galaxies at the characteristic redshift of $z\sim0.235$ and $z\sim0.5422$, respectively. 

\par
For a galaxy at redshift $z$, which falls between the characteristic redshifts $z_a$ and $z_b$ corresponding to bands $a$ and $b$, 
we estimate the rest-frame S\'ersic parameter as
\begin{equation}
    x_{\text{RF}} = w_a x_a + w_b x_b,
    \label{eq:rest-frame-est}
\end{equation}
where $x_{a}$ and $x_b$ are S\'ersic parameter measurements in bands $a$ and $b$ respectively. We add weights $w_a$ and $w_b$ as 
\begin{equation}
    w_a = \left|{\frac{{z-z_a}}{{z_b-z_a}}}\right| \\
    \text{and} \\
    w_b = \left|{\frac{{z-z_b}}{{z_b-z_a}}}\right|.
\end{equation}
If a galaxy is missing S\'ersic parameter measurement in one of the bands (either $a$ or $b$) due to failed fit or magnitude limit, we use the measurement from the available single band.

\subsection{Validation via Comparison with Literature}
\label{sec:Literature-comparison}

\begin{figure*}
    \centering
	\includegraphics[width=2\columnwidth]{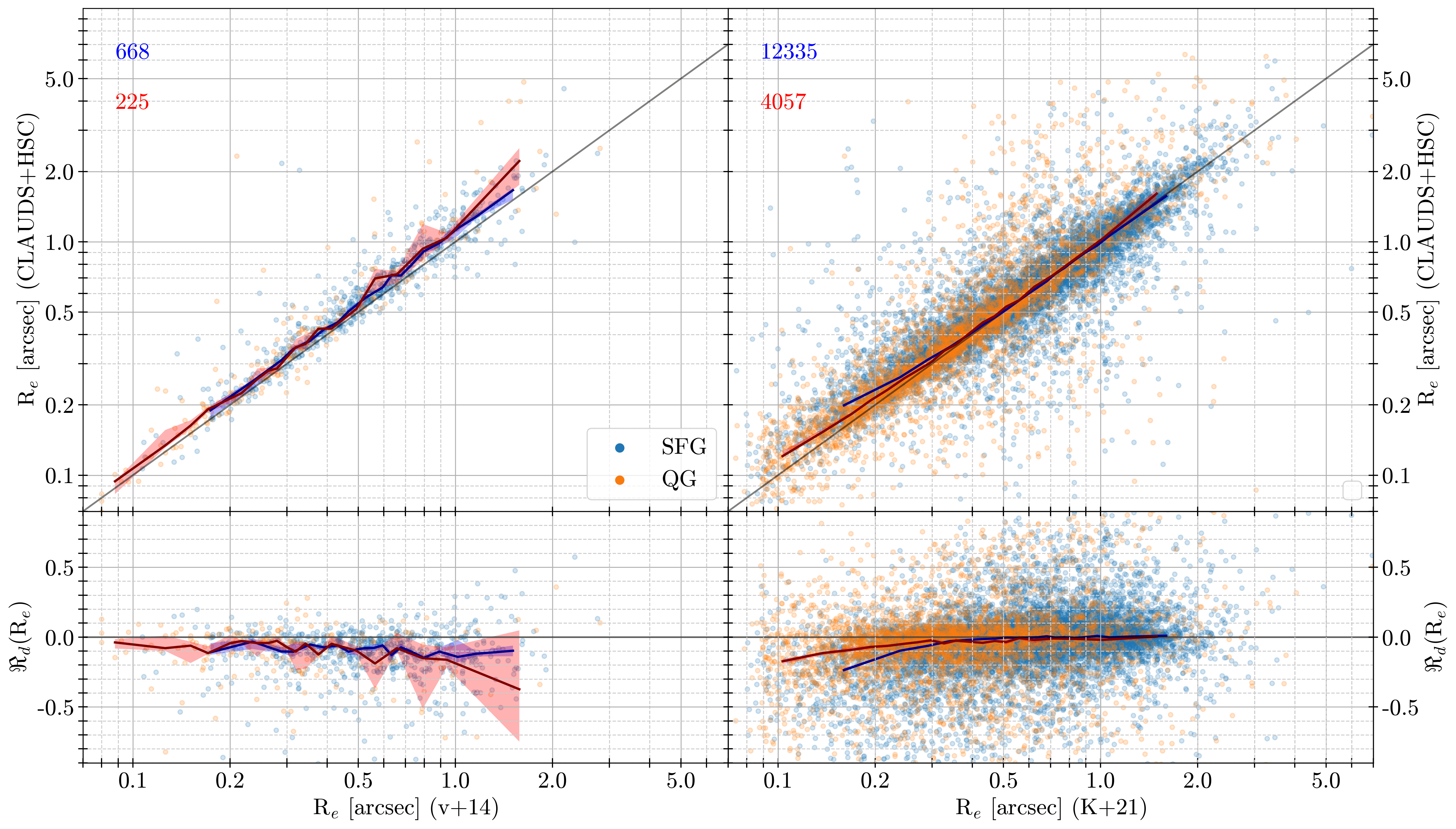} 
    \caption[Comparison between size measurements in the current study and the literature]{Comparison between size measurements in our study and those in the literature. Left panels show the comparison with the size measurements by \citet[v+14]{vanderwel3DHSTCANDELSEvolution2014} and the right panels show the comparison with \citet[K+21]{kawinwanichakijHyperSuprimeCamSubaru2021}, both corrected to rest-frame $5000$~\AA.  We use measurements in \textit{r}, \textit{i} and \textit{z} bands to bracket the rest-frame $5000$~\AA\ depending on the redshift of the galaxy (see Section \ref{sec:2-wavelength-measurement}). Our size measurements are on $y$-axis and those from the literature are on $x$-axis. The blue line shows the median for SFGs and the maroon line is the median relation for QGs. The shaded region indicate the uncertainties in these median sizes. The black line represents the $1:1$ relation. The lower panels show the relative difference in sizes measured as a function of their measurements in the literature. We estimate the relative difference as $\Re_d = (R_e$(literature) - $R_e$(this work$)/$ $R_e$(literature). Similar to the upper panels, the median values with their uncertainties are also shown. There is a good agreement between the measurements in this study and those in the literature.}
    \label{f:van-comp}

\end{figure*}

A number of studies use HST to perform the structural analysis of galaxy light profiles. Since we use ground-based data, our images have (on average) $\sim5$ times larger PSF than HST images. Hence, it is important to investigate how well our ground-based measurements perform when compared to the measurements based on higher-resolution HST data.  

\par
We compare our size measurements with those by \citet{vanderwelStructuralParametersGalaxies2012} based on HST Wide Field Camera 3 (WFC3/IR) images (the two left panels in Figure \ref{f:van-comp}). We perform this comparison on a galaxy-by-galaxy basis because the study by \citet{vanderwelStructuralParametersGalaxies2012} also contains the S\'ersic profile measurements of galaxies in the COSMOS field. Following the prescription described in that study, we correct HST size measurements to the constant rest-frame wavelength of $5000$~\AA. 
In our measurements, we use weighted average of the measurements in two bands that are closest to the rest-frame $5000$~\AA\ at a given redshift (Section \ref{sec:2-wavelength-measurement}).

\par
 
Equivalent to our fitting approach, the measurements by \citet{vanderwelStructuralParametersGalaxies2012} also come from a single S\'ersic profile fitting using \textsc{Galfit}. Because of this similarity in the approaches, a comparison between two size measurements is an independent test of the accuracy of galaxy size measurements based on the ground-based data. A direct comparison on a galaxy-by-galaxy basis shows that the galaxy sizes in our study tend to be on average $\sim5\%$ larger than the measurements based on HST data for both SFGs and QGs (bottom panel in the first column in Figure \ref{f:van-comp}). The fact that ground-based data have worse PSF than the HST data at least partially contributes to the observed offset. At the same time, our data are more sensitive to the low-surface brightness regions than the HST data and therefore our images better incorporate information from outer regions of galaxies. Since the  offset is not very large (within uncertainties of individual measurements) and we do not find any trend in this offset with galaxy size, 
we infer that the size measurements are robust when obtained from the ground-based data with good seeing.

\par 
\citet{kawinwanichakijHyperSuprimeCamSubaru2021} use the same ground-based images as we do (HSC-SSP). 
In the right panel of Figure~\ref{f:van-comp} we compare the two ground-based size measurements, using the size estimates in rest-frame $5000$~\AA\ for the comparison \citep[the only estimates avaiable in][]{kawinwanichakijHyperSuprimeCamSubaru2021}. Although in general we find a good agreement between these two size measurements, there is a significant offset at smaller sizes ($R_e<0.3^{\prime\prime}$): our size measurements are systematically larger than  \citet{kawinwanichakijHyperSuprimeCamSubaru2021}. The offset increases with decreasing galaxy sizes and it is more prominent for SFGs than QGs.

\par 
We speculate that the differences between our size measurements and those of \citet{kawinwanichakijHyperSuprimeCamSubaru2021} arise from a number of differences between the two fitting approaches. \citet{kawinwanichakijHyperSuprimeCamSubaru2021} use \textsc{Lenstronomy} \citep{birrerGravitationalLensModeling2015, birrerLenstronomyMultipurposeGravitational2018} whereas we use \textsc{Galfit}. In contrast to their single-band size measurements corrected to the rest-frame, we estimate the rest-frame $5000$~\AA\ sizes by combining multi-band data that correspond to this rest-frame wavelength directly, without any corrections.
In addition, compared to their cutout images, we select significantly larger cutout images to model the extended profiles of the galaxies. Our simulations show that a large-fitting region is essential to measuring accurately the overall galaxy sizes (Section \ref{sec:cutout}).

%%%%%%%%%%%%%%%%%%%%%%%%%%%%%%%%%%%%%%%%%%%%%%%%%%%%%%%%%%%%%%%%%%
%%%%%%%%%%%%%%%%%%%%%%%%%%%%%%%%%%%%%%%%%%%%%%%%%%%%%%%%%%%%%%%%%%
%%%%%%%%%%%%%%%%%%%%%%%%%%%%%%%%%%%%%%%%%%%%%%%%%%%%%%%%%%%%%%%%%%

\section{Results}
\label{sec:results}

\begin{figure}
    \centering
	\includegraphics[width=0.73\columnwidth]{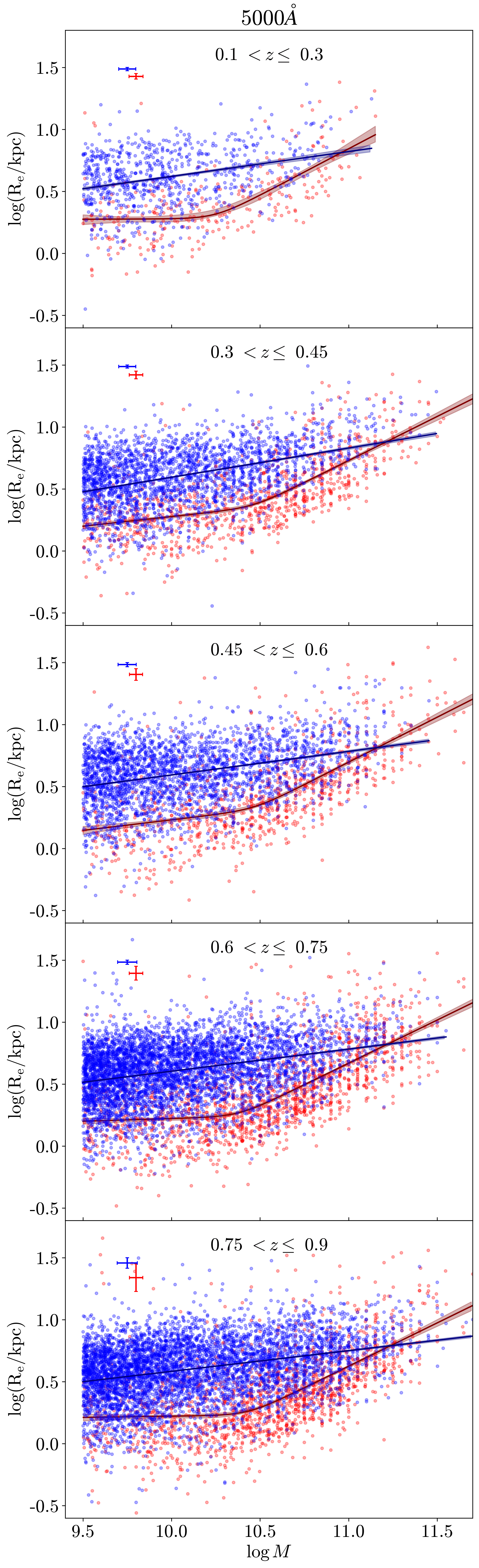} %.png}
    \caption[Fitting double power law in rest-frame $5000$~\AA]{Fitting smoothly broken double power law to the QGs and single power law to the SFGs at rest-frame $5000$~\AA\ in 5 redshift bins. The individual SFGs are shown as blue points; red points represent QGs. The median uncertainties in size measurements for SFGs and QGs are given in the top left corner of each panel. The best-fitting power law relations for SFGs and QGs are shown in navy blue and maroon, respectively, with the uncertainties from Bayesian posterior plotted as shaded region. }
    \label{f:hsc-doube-power-law}
\end{figure}

\begin{table*}
	\begin{center}
		\begin{NiceTabular}{cccccccc}[hvlines]%{|c|c|c|c|c|c|c|c|}
			\hline
                
                \hline
                \Block{1-8}{Rest-frame $3000$~\AA} \\
                \hline

			$z$ & median $z$ & $\alpha$ & $\beta$ & $R_p$ & $\log M_p$ & $\sigma_{\log{R_e}}$ & $R_0$ \\
			 \hline
			$0.1-0.3$ & $0.22\pm0.01$ & $0.05\pm0.14$ & $0.61\pm0.11$ & $2.26\pm0.33$ & $10.16\pm0.17$ & $0.21\pm0.01$ & $4.52\pm0.06$ \\
			
			$0.3-0.45$ & $0.35\pm0.00$ & $0.19\pm0.06$ & $0.65\pm0.08$ & $3.27\pm0.41$ & $10.52\pm0.12$ & $0.23\pm0.01$ & $4.07\pm0.02$ \\
			
			$0.45-0.6$ & $0.53\pm0.00$ & $0.20\pm0.12$ & $0.60\pm0.07$ & $3.18\pm0.56$ & $10.52\pm0.16$ & $0.27\pm0.01$ & $3.91\pm0.03$ \\
			
			$0.6-0.75$ & $0.68\pm0.00$ & $0.12\pm0.08$ & $0.71\pm0.04$ & $2.65\pm0.20$ & $10.49\pm0.07$ & $0.26\pm0.01$ & $3.48\pm0.01$ \\
			
			$0.75-0.9$ & $0.83\pm0.00$ & $0.04\pm0.12$ & $0.59\pm0.06$ & $3.35\pm0.32$ & $10.57\pm0.10$ & $0.28\pm0.01$ & $3.82\pm0.01$ \\
			\hline
                \hline
                
                \hline
                \Block{1-8}{Rest-frame $5000$~\AA} \\
                \hline
                
                \hline
			$z$ & median $z$ & $\alpha$ & $\beta$ & $R_p$ & $\log M_p$ & $\sigma_{\log{R_e}}$ & $R_0$ \\
			\hline
			$0.1-0.3$ & $0.22\pm0.01$ & $0.01\pm0.12$ & $0.75\pm0.14$ & $2.08\pm0.28$ & $10.25\pm0.15$ & $0.22\pm0.01$ & $4.15\pm0.05$ \\
			
			$0.3-0.45$ & $0.36\pm0.00$ & $0.15\pm0.04$ & $0.71\pm0.05$ & $2.38\pm0.19$ & $10.47\pm0.07$ & $0.21\pm0.00$ & $3.28\pm0.01$ \\
			
			$0.45-0.6$ & $0.53\pm0.00$ & $0.17\pm0.06$ & $0.73\pm0.05$ & $2.18\pm0.22$ & $10.47\pm0.09$ & $0.24\pm0.01$ & $3.02\pm0.01$ \\
			
			$0.6-0.75$ & $0.68\pm0.00$ & $0.04\pm0.05$ & $0.70\pm0.04$ & $1.87\pm0.12$ & $10.38\pm0.06$ & $0.24\pm0.00$ & $2.87\pm0.01$ \\
			
			$0.75-0.9$ & $0.84\pm0.00$ & $0.02\pm0.06$ & $0.70\pm0.04$ & $1.83\pm0.12$ & $10.43\pm0.07$ & $0.26\pm0.00$ & $2.60\pm0.00$ \\
			\hline
		\end{NiceTabular}
	\end{center}
	\caption{The best-fit parameters of double power law SMR for QGs in two rest-frame wavelengths: $3000$~\AA\ and $5000$~\AA\ (Equation \ref{eq:dpl}). Here $R_0$ denotes the characteristic size of QGs at $5\times10^{10}$\Msun. Uncertainties smaller than $0.005$ are rounded as $0.00$. Both $R_p$ and $R_0$ are in kpc and $M_p$ is in \Msun.}
	\label{table:DPL}
\end{table*}

\begin{table*}
	\begin{center}
		\begin{NiceTabular}{cccccc}[hvlines]
                \hline
                
                \hline
                \Block{1-6}{Rest-frame $3000$~\AA} \\
                \hline
                
                \hline
			Galaxy & $z$ & median $z$ & $\alpha$ & $R_0$ & $\sigma_{\log{R_0}}$ \\
			\hline
			  \Block{5-1}{SFGs} & 0.1-0.3 & $0.23\pm0.00$ & $0.22\pm0.03$ & $6.35\pm0.30$ & $0.22\pm0.01$ \\
			
			 & 0.3-0.45 & $0.36\pm0.00$ & $0.28\pm0.01$ & $6.49\pm0.15$ & $0.24\pm0.00$ \\
			
			 & 0.45-0.6 & $0.53\pm0.00$ & $0.22\pm0.01$ & $6.02\pm0.14$ & $0.25\pm0.00$ \\
			
			 & 0.6-0.75 & $0.69\pm0.00$ & $0.20\pm0.01$ & $5.95\pm0.09$ & $0.24\pm0.00$ \\
			
			 & 0.75-0.9 & $0.84\pm0.00$ & $0.19\pm0.01$ & $5.83\pm0.08$ & $0.24\pm0.00$ \\
			\hline
			 \Block{5-1}{QGs} & 0.1-0.3 & $0.23\pm0.01$ & $0.58\pm0.08$ & $4.57\pm0.23$ & $0.20\pm0.01$ \\
			
			 & 0.3-0.45 & $0.37\pm0.00$ & $0.61\pm0.05$ & $4.14\pm0.13$ & $0.21\pm0.01$ \\
		
			 & 0.45-0.6 & $0.53\pm0.00$ & $0.59\pm0.06$ & $3.94\pm0.17$ & $0.25\pm0.01$ \\
			
			 & 0.6-0.75 & $0.69\pm0.00$ & $0.70\pm0.03$ & $3.48\pm0.09$ & $0.25\pm0.01$ \\
			
			 & 0.75-0.9 & $0.84\pm0.00$ & $0.59\pm0.04$ & $3.77\pm0.13$ & $0.26\pm0.01$ \\
			\hline
                \hline
                
                \hline
                \Block{1-6}{Rest-frame $5000$~\AA} \\
			\hline
                
                \hline
                Galaxy & $z$ & median $z$ & $\alpha$ & $R_0$ & $\sigma_{\log{R_0}}$ \\
			\hline
			  \Block{5-1}{SFGs} & 0.1-0.3 & $0.23\pm0.00$ & $0.20\pm0.02$ & $5.76\pm0.23$ & $0.21\pm0.01$ \\
			
			 & 0.3-0.45 & $0.36\pm0.00$ & $0.24\pm0.01$ & $5.74\pm0.11$ & $0.21\pm0.00$ \\
			
			 & 0.45-0.6 & $0.53\pm0.00$ & $0.19\pm0.01$ & $5.32\pm0.09$ & $0.22\pm0.00$ \\
			
			 & 0.6-0.75 & $0.69\pm0.00$ & $0.18\pm0.01$ & $5.35\pm0.07$ & $0.22\pm0.00$ \\
			
			 & 0.75-0.9 & $0.84\pm0.00$ & $0.17\pm0.01$ & $5.01\pm0.06$ & $0.22\pm0.00$ \\
			\hline
			  \Block{5-1}{QGs} & 0.1-0.3 & $0.25\pm0.01$ & $0.73\pm0.08$ & $4.18\pm0.19$ & $0.21\pm0.01$ \\
			
			 & 0.3-0.45 & $0.37\pm0.00$ & $0.67\pm0.04$ & $3.36\pm0.08$ & $0.21\pm0.01$ \\
			
			 & 0.45-0.6 & $0.53\pm0.00$ & $0.69\pm0.04$ & $3.10\pm0.09$ & $0.23\pm0.01$ \\
			
			 & 0.6-0.75 & $0.69\pm0.00$ & $0.68\pm0.03$ & $2.90\pm0.06$ & $0.24\pm0.01$ \\
			
			 & 0.75-0.9 & $0.84\pm0.00$ & $0.67\pm0.03$ & $2.64\pm0.05$ & $0.24\pm0.01$ \\
                \hline 

                \hline 
			
		\end{NiceTabular}
	\end{center}
	\caption{The best-fitting parameters from the single power law SMR for SFGs and QGs in two rest-frame wavelengths (Equation \ref{eq:smr}). Uncertainties smaller than $0.005$ are rounded as $0.00$. The unit of $R_0$ is in kpc.}
	\label{table:SPL}
\end{table*}

\subsection{Size-Stellar Mass Relation}
\label{sec:SMR}

Observations show that galaxies exhibit a relation between their stellar mass and size(SMR) at least up to $z\sim3$ \citep[e.g.,][]{shenSizeDistributionGalaxies2003, trujilloLuminositySizeMassSizeRelations2004, guoStructuralPropertiesCentral2009, williamsEvolvingRelationsSize2010, newmanCanMinorMerging2012,  vanderwel3DHSTCANDELSEvolution2014, langeGalaxyMassAssembly2015,  faisstConstraintsQuenchingMassive2017, royEvolutionGalaxySizestellar2018, mowlaCOSMOSDASHEvolutionGalaxy2019, matharuHSTWFC3Grism2019, matharuHSTWFC3Grism2020}.  The form of the SMR differs between the SFGs and QGs 
\citep[e.g.,][]{ vanderwel3DHSTCANDELSEvolution2014,  mowlaCOSMOSDASHEvolutionGalaxy2019, kawinwanichakijHyperSuprimeCamSubaru2021}. 

\par
SFGs exhibit a linear relation between galaxy size and stellar mass in log-log space: 
\begin{equation}
    \log R_e = \log R_0 + \alpha \log \left(\frac{M}{M_0}\right), 
    \label{eq:smr}
\end{equation}
where $\alpha$ is the slope and $R_0$ is the characteristic size at the fiducial mass, $M_0$ \citep[e.g.,][]{vanderwel3DHSTCANDELSEvolution2014, kawinwanichakijHyperSuprimeCamSubaru2021}. In line with the literature, we adopt $M_0=5\times10^{10}$\Msun.
Figure \ref{f:hsc-doube-power-law} illustrates the SMR at rest-frame $5000$~\AA\ for SFGs in our sample divided into five redshift bins (with data points in blue and the best fit relation as blue-shaded line). In Appendix \ref{appendix:SMR}, we show the SMR for our SFGs at rest-frame $3000$~\AA. 

\par
However, QGs have a more complex SMR where the slope changes at the pivot mass $\sim 3\times10^{10}$\Msun\ \citep{langeGalaxyMassAssembly2015,mowlaMassdependentSlopeGalaxy2019, moslehGalaxySizesPerspective2020, kawinwanichakijHyperSuprimeCamSubaru2021,nedkovaExtendingEvolutionStellar2021, damjanovSizeSpectroscopicEvolution2023}. Hence, to explore the trend in size with stellar mass for QGs, we first fit a smoothly broken double power law, 
\begin{equation}
    R(M) = R_p \left(\frac{M}{M_p}\right)^\alpha \left[ \frac{1}{2} {\left(1+ \frac{M}{M_p}\right)^\delta }\right]^{\frac{\beta-\alpha}{\delta}},
\label{eq:dpl}
\end{equation}
where $M_p$ is the pivot stellar mass 
at which the slope changes, $R_p$ is the effective radius at the pivot stellar mass, $\alpha$ is the slope of the SMR at the low-mass end, $\beta$ is the slope at the high-mass end, and $\delta$ is the smoothing factor. Following \citet{mowlaMassdependentSlopeGalaxy2019}, we adopt $\delta=6$ to reduce degeneracy between $\delta$ and the slopes. 

\par 
We fit the SMR for both SFGs and QGs using \textsc{Dynesty}\footnote{\url{https://zenodo.org/record/7832419}}, a \textsc{Python} package designed to fit the data by implementing Bayesian inference \citep{speagleDYNESTYDynamicNested2020}. We 
follow the fitting procedure from \citet[see their Section 3]{vanderwel3DHSTCANDELSEvolution2014}. The differences between our approach and the method of \citet{vanderwel3DHSTCANDELSEvolution2014} are the use of the double power law for QGs and how total likelihood is estimated in our study. \citet{vanderwel3DHSTCANDELSEvolution2014} introduce a $10\%$ contamination in the dataset due to misclassification of SFGs and QGs. Since we use UVJ classification where the separation criteria varies with redshift, we do not account for contamination in this study. Additionally, they also allow for $1\%$ outliers (objects that should not be in the data) in the likelihood estimation.

\par 
Our likelihood function is of the form
\begin{gather}
    \mathcal{L} = \ln{(wp)},
\end{gather}
where $w$ is a weighting factor and $p$ is the probability of observing the given size for a galaxy of mass $M_*$ as described by \citet[their Equation 4]{vanderwel3DHSTCANDELSEvolution2014}.  The weighting factor,  $w$, is  inversely proportional to the galaxy number density at a given stellar mass taken from the stellar mass functions of \citet{muzzinEvolutionStellarMass2013}. By applying this weight, we ensure that each stellar mass range contributes equally to the SMR fit. 

\par
We first fit the double power-law to the SMR of QGs. Figure \ref{f:hsc-doube-power-law} shows (in red) the distribution of galaxies in this parameter space and the best-fit relations in five different redshift bins in the rest-frame $5000$~\AA\  (see Appendix \ref{appendix:SMR} for the results in rest-frame $3000$~\AA). It is clear that QGs have shallower SMR slopes at stellar masses below the pivot mass compared to those of more massive QGs. The SMR for the low mass QGs is horizontal within $2\sigma$ (i.e., $\alpha\sim0$). These shallow slopes of low mass QGs ($M<M_p$) are also consistent, within their uncertainties, with the results of \citet{kawinwanichakijHyperSuprimeCamSubaru2021} and \citet{nedkovaExtendingEvolutionStellar2021}. However, the QGs show a steep SMR above the pivot mass, $M_p(z)$. This pivot mass is around  $\log M \sim 10.5$, except in the lowest redshift bin where it drops to $\sim10.3$. In addition, the high-mass end slopes $\beta$ do not change significantly across the full redshift interval.
Table \ref{table:DPL} gives the best fit parameters of the double power law fitting of QGs in two rest-frame wavelengths. 

\par
Once we estimate the pivot point, we fit the SMR for the QGs with stellar masses larger than $M_p(z)$ and SFGs that are more massive than $10^{9.5} M_\odot$ (mass completeness limit; Section \ref{sec:sample-selection}) using the single-power law from Equation \ref{eq:smr}. We use the Equation \ref{eq:smr} for QGs to be able to directly compare our findings with the results from the literature (Section~\ref{sec:size-evo-visible}). This approach captures the nature of SMR for QGs when we limit our analysis to very massive QGs (above the pivot point)\footnote{We note that although we fit the QGs in two ways (double power law given in Table \ref{table:DPL} and single power law in Table \ref{table:SPL}), the results for massive QGs are very similar.}.  The SMR fitting procedure is the same as above.  
We fit the SMR for SFGs and QGs separately without considering any cross-contamination between the two populations. Unlike \citet{vanderwel3DHSTCANDELSEvolution2014}, we use moving UVJ separation line (with $z$) which further reduces contamination \citep{whitakerNEWFIRMMediumbandSurvey2011}. The best-fit parameters are given in Table \ref{table:SPL}. 

\begin{figure}
    \centering
	\includegraphics[width=0.73\columnwidth]{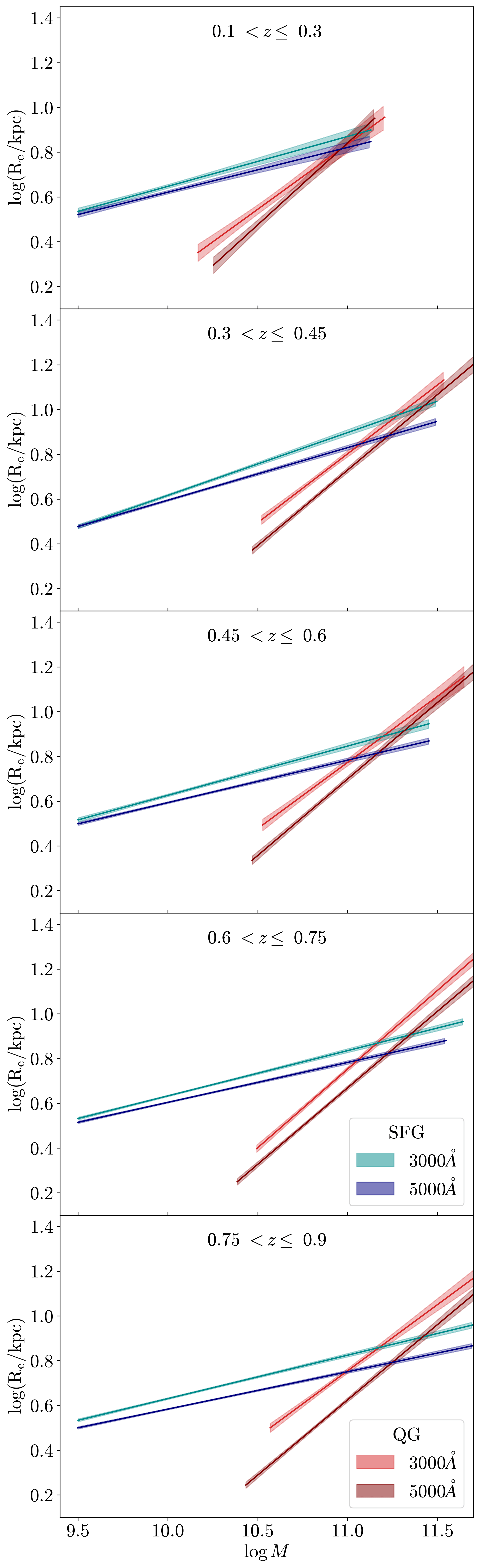} %.pdf}
    \caption[Fitting SMR - 3000A and 5000A]{The best-fitting single power-law SMR for SFGs and massive QGs in two rest-frame wavelengths ($3000$~\AA\ and $5000$~\AA) in five redshift bins. The shaded regions represent the fit uncertainties estimated from the Bayesian posterior for SFGs and QGs respectively. The SMR fits for SFGs and QGs in the shorter wavelength are plotted in cyan and red, respectively. The fits for SFGs and QGs in the longer wavelengths are shown in navy blue and maroon, respectively.}
    \label{f:smr-3000&5000A}
\end{figure}

\par 
Figure \ref{f:smr-3000&5000A} 
shows the resulting SMR in the rest-frame $5000$~\AA\ for SFGs and QGs  (in navy blue and maroon, respectively) separated into five redshift bins. This figure includes the best-fit SMRs for SFGs from Figure \ref{f:hsc-doube-power-law}. For  QGs, the Figure \ref{f:smr-3000&5000A} shows the best-fitting single power-law for massive QGs above the pivot point ($M>M_p$). 

\par
The results are mostly consistent with previous works. Except at very high masses, the sizes of SFGs are larger than those of QGs at a given stellar mass as shown previously by 
\citet{shenSizeDistributionGalaxies2003}, \citet{williamsEvolvingRelationsSize2010}, \citet{vanderwel3DHSTCANDELSEvolution2014}, \citet{moslehGalaxySizesPerspective2020}, \citet{kawinwanichakijHyperSuprimeCamSubaru2021} and others. At very high masses ($\log M\gtrsim11.2$), we find that the average size of QGs becomes comparable to that of SFGs.  
The slopes of the SMR for SFGs ($\alpha\sim 0.2$) and massive QGs ($\alpha\sim 0.7$) are similar to those in the literature \citep{vanderwel3DHSTCANDELSEvolution2014, mowlaCOSMOSDASHEvolutionGalaxy2019, kawinwanichakijHyperSuprimeCamSubaru2021, nedkovaExtendingEvolutionStellar2021}.

\par 
Figure \ref{f:smr-3000&5000A} also shows the resulting best-fit SMRs in the rest-frame $3000$~\AA\ for both SFGs and QGs  (in cyan and red, respectively; see Figure \ref{f:hsc-doube-power-law-3000A} for the distribution of individual data points in this parameter space). 
Except for the clear offset in the zero-points, the results in the rest-frame $3000$~\AA\ are also very similar to the results in the rest-frame  $5000$~\AA.  
For stellar masses $\log M\lesssim11.2$, the SFGs are, on average, larger than QGs. The slopes of the SFGs are comparable between the two wavelengths ($\alpha\sim0.2$). At the same time, the SMR for QGs is systematically $\sim10\%$ steeper in the red light than in the blue light, although this difference is within $2\sigma$ uncertainty on the slope estimates. We return to the difference in the SMR zero points for two rest-frame wavelengths in Section~\ref{sec:size-evo-uv}.

\subsection{Median Size Evolution in the Rest-frame Visible Light}
\label{sec:size-evo-visible}

\begin{figure*}
    \centering
	\includegraphics[scale=0.4]{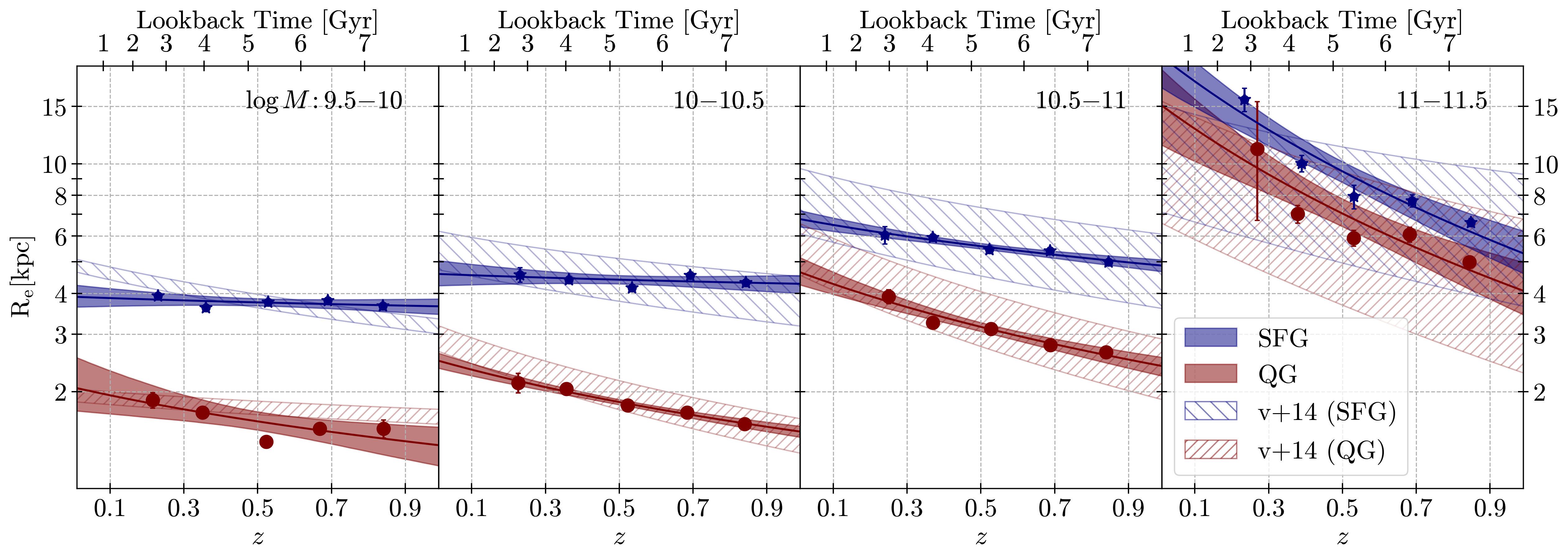}%.png
	
    \caption[Size evolution with v14]{Evolution of the median galaxy sizes at fixed stellar mass in rest-frame $5000$~\AA\ for galaxies segregated in four mass bins. Mass bins are noted in the top right corner of each panel. SFG population is represented by blue colour and the QG population is shown in red. The median sizes in five redshift bins are shown with stars for SFGs and with circles for QGs along with their bootstrapped error bars. The size evolution is represented  by a power-law, $A(1+z)^\beta$ (solid curves with uncertainties from Bayesian posterior as shaded region). For comparison, we show the results from \citet[][v+14]{vanderwel3DHSTCANDELSEvolution2014} as well. Their best fitting evolution curves for SFGs and QGs  at rest-frame $5000$~\AA\ are shown in blue and red hatched regions respectively. In general, there is a good agreement between our results and those of v+14.
    }
    \label{f:size-evo-v}
\end{figure*} 

\begin{figure*}
    \centering
    \includegraphics[scale=0.4]{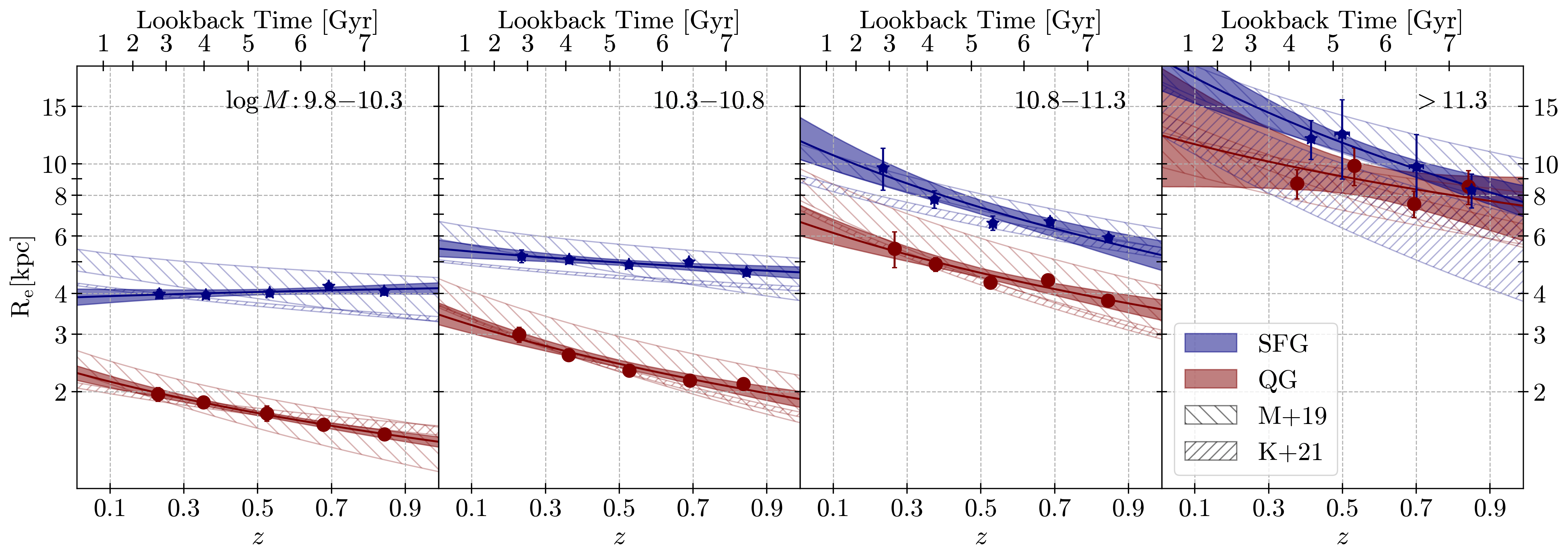} %.png}
	
    \caption[Size evolution with M19 and K21]{Comparison of the size evolution at fixed stellar mass for SFGs and QGs we trace in four stellar mass bins with the results of \citet[][M+19]{mowlaCOSMOSDASHEvolutionGalaxy2019} and \citet[][K+21]{kawinwanichakijHyperSuprimeCamSubaru2021}. The size evolution reported by M+19 and K+21 are shown as hatched regions. Rest of the details in the figure are the same as in Figure \ref{f:size-evo-v}. In general, there is a good agreement between our results and the literature.}
    \label{f:size-evo-KM}
\end{figure*} 

\begin{figure*}
    \centering
    \includegraphics[scale=0.4]{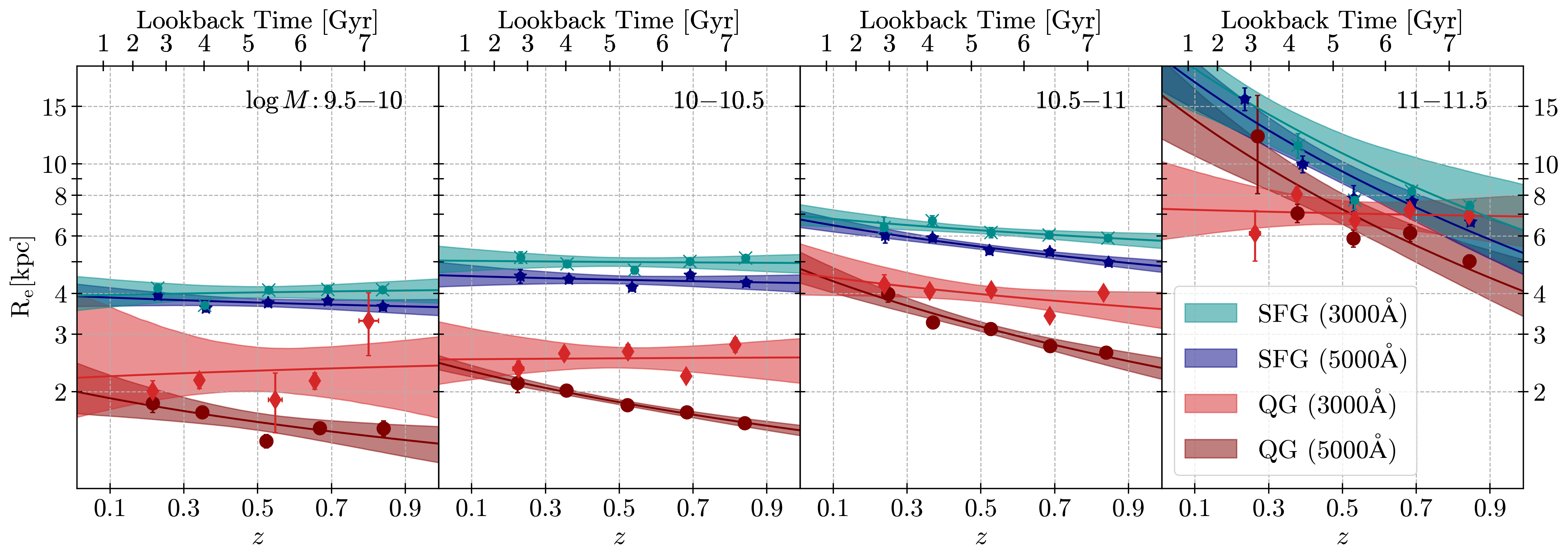} %.pdf}

    \caption[Median size evolution in two rest-frames]{Comparison of the median size evolution for SFGs and QGs at fixed mass between rest-frames UV ($3000$~\AA) and visible ($5000$~\AA) light.  
    As in Figure \ref{f:size-evo-v}, panels show the median size evolution in various mass bins as denoted in the top right corner of each panel. Cyan and navy blue represent SFG population in rest-frames $3000$~\AA\ and $5000$~\AA\ respectively. Similarly, red and maroon colours denote QG population in rest-frame $3000$~\AA\ and $5000$~\AA\ respectively. The best fitting power-law functions (solid curves) for the size evolution of SFGs and QGs are shown with their uncertainties (shaded regions). Galaxies are more extended in shorter wavelength than in longer wavelength.}
    \label{f:size-evo-2wave}
\end{figure*}

We further employ galaxy size measurements to explore the trend in median size with redshift for SFGs and QGs segregated by stellar mass. First, we use median galaxy sizes to explore galaxy size evolution in the rest-frame wavelength of $5000$~\AA.

\par 
Figure \ref{f:size-evo-v} shows the median sizes of SFGs and QGs in this rest-frame wavelength as a function of redshift in four different stellar mass bins. Within each bin, the stellar mass range is fixed across the full redshift interval. 
In addition to SFGs having larger sizes than QGs, median sizes of both galaxy populations at a given fixed stellar mass generally grow with cosmic time. 
\par
We fit the redshift evolution in size of SFGs and QGs with a power law, 
\begin{equation}
\label{eq:evol}
    R_e = R_e^{0} (1+z)^{\beta}, 
\end{equation}
where $R_e^0$ is $R_e$ at $z=0$. When traced in the rest-frame visible light, the pace of size evolution differs between SFGs and QGs (blue vs. red solid lines in Figure~\ref{f:size-evo-v}). In all mass bins below $\log M\sim11$, sizes of QGs evolve faster than those of SFGs. In the mass range $10.5<\log M<11$ and over the redshift interval $0.1<z<0.9$, the median sizes of QGs grow as $R_e \varpropto (1+z)^{-1.0}$ whereas the sizes of SFGs grow as $R_e \varpropto (1+z)^{-0.5}$. For example, the average sizes of $\log M\sim10.75$ QGs grow by $\sim73\%$ in the time interval of 6.1 Gyrs. 
In contrast, similarly massive SFG grows in size by only $\sim31\%$ over the same time interval. 

\par
Furthermore, for both galaxy populations, the pace at which galaxies grow is faster for more massive systems. 
The growth rate $\beta=\frac{d\log{(R_e/\mathrm{kpc})}}{d\log(1+z)}$ for SFGs ranges from $0.10\pm0.18$ for the least massive to $-2.07\pm0.45$ for the most massive SFGs (blue lines in Figure \ref{f:size-evo-v}).
Over the same mass range, the growth rate for QGs changes from $\beta=-0.55\pm0.4$ to $\beta=-1.94\pm0.64$ (red lines in Figure \ref{f:size-evo-v}). QGs grow significantly in size at $z<1$ even at lower masses ($\log M < 10.5$). In contrast, the size evolution of SFGs at these low masses is, within uncertainties, consistent with no growth.

\par  
We note that these growth rates are the median trends for galaxy populations, not individual galaxies. Furthermore, for now we do not consider the fact that stellar mass of SFGs also grows through star formation. Finally, here we do not account for the effect of progenitor bias among QG population. We estimate the amplitude of the effects that mass growth and progenitor bias have on the evolution in average size of SFGs and QGs in Section \ref{sec:discussion}.

\par 
Our findings are in good agreement with results reported in the literature. In Figure \ref{f:size-evo-v}, we compare our results (solid curves and shaded regions) with those from \citet[hatched region]{vanderwel3DHSTCANDELSEvolution2014} at rest-frame $5000$~\AA. Like us, they have fitted galaxy images using \textsc{Galfit}.
However, their fits are based on galaxy images taken with WFC3 onboard HST, which has at least 4 times better resolution than 
our ground-based HSC images (see Section \ref{sec:Literature-comparison}). Furthermore, we note that their best-fitting relation between median size and redshift spans the redshift interval $0<z<3$. Finally, they only have two redshift bins at $z<1$ while we have a finer sampling of this redshift range with $5$ bins. Despite these differences, we find an excellent agreement between our trends and results reported in \citet{vanderwel3DHSTCANDELSEvolution2014}.

\par
In Figure \ref{f:size-evo-KM} we compare our results (solid lines with shaded regions) with \citet[backward hatched regions]{mowlaCOSMOSDASHEvolutionGalaxy2019} and \citet[forward hatched regions]{kawinwanichakijHyperSuprimeCamSubaru2021}. 
\citet{mowlaCOSMOSDASHEvolutionGalaxy2019} follow a similar methodology as ours: they fit the galaxy light profiles using \textsc{Galfit}. 
However, their measurements are based on the same HST imaging as the ones reported in \citet{vanderwel3DHSTCANDELSEvolution2014} and thus do not suffer from atmospheric seeing effects. 
Furthermore, they measure galaxy size growth over the redshift interval $0<z<3$, with only two redshift bins below $z\sim1$. 
At the same time, \citet{kawinwanichakijHyperSuprimeCamSubaru2021} use a ground-based HSC dataset similar to ours, but follow a different methodology (see Section \ref{sec:Literature-comparison} for details). 

\par
Except for the marginal difference for SFGs in the lowest mass bin (first panel of Figure \ref{f:size-evo-KM}), the agreement between our results and the results of  \citet{mowlaCOSMOSDASHEvolutionGalaxy2019} is excellent. With our dataset that covers larger area of the COSMOS field (Section~\ref{sec:clauds}), we are able to obtain a narrower confidence interval (shaded regions in Figure \ref{f:size-evo-KM}). 

\par 
In general, our median size evolution estimates for QGs are in agreement with \citet{kawinwanichakijHyperSuprimeCamSubaru2021}, but we do find some discrepancy between the measurements for SFGs. Overall, the median SFG size measurements of \citet[][blue forward hatched regions in Figure \ref{f:size-evo-KM}]{kawinwanichakijHyperSuprimeCamSubaru2021} are smaller than our measurements and those of \citet{mowlaCOSMOSDASHEvolutionGalaxy2019}. 
This offset could could be a combination of several effects.
As we discuss in Section \ref{sec:Literature-comparison}, the two ground-based studies (this study and that of \citet{kawinwanichakijHyperSuprimeCamSubaru2021}) use different methodology to fit galaxy light profiles. In addition, we select significantly larger cutout images to model the extended galaxy profiles. Furthermore, we have much better estimation of photo-$z$'s and UVJ based SFG-QG classification thanks to $U+grizy+JHK$ data from COSMOS2020, whilst \citet{kawinwanichakijHyperSuprimeCamSubaru2021} use only HSC $grizy$ data to segregate galaxies into SFGs and QGs. On the other hand, our study is somewhat affected by cosmic variance because our dataset covers almost two orders of magnitude smaller area on the sky (1.6~deg$^2$ of the COSMOS field vs $100$~deg$^2$ of the HSC-SSP analyzed in \citealt{kawinwanichakijHyperSuprimeCamSubaru2021}). Using the Equation $3$ from \citet{driverQuantifyingCosmicVariance2010}, we estimate that the cosmic variance will decrease by a factor of $4$ when we expand to the full CLAUDS+HSC survey area (from $12\%$ to $3\%$).

\par
The slower pace of evolution for SFGs compared to QGs for galaxies with $\log M<11$ in the rest-frame $5000$~\AA\ is consistent with the suite of previous studies  \citep[e.g.,][]{lillyHubbleSpaceTelescope1998, ravindranathEvolutionDiskGalaxies2004, bardenGEMSSurfaceBrightness2005, vanderwel3DHSTCANDELSEvolution2014, mowlaCOSMOSDASHEvolutionGalaxy2019, kawinwanichakijHyperSuprimeCamSubaru2021}. Additionally, similar to our findings, \citet{vanderwel3DHSTCANDELSEvolution2014}, \citet{mowlaCOSMOSDASHEvolutionGalaxy2019} and \citet{kawinwanichakijHyperSuprimeCamSubaru2021} find that the rate of median size evolution for both QGs and SFGs depends on galaxy stellar mass. 
This comparison confirms that the  resolution of the HSC+CLAUDS imaging dataset is well suited for detailed probes of galaxy size evolution up to $z\sim1$.

\subsection{Median Size Evolution in Rest-frame UV Light}
\label{sec:size-evo-uv}

In addition to the rest-frame visible light, we also investigate the size evolution of galaxies in rest-frame UV ($3000$~\AA). The comparison of galaxy sizes in two rest-frame wavelengths using a large dataset is unique to our work.
These two rest-frame wavelengths ($3000$ and $5000$~\AA) represent light coming from young and old stars (see Section \ref{sec:2-wavelength-measurement}). 
Hence, comparisons of galaxy morphology in these two rest-frame wavelengths can advance our understanding of the connection between galaxy structural change (i.e., size growth) and the evolution in their stellar content. 

\par 
We combine the median evolutionary trends in galaxy size at fixed stellar mass in rest-frame UV and visible light for both SFGs and QGs  in Figure \ref{f:size-evo-2wave}. The median sizes of SFGs and QGs in UV are represented by cyan crosses and red diamonds respectively and those in visible light are in blue stars and maroon circles. We also show the best-fit size evolution trends (solid curves) along with uncertainties (shaded area).

\par On average, we find that the two galaxy popoulations are larger in rest-frame $3000$~\AA\  than in $5000$~\AA. For example, the median size of SFGs with stellar mass $\log M\sim10.5$ at $z\sim0.5$ is $12\%$ larger in rest-frame blue light than in red light (cyan and blue curves in the second panel of Figure~\ref{f:size-evo-2wave}). The sizes of QGs display even larger difference when measured in two rest-frame wavelengths. At the same mass and redshift ($\log M\sim10.5$,  $z\sim0.5$) QGs are, on average, $33\%$ more extended in the observations at shorter rest-frame wavelength ($3000$~\AA) than in $5000$~\AA\ (red and maroon curves in the second panel of Figure~\ref{f:size-evo-2wave}).

\par Several evolutionary trends in galaxy size at the rest-frame wavelegth of $3000$~\AA\ are similar to those we observe at the rest-frame wavelength of $5000$~\AA.   
Firstly, the median UV-based sizes of both SFGs and QGs evolve with cosmic time. At stellar mass range of $10.8<\log M<11$, the median UV-based sizes of SFGs grow by $\sim50\%$ over the span of $\sim6.1$ Gyrs ($z=0.9$ to $z=0.1$) whilst sizes of similarly massive QGs grow by $21\%$ in the same rest-frame regime and during the same cosmic period. However, the pace of evolution in rest-frame $3000$~\AA\ is slower than in $5000$~\AA. For comparison, in the rest-frame visible light SFGs of similar mass  grow by $90\%$ over the same redshift interval; based on their visible light profiles, similarly massive QGs grow by $78\%$ since $z\sim0.9$.

\par
Finally, the trend in the pace of size evolution  with galaxy mass in the rest-frame $3000$~\AA\ is similar to the trend we observe at $5000$~\AA\ except for QGs in the most massive bin ($\log M>11$).  SFGs in the most massive bin  grow by $\sim165\%$ whilst at lowest mass bin we do not find any significant evidence for size evolution of either SFG or QG population. However, unlike the rest-frame visible light, the most massive QGs in rest-frame UV do not show any significant signs of growth in median size since $z\sim1$. Nevertheless, we note that our sample size is small at these higher mass end and the uncertainties are large.

\par

\par

\begin{table}
	\begin{center}
		\begin{NiceTabular}{cccc}[hvlines]

                \hline
                
                \hline
                \Block{1-4}{Rest-frame $3000$~\AA} \\
                \hline
                
                \hline
			Galaxy & $\log M$ & $R_e^0$ & $\beta$  \\
			\hline
			  \Block{4-1}{SFGs} & $9.7$ & $3.62\pm0.39$ & $0.03\pm0.24$  \\
			
			  & $10.2$ & $4.95\pm0.27$ & $-0.11\pm0.12$  \\
			
			  & $10.7$ & $6.83\pm0.40$ & $-0.26\pm0.13$  \\
			
			  & $11.2$ & $9.57\pm1.17$ & $-0.45\pm0.27$  \\
			
			\hline
			 \Block{2-1}{QGs} & $10.7$ & $5.06\pm0.64$ & $-0.59\pm0.28$  \\
			
			  & $11.2$ & $9.68\pm0.58$ & $-0.44\pm0.14$  \\

			\hline
                \hline
                
                \hline
                \Block{1-4}{Rest-frame $5000$~\AA} \\
			\hline
                
                \hline
                Galaxy & $\log M$ & $R_e^0$ & $\beta$ \\
			\hline
			  \Block{4-1}{SFGs} & $9.7$ & $3.59\pm0.30$ & $-0.08\pm0.19$  \\
			
			  & $10.2$ & $4.75\pm0.28$ & $-0.21\pm0.13$  \\
			
			  & $10.7$ & $6.28\pm0.38$ & $-0.35\pm0.14$  \\
			
			  & $11.2$ & $8.37\pm0.76$ & $-0.50\pm0.20$  \\

			\hline
			  \Block{2-1}{QGs} & $10.7$ & $5.07\pm0.58$ & $-1.10\pm0.26$  \\
			
			  & $11.2$ & $11.95\pm1.96$ & $-1.24\pm0.36$  \\

                \hline 

                \hline 
			
		\end{NiceTabular}
	\end{center}
	\caption{The best-fit parameters for the redshift evolution of characteristic sizes at fixed stellar mass for SFGs and QGs in two rest-frame wavelengths ($R_e = R_e^{0} (1+z)^\beta$). The best-fit curves are displayed in Figures \ref{f:size-evo-sfg} and \ref{f:size-evo-qg} for SFGs and QGs respectively.}
	\label{table:SEvol}
\end{table}

\begin{table}
	\begin{center}
		\begin{NiceTabular}{cccc}[hvlines]

                \hline
                
                \hline
                \Block{1-4}{SFGs} \\
                \hline
                
                \hline
			Restframe $\lambda$ & $\log M_{(z=0.9)}$ & $R_e^0$ & $\beta$  \\
			\hline
			  \Block{4-1}{$3000$~\AA} & $9.7$ & $7.00\pm0.44$ & $-0.91\pm0.14$  \\
			
			  & $10.2$ & $8.06\pm0.63$ & $-0.81\pm0.17$  \\
			
			  & $10.7$ & $9.72\pm1.15$ & $-0.78\pm0.27$  \\
			
			  & $11.2$ & $12.21\pm2.03$ & $-0.80\pm0.37$  \\
			
			\hline
			  \Block{4-1}{$5000$~\AA} & $9.7$ & $6.42\pm0.45$ & $-0.91\pm0.16$  \\
			
			  & $10.2$ & $7.25\pm0.55$ & $-0.82\pm0.17$  \\
			
			  & $10.7$ & $8.52\pm0.80$ & $-0.79\pm0.21$  \\
			
			  & $11.2$ & $10.37\pm1.27$ & $-0.82\pm0.28$  \\

                \hline 

                \hline 
			
		\end{NiceTabular}
	\end{center}
	\caption{The best-fit parameters of redshift evolution of characteristic sizes of SFGs with evolving mass as described in Section \ref{sec:sfg-mevol}. The corresponding evolutionary curves are displayed in Figure \ref{f:size-evo-mevol}.}
	\label{table:SEvol-mevol}
\end{table}

%%%%%%%%%%%%%%%%%%%%%%%%%%%%%%%%%%%%%%%%%%%%%%%%%%%%%%%%%%%%%%%%%%
%%%%%%%%%%%%%%%%%%%%%%%%%%%%%%%%%%%%%%%%%%%%%%%%%%%%%%%%%%%%%%%%%%
%%%%%%%%%%%%%%%%%%%%%%%%%%%%%%%%%%%%%%%%%%%%%%%%%%%%%%%%%%%%%%%%%%

\section{Discussion}
\label{sec:discussion}
\par

So far we have used the median sizes in several mass bin to trace the size evolution of galaxies in our dataset. Although this approach provides direct comparison with the results based on median size measurements from other studies, it does not formally take into account the uncertainties present in the individual size and mass measurements. At the same time, the SMR fitting procedure incorporates these measurement uncertainties and applies weighting based on galaxy stellar mass functions (Section~\ref{sec:SMR}). 
Therefore, we decide to use the size estimates from the best-fitting SMRs for the interpretation of our results. Our conclusions remain the same when we replace the estimates from the best-fit SMRs with the measured median sizes in bins of galaxy stellar mass.

\par
In addition, we also compare this size evolution in two rest-frame wavelengths: $3000$~\AA \, and $5000$~\AA. The analysis of the size growth in two rest-frame wavelengths enables us to trace the distribution of young and old stellar populations in galaxies. We list the parameters of the best-fit function for the redshift evolution of SFG and QG sizes in both wavelengths in Table \ref{table:SEvol}.

\begin{figure*}
    \centering
    
    \includegraphics[scale=0.4]{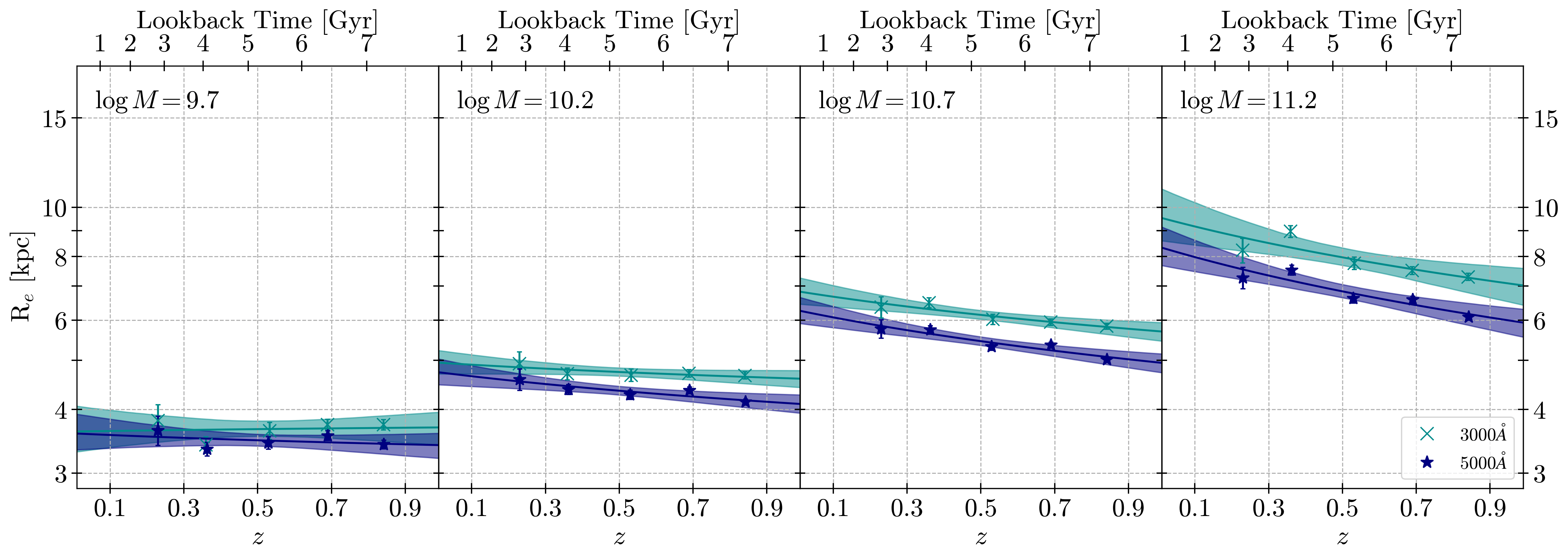} %.pdf}

    \caption[Size evolution of SFGs]{Evolution of characteristic sizes of SFGs at a given fixed stellar mass. Panels show the size evolution of SFGs for a range of characteristic masses (as labeled). The SFG sizes at rest-frame $5000$~\AA\, are shown in navy blue and at rest-frame $3000$~\AA\, in cyan.  The sizes at shorter wavelength are larger than those at longer wavelengths. The size evolution is fit by a power-law $R_e = R_e^{0} (1+z)^\beta$, and the best fitting parameters are given in Table \ref{table:SEvol}.  
    The solid curves show the  evolutionary trends as described by these power-law functions with shaded regions showing the $1\sigma$ confidence intervals.}
    \label{f:size-evo-sfg}
\end{figure*} 

\subsection{Size Evolution of Star-Forming Galaxies}
\label{sec:res-sfg}
 
\subsubsection{Overview of the Observed Trends}
\label{sec:sfg-overview}

Figure \ref{f:size-evo-sfg} shows the size evolution of SFGs in two rest-frame wavelengths for four fixed characteristic stellar masses since $z\sim1$.
As expected, for the evolution in rest-frame $5000$~\AA\ (visible light, navy blue symbols and solid lines with shaded area in Figure~\ref{f:size-evo-sfg}) we find similar trends as in Section \ref{sec:size-evo-visible}. The size growth of SFGs is mass dependent: the higher the mass of SFGs, the faster they grow in size with cosmic time.  
At $\log M = 9.7$, SFGs grow only $\sim5\%$ in size but at $\log M=11.2$, they grow by $31\%$ in $6$ Gyrs from $z\sim0.9$ to $z\sim0.1$.

\par
At the characteristic mass of $5\times10^{10}\text{M}_\odot$, \citet{vanderwel3DHSTCANDELSEvolution2014} report the power-law fit (Equation~\ref{eq:evol}) to the size evolution of SFGs with  $\beta=-0.75$ whereas we find $\beta=-0.35\pm0.14$. When probing the trends in near-IR-based size with redshift for $\log M>10.8$ galaxies  \citet{williamsEvolvingRelationsSize2010} also find a steeper size evolution for SFGs ($\beta=-0.77\pm0.08$). We note that both
\citet{williamsEvolvingRelationsSize2010} and \citet{vanderwel3DHSTCANDELSEvolution2014} studies cover wider redshift ranges: $0.5<z<2$ and $0<z<3$ respectively. Thus, their observed evolution is affected by SFGs at $z>1$. Furthermore, the flatter distribution of median SFG sizes below $z\sim1$ from Figure 4 of \citet{williamsEvolvingRelationsSize2010} and the agreement between our size measurements and those of \citet{vanderwel3DHSTCANDELSEvolution2014} suggest that the size evolution of SFGs weakens at $z<1$. Moreover, \citet[][]{oeschStructureMorphologies782010} and \citet[][]{moslehEvolutionMassSizeRelation2012} report faster evolution of SFG sizes at $z>2$. At the same time, studies that focus at $z<1$ \citep[e.g.,][]{lillyHubbleSpaceTelescope1998, ravindranathEvolutionDiskGalaxies2004, bardenGEMSSurfaceBrightness2005, kawinwanichakijHyperSuprimeCamSubaru2021}  report either a slow or no evolution of the average size of SFGs. Thus, our analysis confirms that the size evolution of SFGs slows down with decreasing redshift and that this change in pace occurs at $z\sim1$.

\par 
As in Figure \ref{f:size-evo-2wave}, Figure \ref{f:size-evo-sfg} also includes the size evolution of SFGs in the rest-frame $3000$~\AA\ (cyan symbols and solid lines with shaded area). In general, we find that the sizes of SFGs are larger at shorter wavelength than at the longer one, as we indicate in Section \ref{sec:size-evo-uv}. 
For example, at $z=0.5$, SFGs appear $\sim5\% - 17\%$ larger in rest-frame UV than in visible light depending on their stellar mass.
Since the shorter wavelength probes the distribution of younger stellar population,
the larger size in $3000$~\AA\, may imply extended (less concentrated) star-forming regions in those star-forming systems. A smaller size at longer wavelength suggests that the older population is more centrally concentrated, in line with the inside-out quenching scenario \citep{tacchellaEvidenceMatureBulges2015, tacchellaDustAttenuationBulge2018, linSDSSIVMaNGAInsideout2019}. 
 
\par
The difference between the two rest-frame wavelengths also depends on galaxy stellar mass. The sizes of SFGs at low mass ($\log M<10$) are almost the same at both wavelengths, and the differences in the sizes between the wavelength regimes increase with their stellar mass. At $z= 0.5$, an average SFG with stellar mass $\log M = 9.7$ has very similar size (within uncertainties) in both red and blue light. However, a massive SFG ($\log M =11.2$) is $\sim17\%$ larger in blue than in red light. 
 
\par
We interpret the lack of difference in two rest-frame sizes of low-mass SFGs as a sign that low-mass galaxies tend to have their younger and older stars well mixed within their light profiles, without any strong signals of either inside-out or outside-in quenching. However, as the galaxies grow in mass and size, the bulges also grow in the centres of these galaxies. These bulges may eventually suppress star formation by stabilizing the gas  within them \citep[e.g.,][]{martigMorphologicalQuenchingStar2009, saintongeImpactInteractionsBars2012, sachdevaGrowthBulgesDisk2017, hashemizadehDeepExtragalacticVisible2022}. As the star formation ceases in the centre, the peak of star formation activity moves   
to the outskirts of disks. Thus, the massive SFGs tend to look bigger in rest-frame UV than in visible light. 

\par
At fixed stellar mass, we find that the pace of SFG size evolution is slightly slower at $3000$~\AA\ than at $5000$~\AA. For example, at $\log M = 10.7$, SFGs grow by $21\%$ ($\beta=-0.35$) in red light and by $15\%$ ($\beta=-0.26$) in blue light between $z\sim0.9$ and $z\sim0.1$. The direction towards faster growth in red light is systematic across all stellar masses of SFGs, but we also note that, within uncertainty, this difference is consistent with the same pace of size evolution in two rest-frame wavelengths.

\par
The slower pace of evolution in blue light than in red light could also indicate the role of bulge growth in the evolution of SFGs. Bulges become more and more significant in the overall galaxy light profile with time \citep[e.g.,][]{sachdevaGrowthBulgesDisk2017}. As the bulges grow, this growth is reflected in the size evolution of SFGs at longer wavelength where we see a slightly faster pace of size evolution. 

\par
There are two scenarios for bulge growth that we can consider: (1) both the bulge and the disk grow in SFGs, and (2) bulge grows into the disk and pushes the peak star-forming regions we can observe at $3000$~\AA\ further out. 
In the first scenario, the slightly faster pace of SFG size evolution in longer wavelength probing older stellar population could suggest that the growth of bulges is faster than the growth of disks at this redshift range. In the second scenario, since only bulges grow significantly, the size growth in visible light is faster as well. 
Although pushing the peak star forming regions outward will make the sizes in UV bigger, growth in visible light is still stronger because it encompasses the light from stars in both the growing bulge and the disk. However, we note that 
we only fit a single S\'ersic profile to observed galaxy light profiles. We will explore the two scenarios  
in more detail in the follow-up study by performing bulge$+$disk decomposition.

\begin{figure*}
    \centering
	\includegraphics[scale=0.4]{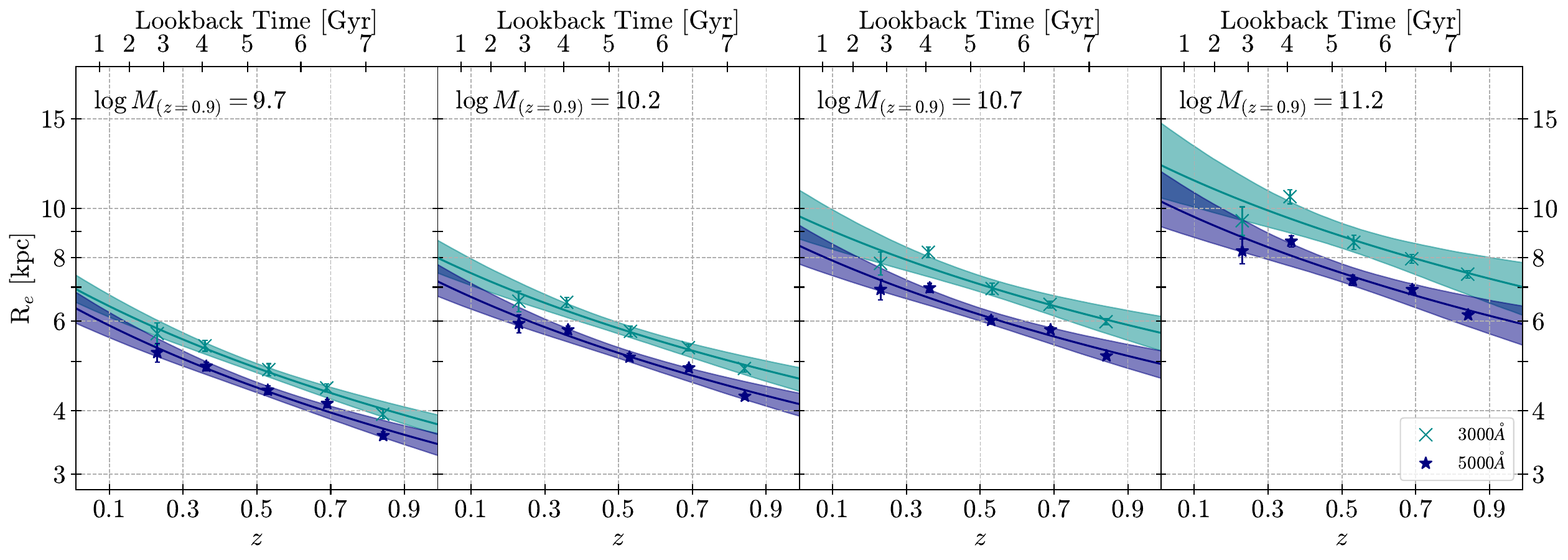}
    \caption[Size evolution of SFGs at evolving mass]{Evolution of characteristic sizes of SFGs assuming evolving stellar mass we calculate based on Equation \ref{eq:mass_est} at $z<1$. Each panel shows the size evolution for a galaxy that has labeled stellar mass at $z\sim 0.9$. This characteristic mass then evolves 
    along the redshift space in each panel via in-situ star formation. We adopt the SFR values of star formation main sequence from \citet{speagleHighlyConsistentFramework2014} to estimate this stellar mass growth. We then adopt galaxy size considering increased mass and observed SMR at each subsequent redshift point. 
    As in Figure~\ref{f:size-evo-sfg}, the navy blue stars denote the size in rest-frame $5000$~\AA\, and the cyan crosses denote those in rest-frame $3000$~\AA, along with their bootstrapped error bars. The best fitting power-laws with their uncertainties are represented by solid curves and shaded areas.  
    The parameters of the best-fit power-laws are given in Table \ref{table:SEvol-mevol}.
    By selecting galaxies with increased stellar mass at lower redshifts we attempt to analyze the size evolution of SFGs by connecting progenitors to their descendants.}
    \label{f:size-evo-mevol}
\end{figure*}

\subsubsection{The Effects of SFG Mass Growth}\label{sec:sfg-mevol}

The earlier studies in general look at the galaxy size evolution at a constant stellar mass across the redshift range that they probe \citep[e.g.,][and many others]{vanderwel3DHSTCANDELSEvolution2014, mowlaCOSMOSDASHEvolutionGalaxy2019,  kawinwanichakijHyperSuprimeCamSubaru2021}. So far, we have also analyzed the size evolution in this manner. However, SFGs are actively forming stars and growing in both size and mass with cosmic time. To study the true size evolution of SFGs, it is important to consider their mass growth.

\par
We use a toy model to trace the size evolution of galaxies having progenitors with similar mass at $z\sim0.9$ by incorporating mass growth via star-formation. 
We assume that the SFGs remain on the star-forming main sequence since $z\sim1$ and that the mass growth is only due to \textit{in situ} star formation in galaxies.
Although galaxies quench and leave the SFG population over cosmic time interval we probe, in the toy model we also assume that this quenching is random and does not affect any sub-population of SFGs specifically.
\par
We adopt the star-forming main sequence relation from \citet{speagleHighlyConsistentFramework2014},
\begin{equation}
    \log \mathrm{SFR}(M,t) = (0.84 - 0.026t) \log M + 0.11t - 6.51, 
\label{eq:SFR-speagle}
\end{equation}
where $t$ is the age of the universe in Gyr and SFR is in $\Msun/$yr.
For SFGs with a specific mass at $z=0.9$, we estimate their mass at various cosmic times between $z=0.9$ and $z=0.1$ as 
\begin{equation}
    M(t_2) = M(t_1) + \int_{t_1}^{t_2} 10^{\log \mathrm{SFR} (M,t)} dt,
    \label{eq:mass_est}
\end{equation}
where $t_1$ is the age of the Universe at $z=0.9$ and $t_2$ is its age at the target redshift. We then estimate the characteristic sizes of these SFGs at lower redshifts  by taking their increased stellar mass as the input for the SMR (Equation \ref{eq:smr}) with redshift-dependent  parameters from Table~\ref{table:SPL}. 
Finally, we fit the size evolution of SFGs growing both in mass and size at two different rest-frame wavelengths. Figure \ref{f:size-evo-mevol} illustrates the results and Table \ref{table:SEvol-mevol} lists the best-fit parameters of the evolutionary trends for the SFGs with growing stellar mass. 

\par
When considering stellar mass growth, for every {\it initial} stellar mass (i.e., stellar mass at $z=0.9$) we find that the size growth is faster than the growth at constant mass. 
In rest-frame visible light, SFGs with initial stellar mass of $\log M=9.7$ grow in size by $\sim65\%$ when we consider mass growth (navy blue curve in the first panel of Figure \ref{f:size-evo-mevol}) against only $\sim 5\%$ growth when the mass growth is ignored in the redshift range we probe (navy blue curve in the first panel of Figure \ref{f:size-evo-sfg}). The observed weak/no evolution at the lowest mass bin in Figure \ref{f:size-evo-sfg} ($\beta=-0.08\pm0.19$ in $5000$~\AA) seems to be due to the selection of different galaxies at different redshifts.
\par
At the same time, the SFGs with initial stellar mass $\log M=11.2$ grow in size by $\sim55\%$ and $\sim32\%$ with and without stellar mass growth, respectively (last panels in Figures~\ref{f:size-evo-mevol} and~\ref{f:size-evo-sfg}). Thus the difference in size growth with or without simultaneous stellar mass growth decreases with stellar mass. The difference we observe 
demonstrates that the size evolution of individual galaxies is faster than what is being suggested in the literature because size growth of SFGs cannot be separated from the increase in their stellar mass.

\par 
Additionally, we do not find any differential size evolution with stellar mass when we consider mass growth. Our analysis shows that the sizes of all massive SFGs ($\log M>9.5$) grow, within uncertainties, at a constant pace ($\beta\sim -0.8$ to $-0.9$) (Figure \ref{f:size-evo-mevol} and Table \ref{table:SEvol-mevol}).

\par
Furthermore, the differences in the rate of size growth between two wavelength regimes that we report in Section~\ref{sec:sfg-overview} disappear when we include stellar mass growth. Both cyan and navy blue curves represent power-law functions with very similar exponents in all panels of Figure \ref{f:size-evo-mevol}. 
However, throughout the redshift range we probe, the SFGs remain more extended in the shorter wavelength than in the longer wavelength in all initial stellar mass bins.

\par 
\subsubsection{Slowing down of Size Evolution for SFGs}
\begin{figure}
    \centering
    \includegraphics[width=1\columnwidth]{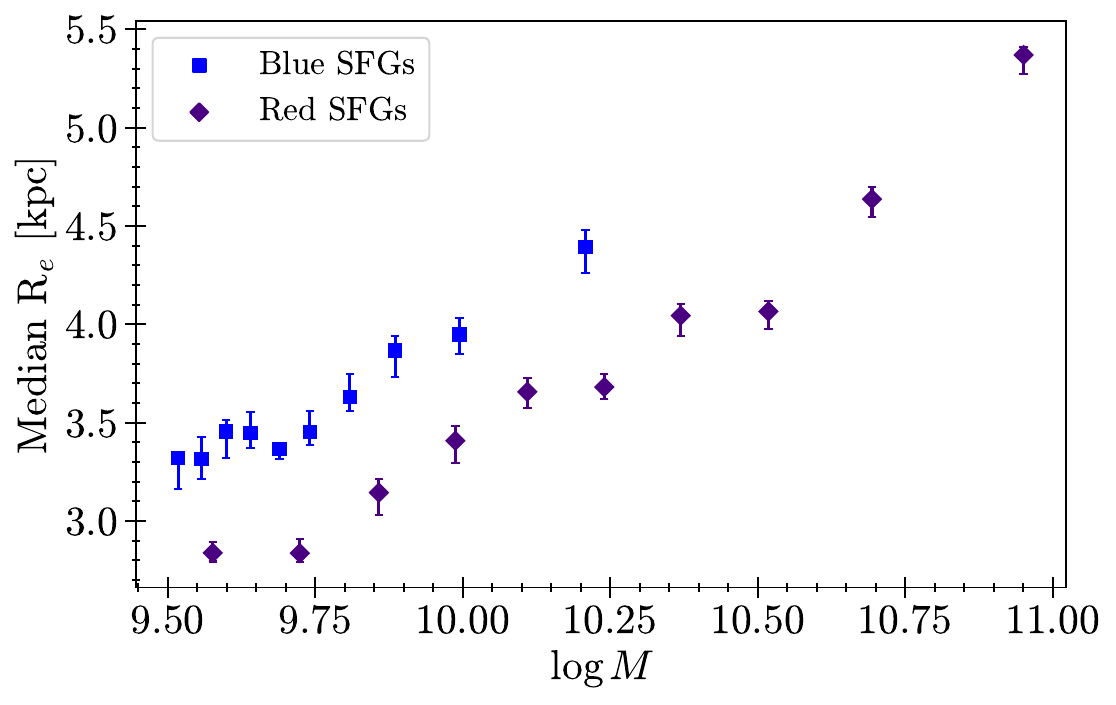}
    \caption[]{The median rest-frame $5000$~\AA\ sizes of red SFGs ($U-V>1$) and blue SFGs ($U-V\leq1$) in ten equally populated mass bins. The red SFGs are shown as indigo diamonds and blue SFGs as blue squares. All errors are bootstrapped. Blue SFGs tend to be less massive than red SFGs and thus mass bins are concentrated at lower masses for blue SFGs. At a given mass, blue SFGs are $20\%$ larger than red SFGs.}
    \label{f:sfg-rb-smr}
\end{figure}

In combination with the observed rates of size growth at $z>1$ from the literature, our results for the star-forming population point to slowing down of SFG size evolution with cosmic time (\ref{sec:sfg-overview}). To explore this further, we separate SFGs into two groups based on $U-V$ colour: red SFGs ($U-V>1$) and blue SFGs ($U-V\leq1$). 
In Figure \ref{f:sfg-rb-smr} we show the characteristics of these two populations in size-stellar mass parameter space for sizes measured at the rest-frame $5000$~\AA. The figure shows that (1) blue SFGs (blue squares) cover a smaller range of masses (up to $10^{10.5}$\Msun) than red SFGs (purple diamonds); (2) at the same stellar mass, blue SFGs are around $20\%$ larger than red SFGs. 

\par
The lack of massive blue SFGs may be related to the extent of the star formation episodes. Blue SFGs are forming stars at high rates and thus are still "catching up" with the SFGs that are in slightly more advanced stage of the evolution (but still on the main sequence). 
As the SFGs evolve, they become redder in colour because low-mass stars become dominant light source. A fraction of these stars are in central bulges. 
The presence of bulges in red SFGs makes them more centrally concentrated in the rest-frame $5000$~\AA\ than blue SFGs of the same mass.  
Hence, on average, these red SFGs have smaller sizes in visible light than blue SFGs.

\begin{figure}
    \centering
    \includegraphics[width=1\columnwidth]{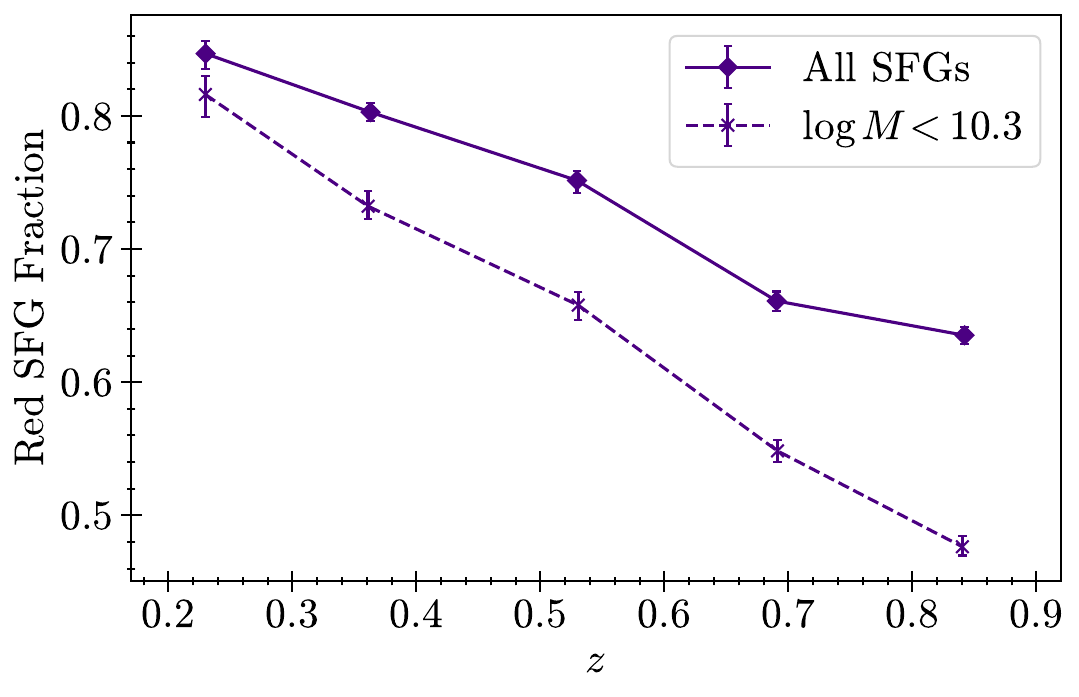}
    \caption[]{The redshift evolution of the SFG red fraction: fraction of red SFGs in the complete SFG population along with bootstrapped error bars (diamonds with solid line). Conversely, the fraction of blue SFGs decrease with cosmic time. We find that the fraction of red SFGs increases with cosmic time. Since the majority of blue SFGs have mass $\log M<10.3$ (see Figure \ref{f:sfg-rb-smr}), we also show the red SFG fraction in this mass range (crosses with dashed line) to demonstrate that the trend remains the same. This fractional increase in the smaller and redder SFGs with cosmic time could explain the slowing pace of the characteristic size evolution of SFGs since $z\sim1$.}
    \label{f:sfg-rbfrac}
\end{figure}

\begin{figure}
    \centering
    \includegraphics[width=1\columnwidth]{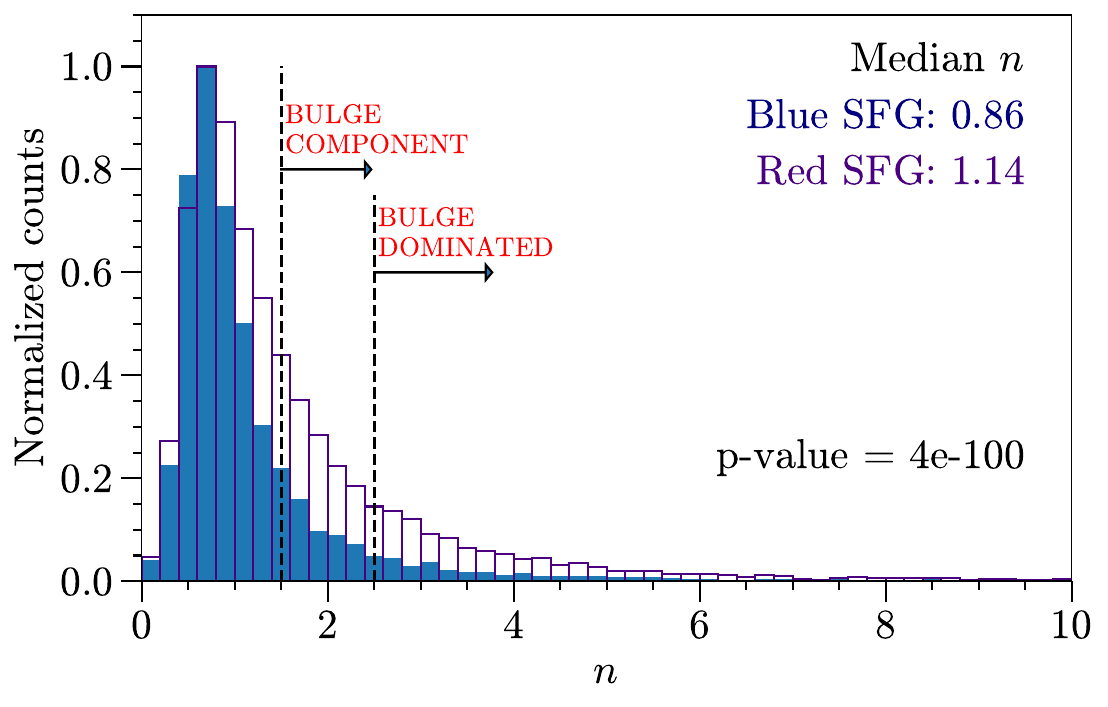}
    \caption[]{The distribution of S\'ersic indices of blue SFGs (blue filled histograms) and red SFGs (purple unfilled histograms). Galaxies with $n>1.5$ are often considered to have a significant bulge component whilst galaxies with $n>2.5$ are considered to bulge dominated. Among bulge-dominated SFGs, $83\%$ are red. We also show the p-value from the K-S test in the figure legend.}
    \label{f:sfg-rb-nhist}
\end{figure}

\par 
We also find that the fraction of red galaxies in the SFG population is steadily increasing  with cosmic time across the redshift range we explore (Figure \ref{f:sfg-rbfrac}). At $0.75<z<0.9$, red SFGs constitute $\sim64\%$ of the SFG sample (purple diamonds with solid curve). This percentage increases to $\sim80\%$ in the lowest redshift bin. Conversely, the fraction of blue SFGs decreases with redshift because the sum of fractions of red and blue SFGs equals to $1$ in each redshift bin. 

\par
Since blue SFGs have low masses, we investigate how the fraction of red SFGs change with redshift for $\log M<10.3$ (crosses with dashed curve in Figure \ref{f:sfg-rbfrac}). We find a similar trend: the fraction of red SFGs increases with decreasing redshift. Low-mass red SFGs constitute $\sim48\%$ of the low-mass SFG sample in the highest redshift bin but this percentage increases to $\sim82\%$ in the lowest redshift bin.
This increase in the fraction of smaller red galaxies in the SF population could explain the observed slower pace of sizes growth for SFGs at $z<1$ with respect to higher redshifts.

\par
Based on the observational studies, galaxies with S\'ersic index $n\gtrsim1.5$ are generally considered to have a significant bulge component and those with $n\gtrsim2.5$ are considered bulge-dominated or spheroidal galaxies \citep[e.g.,][]{duttonOriginExponentialGalaxy2009, sachdevaMagnitudeSizeEvolution2013}.  Figure \ref{f:sfg-rb-nhist} shows that in our sample red SFGs dominate both the $n>1.5$ (galaxies with significant bulge component) and $n>2.5$ (bulge-dominated galaxies) subsamples with 81\% and 83\%, respectively.  The average S\'ersic indices of red SFGs are higher than their blue counterparts. The median $n$ of red SFGs ($n=1.14$) is $33\%$ higher than blue SFGs ($n=0.86$). We perform a Kolmogorov–Smirnov test \citep[K-S test,][]{kolmogorov-smirnovSullaDeterminazioneEmpirica1933, smirnovTableEstimatingGoodness1948} to check whether the two distributions differ statistically from each other. An extremely small p-value
implies that the two samples do not originate from the same parent (sub)population.

\par 
Even when we remove massive red SFGs and restrict the analysis of S\'ersic indices to $\log M<10.3$, we find similar trends as in Figure \ref{f:sfg-rb-nhist}. Among low-mass SFGs with significant bulge component, $\sim69\%$ are red SFGs; among bulge dominated SFGs, this percentage is $\sim72\%$.

\par

To summarize, we see that blue SFGs are more extended but lower in mass than red SFGs. The fraction of red SFGs in the parent SFG population is increasing steadily with cosmic time. Red SFGs tend to be more concentrated and host significant bulge component in them compared to their bluer SFG counterparts.
Together with the difference we see between sizes of SFGs at shorter and longer rest-frame wavelengths (Figures \ref{f:smr-3000&5000A}, \ref{f:size-evo-2wave}, \ref{f:size-evo-sfg}, and \ref{f:size-evo-mevol}) and the scenarios that can explain it (Section~\ref{sec:sfg-overview}), this analysis suggests that the emergence of bulges is driving the observed average size evolution of SFGs.

\par

Several physical processes may be responsible for the emergence and growth of bulges in galaxies. Bulge growth can happen internally (secular evolution) or it can be induced externally. 

\par
In secular evolution, bulge formation and growth is often a result of perturbations in the galaxy disks due to internal substructures such as spiral arms and bars \citep{ kormendySecularEvolutionFormation2004,athanassoulaNatureBulgesGeneral2005, kormendyInternalSecularEvolution2008, gadottiStructuralPropertiesPseudobulges2009, sellwoodSecularEvolutionDisk2014}. This secular growth of bulges (also called pseudo-bulges) is global to all disk galaxies. The pseudo-bulges are characterised by disk like profiles ($n\sim1$) and thus do not increase galaxy S\'ersic index. 

\par
In contrast, bulge formation and growth induced externally (e.g., through mergers) by displacing the stars and gas in the disks towards galaxy centres, can contribute to the increase in galaxy S\'ersic index \citep{naabInfluenceGasStructure2006, gadottiStructuralPropertiesPseudobulges2009, hopkinsMergersBulgeFormation2010, oserTwoPhasesGalaxy2010, tacchellaMorphologyStarFormation2019}. 
Bulges formed in this manner (classical bulges) have light profiles and properties similar to elliptical galaxies. These bulges have higher fraction of old red stars than in the disks and appear redder in colour \citep[e.g.,][]{bredaStellarAgeGradients2020}. They generally have centrally concentrated surface brightness distributions, which are reflected in their higher $n$. 

\par
Several processes can cause a compaction  towards galaxy centre resulting in central bulge formation and growth. Those mechanisms include galaxy wet mergers \citep{zolotovCompactionQuenchingHighz2015, inoueNonlinearViolentDisc2016} and collisions of counter-rotating gas streams that feed the galaxy \citep{danovichFourPhasesAngularmomentum2015}. In both cases, dissipative gaseous accretion promotes bursts of star formation and growth of stellar mass towards galaxy centre \citep[e.g.,][]{dekelFormationMassiveGalaxies2009, lapinerWetCompactionBlue2023}. The absence of further gas inflow results in gas depletion and eventual quenching of the core regions \citep[e.g.,][]{ceverinoHighredshiftClumpyDiscs2010, zolotovCompactionQuenchingHighz2015, tacchellaConfinementStarformingGalaxies2016}. 
Thus, the bulges in SFGs tend to have high central mass concentrations and appear redder in colour as they age. We note, however, that wet compaction events are more common at earlier cosmic times than within the redshift interval we probe here.

\begin{figure*}
    \centering

	\includegraphics[scale=0.5]{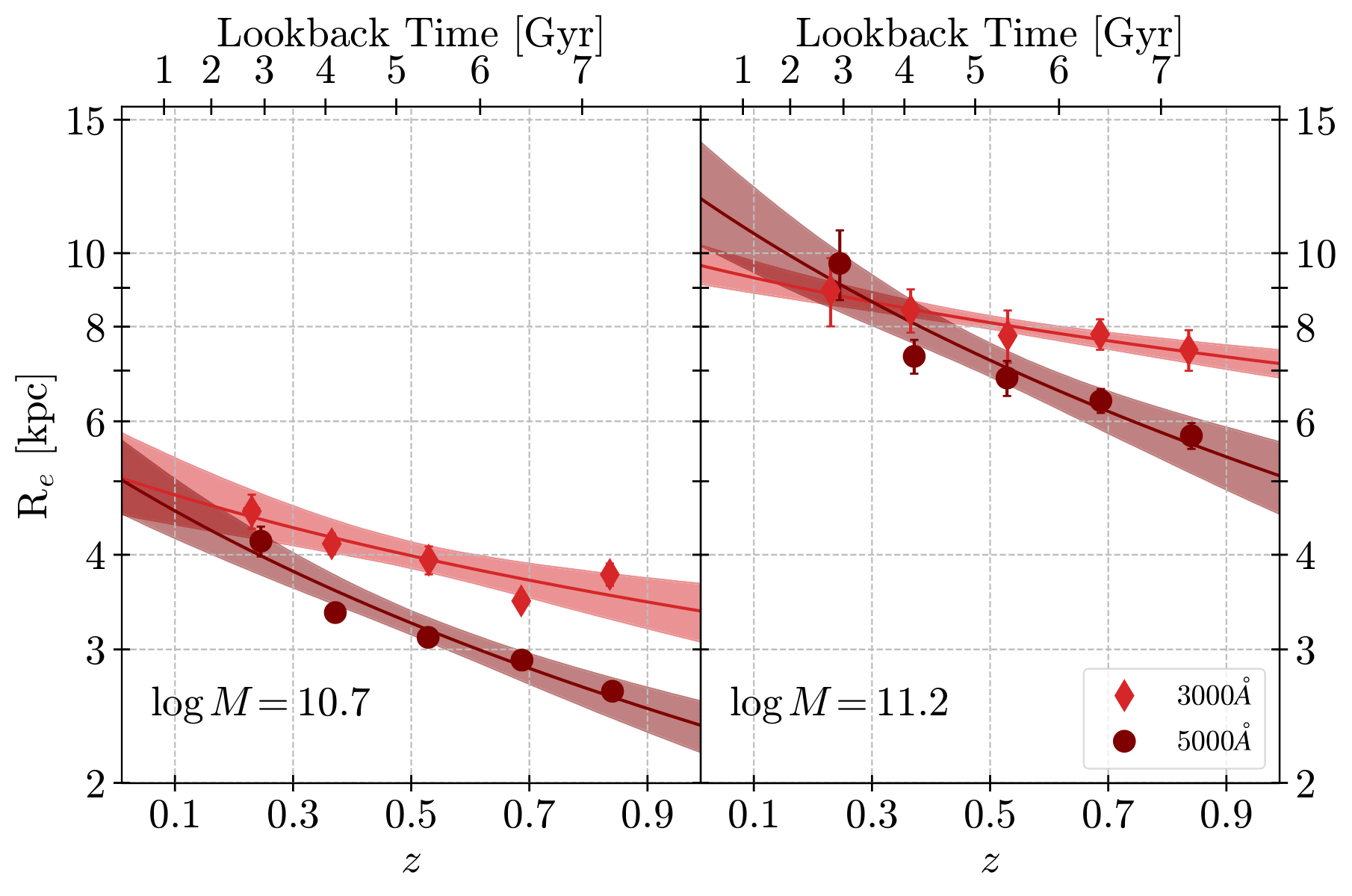} %.pdf}
     \caption[Size evolution of QGs]{Evolution of median sizes of QGs. Panels show the size evolution of SFGs at two different characteristic masses.  The QG sizes at rest-frame $3000$~\AA\ is shown in red and at rest-frame $5000$~\AA\ in maroon.  The sizes at shorter wavelength is larger than those at longer wavelengths. The size evolution is fit using a power-law: $A(1+z)^\beta$ and the best fitting parameters are given in Table \ref{table:SEvol}. 
     The solid curves show the best-fit evolutionary curves with shaded regions showing their $1\sigma$ confidence intervals.}
    \label{f:size-evo-qg}
\end{figure*}

\par
Studies based on cosmological hydrodynamic simulations (e.g., IllustrisTNG) predict that the emergence of centrally concentrated spheroidal components is a common phenomena in SFGs \citep[e.g.,][]{tacchellaMorphologyStarFormation2019}. 
Observational studies show that the overall S\'ersic index and bulge-to-total ratios increase with cosmic time for SFGs which agrees with the predictions of simulations that the bulges are growing in SFGs \citep[e.g.,][]{langBulgeGrowthQuenching2014}.   In light of these results, the emergence of classical bulges must be responsible for the high $n$ tail of the S\'ersic index distribution for SFGs in our sample.

\par

\subsubsection{Limitations}
\label{sec:sfg_limitations}

Effects of age and metallicity on galaxy light profiles cannot be disentangled without spatially-resolved spectroscopic data. Observed larger galaxy sizes in shorter wavelength can  result from either young or low-metallicity stars in their outskirts as both stellar population properties contribute to galaxy colours \citep[e.g.,][]{gustafssonChemicalAnalysesCool1989, streichRelationMetallicityRGB2014}. If light profiles of our  $\log M > 9.5$ SFGs in bluer rest-frame wavelength are at least in part affected by the presence of low-metallicity stars, measured larger sizes in $3000$~\AA\ can point towards accretion (in addition to star-formation) in the outskirts of these galaxies. Low mass galaxies generally have lower metallicities than massive ones \citep[e.g.,][]{tremontiOriginMassMetallicityRelation2004, fosterGalaxyMassAssembly2012, zahidFMOSCOSMOSSurveyStarforming2014, maOriginEvolutionGalaxy2016}. Thus the accretion of low-mass satellites can lower the average metallicity and increase the surface brightness of galaxy outskirts in bluer light.

\par
The presence of dust can also cause observed differences between rest-frame $3000$~\AA\, and $5000$~\AA. 
The rest-frame $3000$~\AA\ is more impacted by dust absorption than $5000$~\AA. Some studies show that the dust attenuation is stronger in bulges than in disks \citep[e.g.,][]{driverMillenniumGalaxyCatalogue2007}. 
This radial dependence of dust reddening could impact the fitting of S\'ersic profile at $3000$~\AA. As the contribution of unobscured blue light from the outskirts dominates the overall light profile of a galaxy, its $R_e$ is overestimated and $n$ is underestimated. However, we find that the red SFGs  tend to be smaller and have higher S\'ersic indices than blue SFGs in the rest-frame $5000$~\AA\ where galaxy light is less affected by dust compared to the rest-frame $3000$~\AA\ (Figures \ref{f:sfg-rb-smr} and \ref{f:sfg-rb-nhist}). Hence, we conclude that the observed differences between red and blue SFGs in our study are not driven by dust reddening alone.

\par
With this study, we do not perform detailed analysis of galaxy components (bulges v/s disks). This is a necessary next step in testing the bulge growth scenario for the size evolution of SFGs.  
In the follow-up study we plan to perform bulge-disk decomposition and estimate bulge-to-total (B/T) ratios for all CLAUDS+HSC galaxies using a multi-component model fitting approach.

\subsection{Size Evolution of Quiescent Galaxies}
\label{sec:res-qg}

\subsubsection{Overview of the Observed Trends}

As in Section \ref{sec:res-sfg}, we use characteristic sizes estimated from the SMR to trace the evolution in size for QG population in two rest-frame waveleghts (Figure \ref{f:size-evo-qg}). 
We limit the analysis to two different characteristic masses ($\log M = 10.7, 11.2$) that are above the pivot point of the SMR (Section \ref{sec:SMR}) for quiescent population. 

\par 
As expected, the results we show in Figure \ref{f:size-evo-qg} are in line with our findings in Section \ref{sec:size-evo-uv}. Similar to SFGs, we find that the sizes of QGs are larger in rest-frame $3000$~\AA\, than in $5000$~\AA\ (red and maroon curves in Figure \ref{f:size-evo-qg}). At $z\sim0.5$, QGs with mass $\log M=10.7$ are $\sim23\%$ larger in size in shorter wavelength than in longer wavelength. This offset in size decreases with increase in stellar mass. At the same redshift, QGs with mass $\log M\sim11.2$ are $\sim12\%$ more extended in the rest-frame UV light than in visible light.

\par 
 With respect to SFGs in our sample, QGs exhibit much stronger size evolution in rest-frame $5000$~\AA\ than in $3000$~\AA. For example, in the rest-frame $3000$~\AA, QGs with mass $\log M \sim 10.7$ grow in size by $\sim38\%$ in the rest-frame UV and by $\sim82\%$ in the rest-frame visible light over the span of 6 Gyrs that we probe. For very massive QGs ($\log M \sim 11.2$), this growth is by $\sim 27\%$ and $\sim 97\%$, respectively, for sizes measured in rest-frame UV and visible light.

\par 
Furthermore, the average size of a QG is significantly smaller than an SFG at a given mass and redshift (Figure \ref{f:size-evo-2wave}). Comparison of the last two panels of Figure \ref{f:size-evo-sfg} with Figure \ref{f:size-evo-qg} shows that at $z=0.5$ and rest-frame wavelength $5000$~\AA, $\log M=10.7$ SFG is $\sim70\%$ larger than QG of the same stellar mass. However, SFGs and QGs with $\log M=11.2$ have similar sizes. 
The trends in the bluer wavelength are also similar to those in the redder wavelength. 
This is interpreted to be due to the fading of disk when a galaxy fully quenches \citep[e.g.,][]{christleinCanEarlyTypeGalaxies2004, carolloZENSIVSimilar2016, matharuHSTWFC3Grism2020, estrada-carpenterCLEARMorphologicalEvolution2023}. Since the disks are more extended than their more centrally concentrated bulges/spheroids, the overall size of a galaxy shrinks when the disk completely fades. 

\par 

Alternatively, the smaller sizes of QGs relative to SFGs of comparable stellar mass could be due to the preferential growth of central regions over the outer ones during and after quenching. 
We can investigate the effects of both disk fading and growth of bulges through bulge$+$disk decomposition in two rest-frame wavelengths. When the star formation in galaxy disk is quenched, the disk become fainter, especially at UV wavelengths (although it does not disappear completely). The surface brightness of the disk diminishes significantly in the  rest-frame UV  regime and less so at longer (rest-frame optical) wavelengths. However, this fading should not affect the central regions (galaxy bulges). We will explore this effect in our future work through bulge$+$disk decomposition.

\par
Finally, the QGs tend to grow in size at a more rapid pace compared to SFGs even though they do not actively form stars. This observation is in line with the literature
\citep[e.g.,][]{conseliceEvolutionGalaxyStructure2014, vanderwel3DHSTCANDELSEvolution2014, straatmanSizesMassiveQuiescent2015, damjanovQuiescentGalaxySize2019, kawinwanichakijHyperSuprimeCamSubaru2021}. 
This strong size evolution could be driven by several physical processes affecting individual QGs, such as major mergers \citep{conseliceDirectMeasurementMajor2003, conseliceDirectMeasurementGalaxy2022, bluckStructuresTotalMinor2012, kavirajRoleMajorMergers2014, manResolvingDiscrepancyGalaxy2016, manthaMajorMergingHistory2018, huskoStatisticsGalaxyMergers2022, huskoBuildupGalaxiesTheir2023}, minor mergers and accretion \citep{khochfarSimpleModelSize2006,naabMinorMergersSize2009, bluckStructuresTotalMinor2012, newmanCanMinorMerging2012, fagioliMinorMergersProgenitor2016, sawickiLARgESurveyIII2020, conseliceDirectMeasurementGalaxy2022, huskoBuildupGalaxiesTheir2023, suessMinorMergerGrowth2023}, and adiabatic expansion \citep{boilyImpactMassLoss2003, goodwinGasExpulsionDestruction2006, baumgardtComprehensiveSetSimulations2007, fanDramaticSizeEvolution2008, fanCosmicEvolutionSize2010, damjanovRedNuggetsCompact2009, ragone-figueroaPuffingEarlytypeGalaxies2011, lapiDramaticSizeKinematic2018}. In addition, progenitor bias due to newcomers 
affects the average QG size evolution \citep{franxStructureStarFormation2008, carolloNewlyQuenchedGalaxies2013, fagioliMinorMergersProgenitor2016, damjanovQuiescentGalaxySize2019, damjanovSizeSpectroscopicEvolution2023}.

\par
Galaxy mergers can contribute significantly to the size evolution of QGs. Although major dry mergers can yield an increase in size of QGs proportional to the increase in mass \citep[eg.,][]{boylan-kolchinRedMergersAssembly2006, naabFormationEarlyTypeGalaxies2007}, the average number of major mergers per galaxy since $z\sim1$ is low \citep[e.g.,][]{bundyGreaterImpactMergers2009,  lopez-sanjuanMinorRoleGasRich2010, kavirajRoleMajorMergers2014, thibertAutomatedDetectionMerging2018}. However, both observational studies \citep[e.g.,][]{bezansonRelationCompactQuiescent2009, vandokkumGrowthMassiveGalaxies2010, trujilloDissectingSizeEvolution2011, fagioliMinorMergersProgenitor2016, zahidCoevolutionMassiveQuiescent2019} and theoretical/simulation based studies \citep[e.g.,][]{khochfarSimpleModelSize2006, naabMinorMergersSize2009, oserTwoPhasesGalaxy2010, oserCosmologicalSizeVelocity2012, hilzRelaxationStrippingEvolution2012} show that minor mergers and accretion can support size growth in QGs. 
Theoretical studies \citep[e.g.,][]{naabMinorMergersSize2009} and observational studies that incorporate these theoretical predictions \citep[e.g.,][]{bezansonRelationCompactQuiescent2009, zahidCoevolutionMassiveQuiescent2019}  
indicate that a series of minor mergers, with the fractional growth in $R_e \propto M_*^2$ (where $M_*$ is the fractional growth in stellar mass) can explain QG size growth since $z\sim 2-3$. 

\par  
The fact that the QGs have larger sizes in shorter wavelength (red curves in Figure \ref{f:size-evo-qg}) than the longer one (maroon curves in Figure \ref{f:size-evo-qg}) could indicate growth through minor mergers and accretion \citep[e.g.,][]{bezansonRelationCompactQuiescent2009, naabMinorMergersSize2009, suessColorGradientsQuiescent2020, suessMinorMergerGrowth2023}. As these are passive galaxies, they are not actively forming stars, but the bulk of the young and/or low-metallicity stars traced by shorter wavelength could be added \textit{ex-situ} or formed from \textit{ex-situ} added gas through accretion and minor mergers. These materials are added to the outskirts puffing up galaxy light profile. 

\par
On the other hand, the longer wavelength profiles are dominated by the \textit{in-situ} old stars which are concentrated towards galaxy centre. As time progresses, the \textit{ex-situ} stars can also migrate inward and add to the galaxy profile in the longer wavelength as they age \citep[e.g.,][]{boeckerOriginStarsInner2023}. Since the longer wavelength probes newly added old \textit{ex-situ} stars along with the already existing \textit{in-situ} aged stellar population in galaxies, we see the stronger evolution in rest-frame $5000$~\AA.

\begin{figure*}
    \centering
	\includegraphics[scale=0.5]{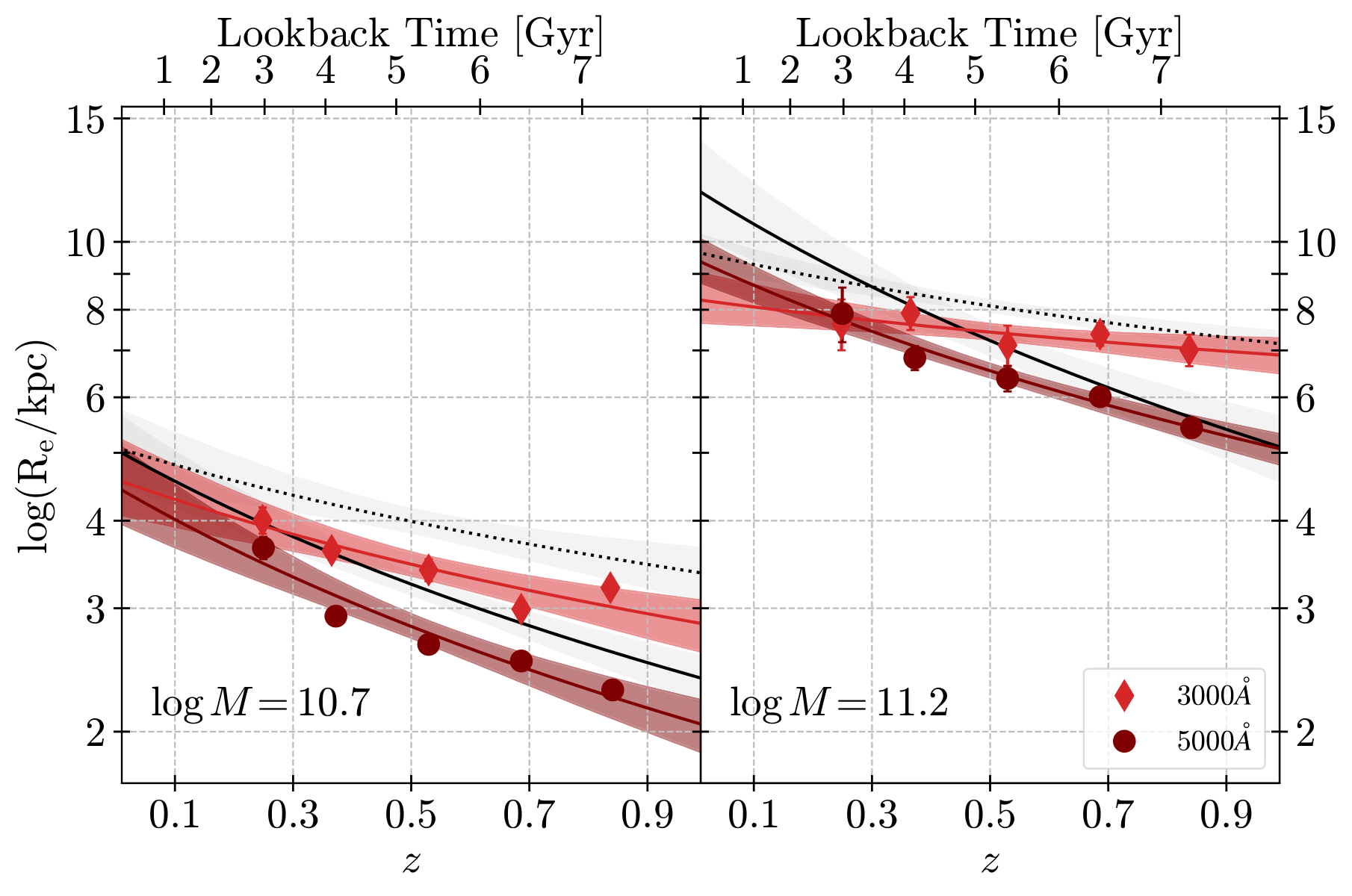} %.pdf}
    \caption[Relation between S\'ersic index and stellar mass of QGs without progenitor bias]{Size Evolution of QGs in two rest-frame wavelengths when progenitor bias is removed. The size evolution curves of the entire QG population are shown in gray (dotted and solid curves for rest-frame UV and visible light respectively). The best-fit parameters are given in Table \ref{table:SEvol-nopb}. Other details are same as for Figure \ref{f:size-evo-qg}.}
    \label{f:qg-se-nopb}
\end{figure*}

\subsubsection{Removing the Effect of Progenitor Bias}
\label{sec:pb-removal}

\par 
Newly quenched galaxies (newcomers) can affect the distribution of QG sizes at a given mass because they are, on average, more extended than the rest of the QG population \citep[progenitor bias;][]{sagliaFundamentalPlaneEDisCS2010, damjanovSizeSpectroscopicEvolution2023}.  
Progenitor bias should not contribute significantly to the size evolution of QGs.   
For example, \citet{damjanovSizeSpectroscopicEvolution2023} find that the newcomers contributed only $11\%$ to the growth of QGs at $z<0.6$. Nonetheless, it is important to remove the impact of progenitor bias from the QG population to understand how aging QGs may evolve with time.
Here we account for the effect of progenitor bias by removing the most extended galaxies and re-fitting the ``corrected" QG SMR in five redshift bins that cover the $0.1<z<0.9$ interval.

\par
We assume that in total $\sim10\%$ of the QG population above the SMR pivot point in each redshift bin are newcomers \citep{moutardVIPERSMultiLambdaSurvey2016}.
However, the probability for an SFG to quench increases with stellar mass and is given by 

\begin{equation}
\begin{split}
    p_Q &\varpropto \left(1-e^{-\frac{M}{M^*}}\right), \\
    & = c\left(1-e^{-\frac{M}{M^*}}\right),
\end{split}
\label{eq:quench-fraction}
\end{equation}
where $M$ is the stellar mass of the SFG in solar masses, $c$ is the normalization constant and $M^*$ is the characteristic mass from stellar mass (Schechter) function for SFGs in \Msun\ \citep{pengMassEnvironmentDrivers2010}. $M^*$ varies with redshift and we adopt the values from \citet{muzzinEvolutionStellarMass2013}. In addition, the number densities of SFGs and QGs also vary with stellar mass and redshift \citep[e.g.,][]{muzzinEvolutionStellarMass2013}.

\par 
To incorporate this information into a toy model of QG size growth,  we first divide massive QGs ($M>M_p$) in each redshift bin into $10$ equal frequency bins of stellar mass. Then we determine the fraction of galaxies to be removed from each mass bin as 

\begin{equation}
    f_{pb} =  p_Q   \frac{n_Q(M)}{n_T(M)} , 
\label{eq:pb-fraction}
\end{equation}
where  $n_Q(M)$ and $n_T(M)$ are the number densities of QGs and all galaxies respectively. We adopt the number densities from \citet{muzzinEvolutionStellarMass2013}. We use the normalization constant $c$ in the equation for $p_Q$ (Equation \ref{eq:quench-fraction}) to keep the percentage of the newly quenched galaxies to be $\sim10\%$ . We adopt $c=3/5$ which removes $\sim8-13\%$ of  the most extended galaxies in each redshift bin above the SMR pivot point (see also Appendix \ref{appendix:PB-Removal}). Across the stellar mass bins, this percentage varies from $\sim0\%$ to $\sim20\%$ of the QG population in that bin. We then perform the SMR fitting  and examine the evolution of characteristic sizes of remaining (aging) QGs at two different stellar masses in $3000$~\AA\ and $5000$~\AA\ (Figure \ref{f:qg-se-nopb}; the best-fit parameters of the size evolution are given in Table \ref{table:SEvol-nopb}). 

\begin{table}
	\begin{center}
		\begin{NiceTabular}{cccc}[hvlines]

                \hline
                
                \hline
                \Block{1-4}{QGs} \\
                \hline
                
                \hline
			Restframe $\lambda$ & $\log M_{(z=0.9)}$ & $R_e^0$ & $\beta$  \\
			\hline
			  \Block{2-1}{$3000$~\AA} & $10.7$ & $4.58\pm0.60$ & $-0.69\pm0.28$  \\
			
			  & $11.2$ & $8.28\pm0.70$ & $-0.27\pm0.19$  \\
			
			\hline
			  \Block{2-1}{$5000$~\AA} & $10.7$ & $4.48\pm0.59$ & $-1.13\pm0.29$  \\
			
			  & $11.2$ & $9.45\pm0.70$ & $-0.91\pm0.17$  \\

                \hline 

                \hline 
			
		\end{NiceTabular}
	\end{center}
	\caption{The parameters of the best-fit power-law 
 (Equation \ref{eq:evol}) to the redshift evolution of characteristic QG sizes after the removal of the the progenitor bias (Section\ref{sec:res-qg}).}
	\label{table:SEvol-nopb}
\end{table}

\par 

When we remove the supposed newcomers from the QG sample, the average characteristic sizes of QGs become smaller, as expected. The decrease in the average size of QGs with stellar mass $\log M=10.7$ is similar across the redshift range at both wavelength regimes ($11-15\%$). However, at high stellar mass ($\log M=11.2$), this decrease ranges from $3\%$ ($5\%$) at $z=0.9$ to $18\%$ ($13\%$) at $z=0.1$ in rest-frame visible (UV) light.

\par

We do not find any significant difference in the pace of size growth for QGs with mass $\log M=10.7$ when we account for progenitor bias in both wavelength regimes (compare coloured curves against black curves in the first panel of Figure \ref{f:qg-se-nopb}). However, there are indications that the size evolution becomes weaker when we remove the progenitor bias for QGs with mass $\log M =11.2$ (second panel of Figure \ref{f:qg-se-nopb}). Nonetheless, the difference is within uncertainties for both UV and visible light. Thus, the progenitor bias only causes a significant vertical offset in the redshift evolution of average QG sizes without strong signals of change in the exponent of the evolutionary trends for stellar masses higher than the pivot mass ($\log M\sim10.5$, Section~\ref{sec:SMR}).

\par 
Progenitor bias can affect the size evolution of QGs in two ways. First, the sizes of SFGs are larger than QGs at the same stellar mass and the size of SFGs increase with time (e.g., Figures \ref{f:hsc-doube-power-law} and \ref{f:size-evo-2wave}). Thus, the newcomers are larger than the those quenched earlier in time. This causes a vertical shift in the size evolution curve by increasing the average size of QGs at all redshifts. We see this effect in our sample (the black curves are higher than coloured curves in Figure \ref{f:qg-se-nopb}). 

\par
Second, disks of galaxies fade when they quench \citep{christleinCanEarlyTypeGalaxies2004, carolloZENSIVSimilar2016, estrada-carpenterCLEARMorphologicalEvolution2023}. Bulge becomes more prominent in a galaxy light profile resulting in smaller effective sizes of galaxies.  Since the disk fading affects light profiles in rest-frame $3000$~\AA\ more than $5000$~\AA\ and the quenched fraction increases with cosmic time, the progenitor bias effect should slow down the pace of size evolution in the shorter wavelength (red curves in Figure \ref{f:qg-se-nopb}). We expect to see a strengthening of the size evolution for QGs in rest-frame UV light after we remove the progenitor bias. Instead, we find that the pace of the size evolution is unaffected  indicating that the process of disk fading does not affect the average size evolution of massive QGs. 

\par
Thus, we conclude that the effects of progenitor bias on the massive QG population is mainly on the vertical offset of the evolutionary trend. The strength of the redshift evolution in size for massive QGs is determined by other physical processes such as minor mergers and accretion adding low-mass and/or low-metallicity stellar population (Section \ref{sec:sfg_limitations}) to galaxy outskirts.

\subsection{Evolution in Central Stellar Mass Surface Density}
\label{sec:central-mass-density}

Central stellar mass surface density, the projected stellar mass surface density within central $1$ kpc radius, is another physical parameter that is widely used in the analysis of galaxy evolution  
\citep[e.g.,][]{cheungDependenceQuenchingInner2012, fangLinkStarFormation2013, barroCANDELSProgenitorsCompact2013, barroCANDELS3DHSTCompact2014, barroStructuralStarformingRelations2017, suessDissectingSizeMassS1Mass2021, suessColorGradientsQuiescent2020, xuCriticalStellarCentral2021, jiAGNSelectionMethods2022, jiReconstructingAssemblyMassive2022, jiReconstructingAssemblyMassive2023, estrada-carpenterCLEARMorphologicalEvolution2023}. Stellar mass surface density is derived using both $R_e$ and $n$ and thus it does not include uncertainties due to covariance between $R_e$ and $n$ making it a more robust parameter than these two parameters individually \citep{jiHSTImagingIonizing2020, jiAGNSelectionMethods2022}.

\par

\begin{figure}
    \centering
	\includegraphics[width=0.99\columnwidth]{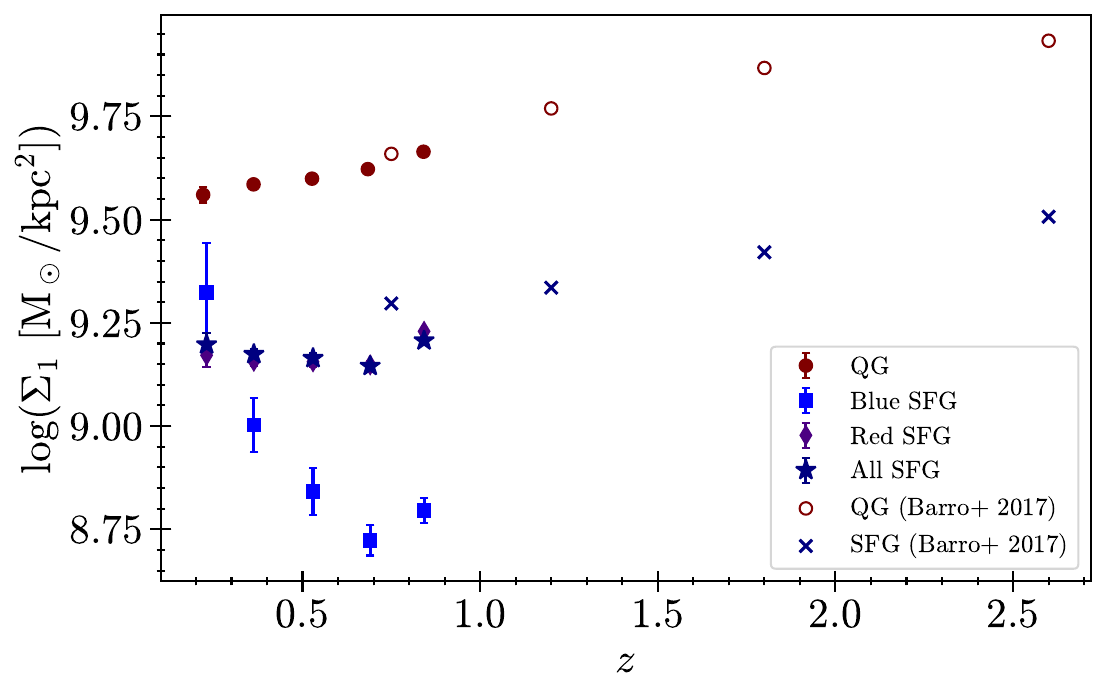} 
    \caption[Sigma1-evolution]{The evolution of characteristic $\Sigma_1$ for galaxies with mass $5\times 10^{10}$ \Msun. We show these values for all SFGs (dark blue stars) and QGs (maroon dots) along with their uncertainties. We also show the evolution in $\Sigma_1$ in two subpopulations of SFGs - blue (blue diamonds) and red (purple squares).  We compare our results with the estimated characteristic $\Sigma_1$ values at higher redshift for SFGs (blue crosses) and QGs (maroon open circles)  from  \citet{barroStructuralStarformingRelations2017}.  }
    \label{f:sigma1-evolution}
\end{figure}

We calculate $\Sigma_1$ in units of $\Msun/\text{kpc}^2$ and fit the relation between $\Sigma_1$ and galaxy stellar mass in  Appendix \ref{appendix:sigma1-mass-relation}. 
We first fit the $\Sigma_1$-stellar mass relation in log-log space for all SFGs and QGs (Figure \ref{f:sigma1-logM}) and also for SFGs split into blue and red subpopulations (Figure \ref{f:sigma1-logM-SFG}). One of the parameters of the fits we perform in Appendix \ref{appendix:sigma1-mass-relation} is the characteristic $\Sigma_1$ value\footnote{Throughout this work, we use its logarithmic value for analysis, i.e., $\log(\Sigma_1/\Msun/\text{kpc}^2 )$.} for galaxies  with stellar mass of $5\times10^{10}$ \Msun. Figure \ref{f:sigma1-evolution} shows the redshift evolution of this characteristic value for SFGs (dark blue stars) and QGs (maroon filled circles).

The QGs have $\sim 2$ ($3$) times higher $\Sigma_1$ than SFGs at $0.1<z<0.3$ ($0.75<z<0.9$). As the majority of the SFGs are red (Figure \ref{f:sfg-rbfrac}), the average $\Sigma_1$ of red SFGs (purple diamonds) are almost identical to those of all SFGs. However, blue SFGs (blue squares) have significantly lower $\Sigma_1$ values than red SFGs, except at $0.1<z<0.3$ where the sample size is small. Red SFGs have $\sim 1.5$ ($2.7$) times higher central stellar mass density than blue SFGs at $0.3<z<0.45$ ($0.75<z<0.9$). This difference is in line with SFGs growing their central regions and increasing $\Sigma_1$ during their evolution towards the red sequence 
\citep[as also reported by][]{guoImplicationsIncreasedCentral2021, quilleyAgingGalaxiesMorphological2022}.

\par

We compare our results with the evolution in $\Sigma_1$ for SFG and QG at $0.5<z<3$ \citep[maroon open circles (QGs) and blue crosses (SFGs) in Figure \ref{f:sigma1-evolution}]{barroStructuralStarformingRelations2017}. High-redshift  $\Sigma_1$ are estimated based on spatially resolved SED fits derived from surface brightness profiles in nine HST bands. We do not have spatially resolved SED fits for $z<1$ galaxies. By incorporating spatially resolved SEDs, \citet{barroStructuralStarformingRelations2017} are potentially able to incorporate the effects of colour gradients in the $\Sigma_1$ calculation. 
However, these gradients in both SFGs and QGs do not evolve at $z<1$ significantly \citep{suessHalfmassRadiiQuiescent2019}. Hence, the fact that we  consider only global rather than spatially resolved colour gradients does not affect the difference in the evolutionary trends for $\Sigma_1$ between two studies.

\par 

Figure \ref{f:sigma1-evolution} shows decreasing trend in $\Sigma_1$ of SFGs at $z>1$ with cosmic time. At $z<1$, we find that the SFGs with mass $5\times10^{10}$\Msun\ have, on average, a constant $\Sigma_1$ value. We interpret this lack of evolution in $\Sigma_1$ for SFGs in our sample as another aspect of prominent bulge growth in them. The growth of centrally concentrated bulges increases the $\Sigma_1$ values at lower redshift, resulting in flattening of the observed evolutionary trend.

\par

The characteristic $\Sigma_1$ values for QGs also decrease with time since $z\sim3$. Potential mechanisms driving this decrease in $\Sigma_1$ include minor mergers \citep[resulting in much larger size than stellar mass increase; e.g.,][]{bezansonRelationCompactQuiescent2009, hopkinsCompactHighredshiftGalaxies2009, naabAreDiskGalaxies2009, vandokkumGrowthMassiveGalaxies2010, fagioliMinorMergersProgenitor2016, zahidCoevolutionMassiveQuiescent2019}, adiabatic expansion due to mass loss and the rearrangement of stellar orbits in shallower gravitational potential \citep[e.g.,][]{damjanovRedNuggetsCompact2009, damjanovRedNuggetsHigh2011, vandokkumGrowthMassiveGalaxies2010, vandokkumDenseCoresGalaxies2014, lapiDramaticSizeKinematic2018}, and the influx of larger newcomers \citep[Section \ref{sec:pb-removal};][]{carolloNewlyQuenchedGalaxies2013, damjanovQuiescentGalaxySize2019, damjanovSizeSpectroscopicEvolution2023, jiReconstructingAssemblyMassive2022}.

\par 

However, QGs at $z<1$ show that the pace of redshift evolution in $\Sigma_1$ slows down at $z<0.9$. For example, $\Sigma_1$ of QGs decreases by $\sim10\%$ from $z\sim0.7$ to $z\sim0.2$ (or over 4~Gyrs of cosmic time), while the decline is somewhat more dramatic at higher $z$ ($20\%$ between the redshift bins $1.5<z<2$ and $1<z<1.5$, corresponding to $\sim2$~Gyrs of cosmic time). This weakening of $\Sigma_1$ evolution reflects the lack of evolution in $\Sigma_1$ for SFGs (especially red SFGs shown in purple diamonds) of the same mass. This similarity in trends between SFGs and QGs is driven partly by the quenching processes that transfer a fraction of SFGs in one redshift bin to the the QG population in the next lower redshift bin.

\begin{figure}
    \centering
	\includegraphics[width=1\columnwidth]{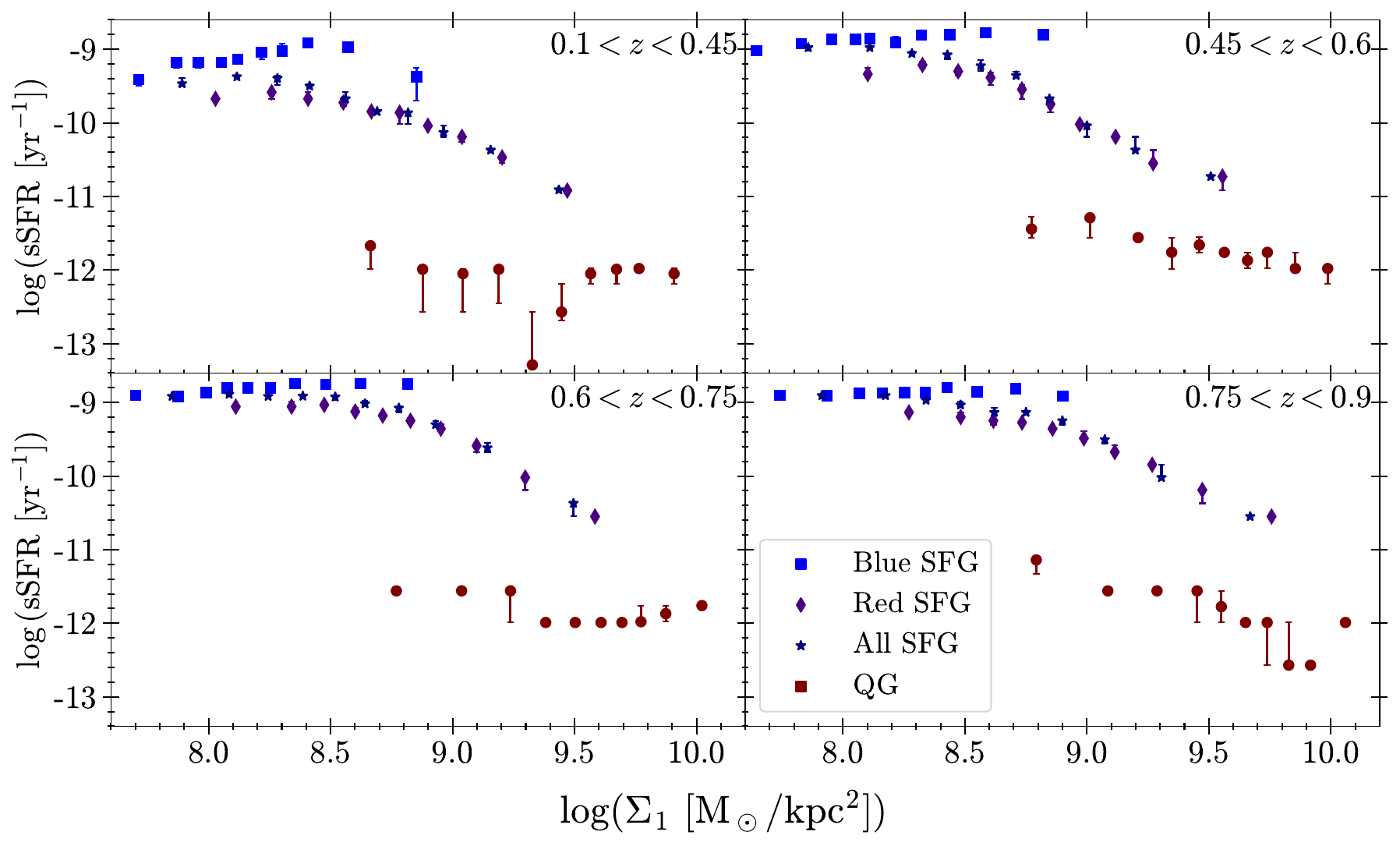}
    \caption[sSFR-Sigma Relation]{The relation between specific star formation rate (sSFR) and $\Sigma_1$ in four redshift bins for SFGs (red, blue and all) and QGs. We divide each galaxy groups into $10$ equally populated bins of $\Sigma_1$ and estimate median sSFR. The errorbars correspond to the bootstrapped errors for the medians. }
    \label{f:sSFR-sigma1}
\end{figure}

\par 

Earlier studies show that galaxy star formation activity is correlated with their central stellar mass surface density \citep[e.g.,][]{barroStructuralStarformingRelations2017, whitakerPredictingQuiescenceDependence2017,  suessDissectingSizeMassS1Mass2021}. These studies demonstrate that galaxies follow an "L-shaped track" in the specific star formation rate (sSFR) vs $\Sigma_1$ parameter space.
We explore this relation in Figure \ref{f:sSFR-sigma1}, where  
we show median sSFR values for both SFGs (dark blue stars) and QGs (maroon filled circles) in ten equally populated $\Sigma_1$ bins.
The sSFR values for SFGs are roughly stable at $\log \Sigma_1\lesssim9$. After that point, the sSFR starts to decrease with the increase in $\Sigma_1$. As the SFGs build up their central regions, their star-formation activity declines. However, we note that the observed correlation between $\Sigma_1$ and the sSFR may not necessarily imply a causal relation between the two parameters since a third, hidden parameter may be involved \citep[as also noted by, e.g.,][]{lillySurfaceDensityEffects2016, chenQuenchingContestGalaxy2020, jiReconstructingAssemblyMassive2022, jiReconstructingAssemblyMassive2023}.

\par 

When we divide the SFG population into red and blue SFGs (purple diamonds and blue squares, respectively, in Figure \ref{f:sSFR-sigma1}), red SFGs show somewhat lower sSFR at any given $\Sigma_1$. 
Furthermore, there is a maximum $\Sigma_1$ above which only red SFGs exist.

\par

The sSFR values for blue SFGs do not change with $\Sigma_1$, indicating that these young SFGs are actively forming stars at all galactocentric radii. The sSFR for red SFGs declines with $\Sigma_1$ above the threshold value where blue SFGs are absent. 
This trend suggests that by probing the red SFG subpopulation, we trace how these galaxies build up their central bulges and undergo inside-out quenching.

\par 

The threshold $\Sigma_1$ value above which all SFGs are red is $\log \Sigma_1 \sim 9$. This threshold value does not evolve with redshift. In line with the values reported by \citet{whitakerPredictingQuiescenceDependence2017}, our result suggests that as SFGs build up sufficient central stellar mass density, quenching process becomes prominent above this universal minimum value.

\par 

In addition, we find that there is a threshold $\Sigma_1$ ($\log\Sigma_1\sim9.7$) beyond which a galaxy cannot sustain star formation. This finding broadly agrees with previous results of \citet{whitakerPredictingQuiescenceDependence2017}, \citet{suessDissectingSizeMassS1Mass2021} and \citet{estrada-carpenterCLEARMorphologicalEvolution2023}.

\par
Thus, we observe the minimum value for $\Sigma_1$ that corresponds to the point when quenching becomes prominent globally (throughout the galaxy) and the maximum value above which all galaxies are quiescent. Between these two threshold values of $\Sigma_1$, red bulges continue to grow in SFGs as indicated by the negative correlation between $\Sigma_1$ and sSFR (Figure \ref{f:sSFR-sigma1}).

\par 

The range of $\Sigma_1$ values that red SFGs can have once they drop below sSFR for the blue subpopulation suggests that the process we trace is the slow process of bulge growth that drives this decline in sSFR.  
A number of studies  support the emergence/significant growth of centrally concentrated bulges in galaxies before and while they transition through the green valley and quench \citep{barroStructuralStarformingRelations2017, wuColorsSizesRecently2020, guoImplicationsIncreasedCentral2021, quilleyAgingGalaxiesMorphological2022, jiReconstructingAssemblyMassive2023}. Therefore, we conclude that the bulge growth is the major driving force behind the evolution of SFGs towards the red sequence at $z<0.9$.

\section{Conclusions}
\label{sec:conclusions}

We measure morphological parameters for 21,200 massive ($\log M>9.5$) star-forming and quiescent galaxies (SFGs and QGs) in the redshift range $0.1<z<0.9$ and in two rest-frame wavelengths, $3000$~\AA\ and $5000$~\AA. We use the multi-band images from the central region of the COSMOS field in the CLAUDS+HSC Survey and ancillary data from COSMOS2020 catalogue for the estimates of galaxy redshifts and stellar masses. We measure galaxy sizes by fitting a single-component S\'ersic profile to their light profiles using \textsc{Galfit} software.

\par 
We fit the size-mass relation (SMR) for SFGs and QGs separately in five redshift bins in the two rest-frame wavelengths (Equations \ref{eq:smr} and \ref{eq:dpl}. We then analyse the redshift evolution of galaxy sizes by fitting a power-law (Equation \ref{eq:evol}). 

\par 
The main conclusions of our study are:
\begin{enumerate}
    \item Both SFGs and QGs on average appear more extended in blue light (rest-frame $3000$~\AA) than in red light (rest-frame $5000$~\AA) at all redshifts and masses (Figures \ref{f:smr-3000&5000A}, \ref{f:size-evo-2wave}, \ref{f:size-evo-sfg} and \ref{f:size-evo-qg}). We find that the pace of evolution at fixed stellar mass is stronger in the rest-frame visible light than in the rest-frame UV light and this difference is more prominent for QGs (Figure \ref{f:size-evo-qg}). In addition, the strength of the evolution in both wavelengths in general increases with the stellar mass of galaxies (Figures \ref{f:size-evo-2wave} and \ref{f:size-evo-sfg}).  
    
    \item  When we incorporate mass growth for SFGs assuming the evolving star formation main sequence, the pace of the  SFG size evolution (i.e., the exponent of the best-fitting power-law) becomes independent of both galaxy stellar mass and the rest-frame wavelength (Figure \ref{f:size-evo-mevol}).
    
    \item The strength of the size evolution at constant stellar mass for SFGs is weaker than the strength reported in studies that cover wider redshift ranges. We explore this slowing down of the size evolution since $z\sim1$ by dividing the SFG population into two groups: blue SFGs ($U-V\leq1$) and red SFGs ($U-V>1$). We find that red SFGs are smaller and more massive than blue SFGs and that their fraction increases by $25\%$ during the $6$ billion years of cosmic time that we explore. In addition, red SFGs constitute the majority ($83\%$) of the bulge-dominated SFGs ($n>2.5$). We conclude that the emergence of bulges is a primary driver of size evolution for SFGs. 
    
    \item We use a a toy model to remove the effect of newcomers (progenitor bias) from the observed trend in size evolution for QGs.
    The removal of the largest QGs changes the characteristic size of QGs at $z=0$ ($R_e^0$ in Equation \ref{eq:evol}). 
    This change in $R_e^0$ is systematically negative for all masses and in both rest-frame wavelengths. Our analysis shows that newcomers increase the average size of QGs by up to $\sim20\%$ since $z\sim1$. In contrast, the pace of QG size evolution is not affected systematically by this effect. We conclude that the strength of the evolution is not determined by disk fading during the process of quenching but rather by physical processes affecting individual galaxies after they become quiescent (e.g., mergers and accretion).

    \item 
    We find that the redshift evolution of $\Sigma_1$ stops (slows down) for SFGs (QGs) at $z<1$. This change in the evolutionary pace compared to higher $z$ is consistent with the growth of bulges in SFGs and the influx of newcomers to the QG population. We also find that all SFGs with $\log \Sigma_1\gtrsim9$ are red ($U-V>1$) and that all galaxies with $\log \Sigma_1\gtrsim9.7$ are quiescent. Finally, the sSFR of red SFGs within these two threshold values of $\Sigma_1$ declines with $\Sigma_1$, suggesting that the growth of bulges drives the evolution of SFGs towards the QG population.

\end{enumerate}

\par 
To explore further the role of bulges in the average size growth of the SFG population and the impact of various physical processes on the QG size evolution, we will perform bulge$+$disk decomposition. We plan to implement this procedure in the future by fitting a double-component S\'ersic profile to galaxy light profiles. Furthermore, here we do not investigate the impact of environment on the galaxy size evolution due to the small volume we probe. To take into account environmental effects, we will expand our analysis based both on a single- and double-component S\'ersic profile to the entire $18.6$ deg$^2$ of the CLAUDS+HSC survey. Our pilot study demonstrates the power of deep ground-based photometric surveys and galaxy light profiles fitting in multiple rest-frame wavelengths to probe in detail the strength of galaxy structural evolution at $z<1$ and constrain its driving mechanisms.

\section*{Acknowledgements}

We thank the anonymous referee for their thoughtful suggestions that helped us improve the clarity and completeness of the manuscript. We gratefully acknowledge the contributions of Anneya Golob, Bobby Sorba, Harrison Souchereau and Ikuru Iwata. We thank the extra-galactic research group at Saint Mary's University, Canada, especially Vicente Estrada-Carpenter, Lingjian Chen and Devin Williams for their valuable inputs and support. We also thank Lalitwadee Kawinwanichakij, Dan Taranu, Connor Bottrell, Arjen van der Wel, Chien Peng, Lamiya Mowla and Jasleen Matharu for helpful discussions. The research of A.G., I.D. and M.S. is supported by the National Sciences and Engineering Council (NSERC) of Canada.

\par
These data were obtained and processed as part of the CFHT Large Area U-band Deep Survey (CLAUDS), which is a collaboration between astronomers from Canada, France, and China described in \cite{sawickiCFHTLargeArea2019}.   CLAUDS is based on observations obtained with MegaPrime/ MegaCam, a joint project of CFHT and CEA/DAPNIA, at the CFHT which is operated by the National Research Council (NRC) of Canada, the Institut National des Science de l'Univers of the Centre National de la Recherche Scientifique (CNRS) of France, and the University of Hawaii. CLAUDS uses data obtained in part through the Telescope Access Program (TAP), which has been funded by the National Astronomical Observatories, Chinese Academy of Sciences, and the Special Fund for Astronomy from the Ministry of Finance of China. CLAUDS uses data products from TERAPIX and the Canadian Astronomy Data Centre (CADC) and was carried out using resources from Compute Canada and Canadian Advanced Network For Astrophysical Research (CANFAR).

\par
The Hyper Suprime-Cam (HSC) collaboration includes the astronomical communities of Japan and Taiwan, and Princeton University, USA. The Hyper Suprime-Cam (HSC) collaboration includes the astronomical communities of Japan and Taiwan, and Princeton University. The HSC instrumentation and software were developed by the National Astronomical Observatory of Japan (NAOJ), the Kavli Institute for the Physics and Mathematics of the Universe (Kavli IPMU), the University of Tokyo, the High Energy Accelerator Research Organization (KEK), the Academia Sinica Institute for Astronomy and Astrophysics in Taiwan (ASIAA), and Princeton University.

\par
This paper is based (in part) on data collected at the Subaru Telescope and retrieved from the HSC data archive system, which is operated by Subaru Telescope and Astronomy Data Center (ADC) at National Astronomical Observatory of Japan. Data analysis was in part carried out with the cooperation of Center for Computational Astrophysics (CfCA), National Astronomical Observatory of Japan.

\par
\textit{Software:} \textsc{NumPy} \citep{vanderwaltNumPyArrayStructure2011, harrisArrayProgrammingNumPy2020}, \textsc{SciPy} \citep{virtanenSciPyFundamentalAlgorithms2020}, \textsc{Astropy} \citep{astropycollaborationAstropyCommunityPython2013, astropycollaborationAstropyProjectBuilding2018, astropycollaborationAstropyProjectSustaining2022}, \textsc{Photutils} \citep{bradleyAstropyPhotutils2022},  \textsc{PetroFit}, \textsc{Scikit-learn} \citep{pedregosaScikitlearnMachineLearning2011}, \textsc{Matplotlib} \citep{hunterMatplotlib2DGraphics2007}, \textsc{GalfitPyWrap}\footnote{\url{https://github.com/Grillard/GalfitPyWrap}}, \textsc{SExtractor} \citep{bertinSExtractorSoftwareSource1996},  \textsc{PSFEx} \citep{bertinAutomatedMorphometrySExtractor2011}, \textsc{Galfit} \citep{pengDetailedStructuralDecomposition2002, pengDetailedDecompositionGalaxy2010}, \textsc{Dynesty} \citep{speagleDYNESTYDynamicNested2020}.

\section*{Data Availability}
Hyper Suprime-Cam Subaru Strategic Program (HSC-SSP) Public Data are available at \url{https://hsc.mtk.nao.ac.jp/ssp}. CLAUDS \textit{u}-band images of Deep and UltraDeep layers of the HSC-SSP and a catalogue that includes the parameters of the best-fit 2D S\'ersic models we produced for this paper as well as matched COSMOS2020 photometry, stellar masses, and SFRs are available at \url{https://www.clauds.net}.

%%%%%%%%%%%%%%%%%%%%%%%%%%%%%%%%%%%%%%%%%%%%%%%%%%

%%%%%%%%%%%%%%%%%%%%%%%%%%%%%%%%%%%%%%%%%%%%%%%%%%

\bibliographystyle{mnras}
\bibliography{main}

%%%%%%%%%%%%%%%%% APPENDICES %%%%%%%%%%%%%%%%%%%%%

\appendix

\section{Robustness of the Fitting Pipeline and S\'ersic parameters}
\label{appendix:Simulations}

\begin{figure*}
    \centering
	\includegraphics[width=1.5\columnwidth]{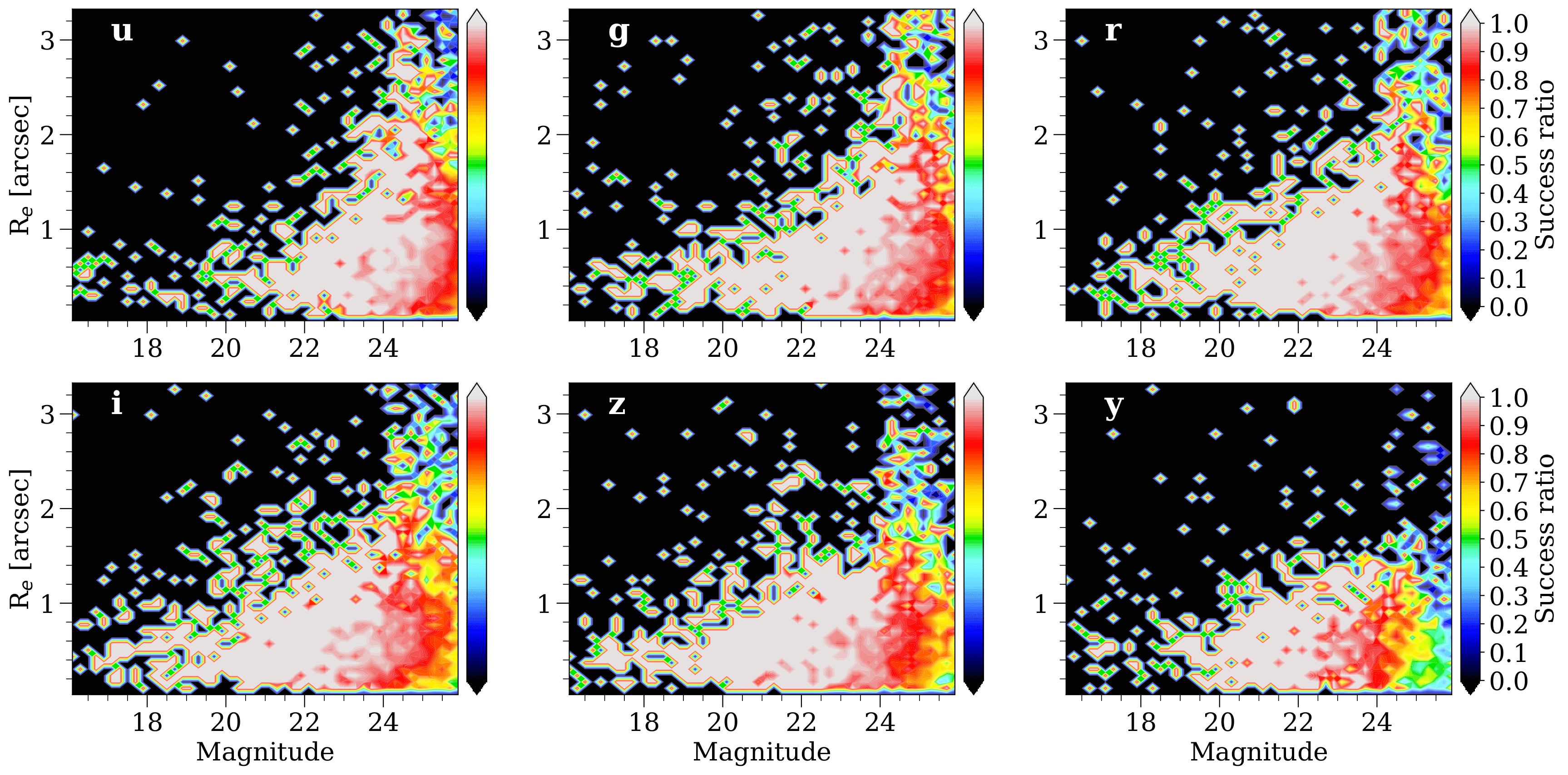} %.pdf}
    \caption[Size Heatmap]{Heat maps showing the simulation success rate (Equation \ref{eq:sim-sr}) as a function of input magnitude and input size. Galaxy sizes do not affect the success rate significantly in a given magnitude bin as indicated by the (almost) vertical colour stripes.}
    \label{f:heatmap-re}
\end{figure*}

\begin{figure*}
    \centering
	\includegraphics[width=1.5\columnwidth]{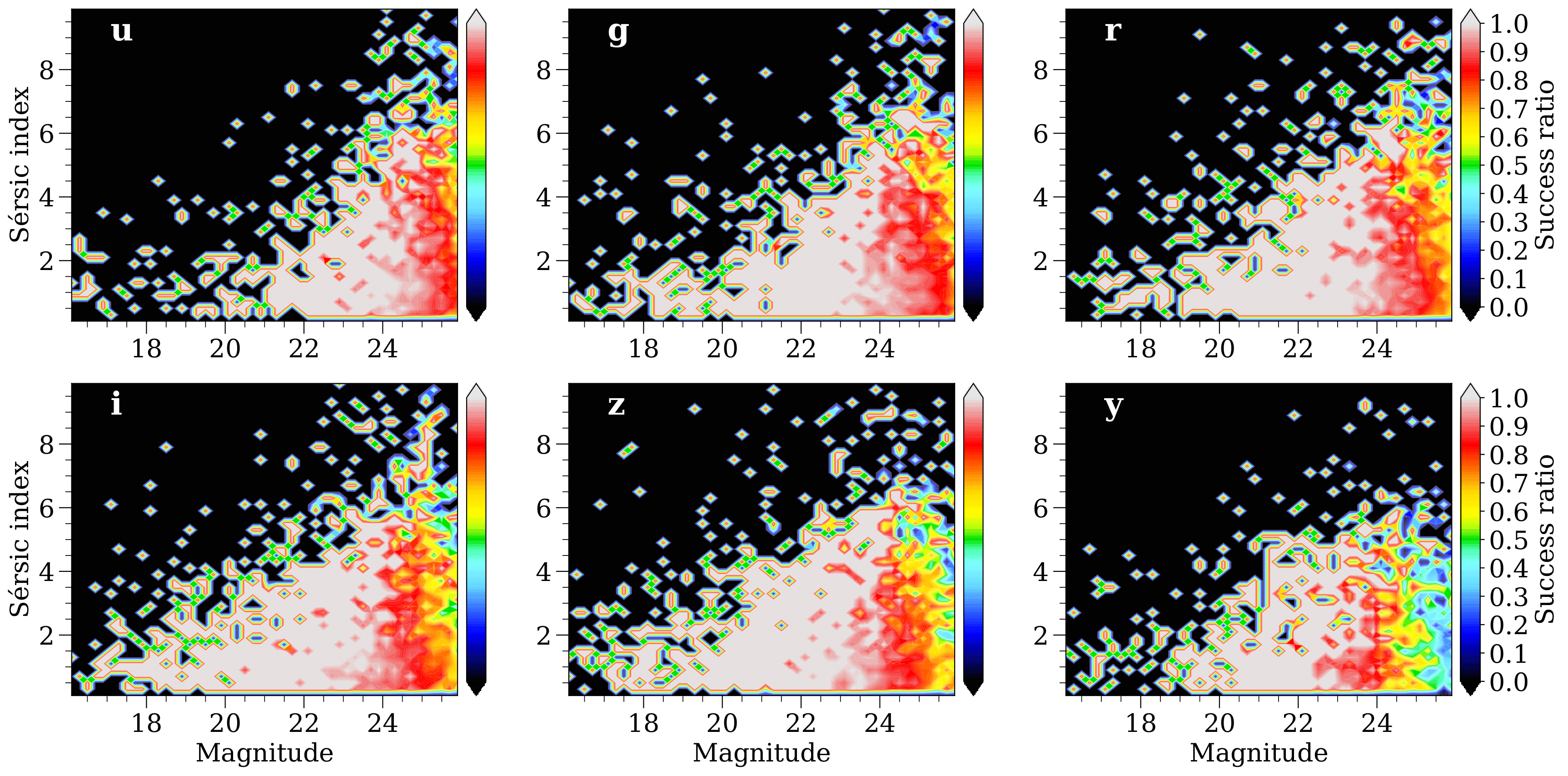} %.pdf}
    \caption[Size Heatmap]{Heat maps showing the simulation success rate (Equation \ref{eq:sim-sr}) as a function of input magnitude and input S\'ersic index. S\'ersic indices of galaxies do not affect the success rate significantly in a given magnitude bin as indicated by the vertical colour stripes.}
    \label{f:heatmap-n}
\end{figure*}

\begin{figure*}
    \centering
	\includegraphics[width=1.5\columnwidth]{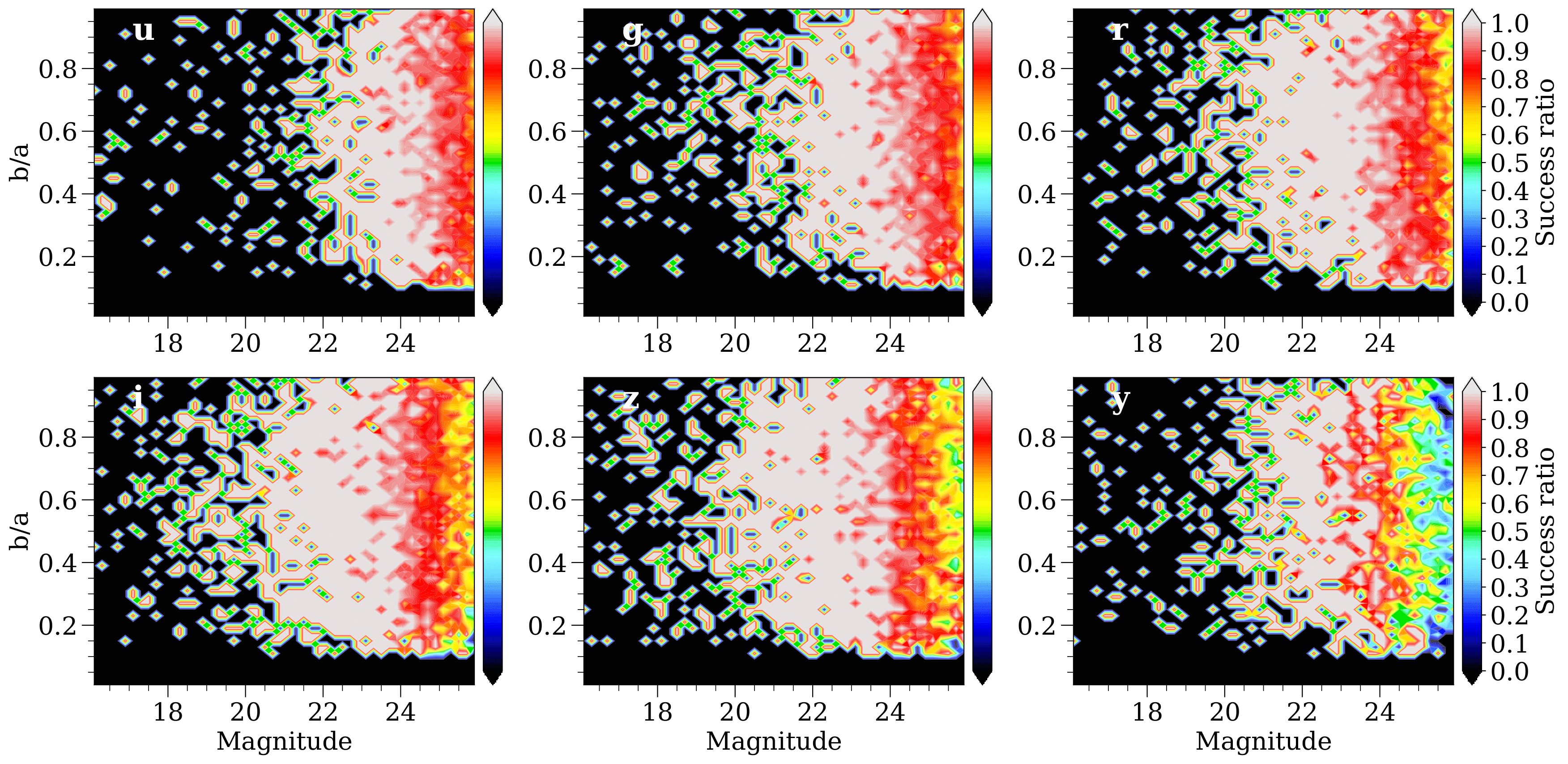} %.pdf}
    \caption[Size Heatmap]{Heat maps showing the simulation success rate (Equation \ref{eq:sim-sr}) as a function of input magnitude and input axis ratio. Axis ratios of galaxies do not affect the success rate significantly in a given magnitude bin as indicated by the vertical colour stripes.}
    \label{f:heatmap-ba}
\end{figure*}

In Section \ref{sec:Fitting}, we mention the magnitude limits that we adopt in each band while fitting the light profiles of galaxies. These magnitude limits are as follows: \textit{u}$<26$; \textit{g}$<25.5$; \textit{r}, \textit{i}, \textit{z}$<25$. We introduce these limits based on the robustness analysis of our fitting pipeline through the simulations we perform (Section \ref{sec:Simulations}). We adopt the simulation success rate described in Equation \ref{eq:sim-sr} as a proxy for the robustness of the pipeline. This rate tells us what fraction of galaxies with a certain properties we can fit successfully. 

\par 
Through simulations we investigate how this success rate changes with intrinsic S\'ersic parameters of the simulated galaxies. In Figure \ref{f:hsc-sim-sratio}, we show how success rates in each CLAUDS+HSC bands change with the band magnitude of galaxies. It is clear from the figure that the success rate is sensitive to the brightness of a galaxy beyond $24^\text{th}$ magnitude in the bands used in this study. 

\par
In Figures \ref{f:heatmap-re}, \ref{f:heatmap-n} and \ref{f:heatmap-ba}, we investigate how the success rate depend on other S\'ersic parameters - size ($R_e$), S\'ersic index ($n$) and axis ratio ($b/a$) respectively through heat maps. These figures show the success rate as function of both the input/intrinsic values of these S\'ersic parameters and magnitude in a given band. We find that the colours in the heat map in every panel in each figure change horizontally but remain more or less constant vertically. This observed colour pattern in the heat map shows that the success rate and thereby the robustness of the fitting algorithm does not depend on the intrinsic shape of the galaxy ($R_e$, $n$ and $b/a$) but the brightness alone. Therefore, we do not adopt any cuts on the data based on $R_e$, $n$ and $b/a$. 

\section{Size-Mass Relation Fitting in Rest-frame UV Light }
\label{appendix:SMR}

\begin{figure}
    \centering
	\includegraphics[width=0.73\columnwidth]{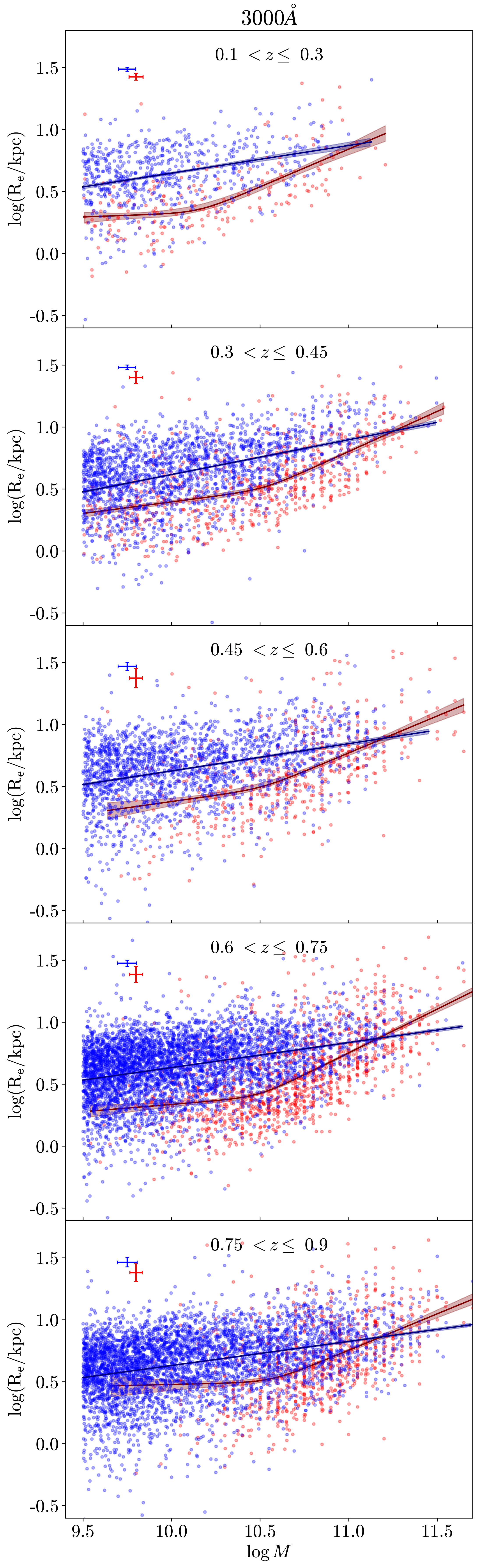}
    \caption[Fitting double power law in rest-frame $3000$~\AA]{Fitting smoothly broken double power law to the QGs and single power law to the SFGs at rest-frame $3000$~\AA\ in $5$ redshift bins. The individual SFGs are shown as blue points and QGs in red. The median uncertainties in their measurements are given in the top left corner of each panel. The best-fitting power law relations for SFGs and QGs are shown in navy blue and maroon respectively with their uncertainties from Bayesian posterior plotted as shaded region. }
    \label{f:hsc-doube-power-law-3000A}
\end{figure}

In Figure \ref{f:hsc-doube-power-law-3000A}, we show the size-mass relation (SMR) for star-forming galaxies (SFGs; blue) and quiescent galaxies (QGs; red) in rest-frame UV light ($3000$~\AA). We fit a single power-law for SFGs but a double power-law for QGs initially followed by a single power-law fit for QGs above the pivot mass ($M_p$; mass at pivot point) in the double power-law SMR (see Section \ref{sec:SMR}). We provide the best-fit parameters of double power-law SMR in Table \ref{table:DPL} and single-power law SMR in Table \ref{table:SPL}. 

\par 
As we see in Figure \ref{f:hsc-doube-power-law-3000A}, the SFGs are on average more extended than QGs in rest-frame UV light in every redshift bin at $0.1<z<0.9$. The SFGs and low mass QGs ($M<M_p$) have shallow SMR with an exponent around $0.2$ but the high mass QGs have a steep exponent ($\sim0.6$). These differences in SFG and QG SMRs are similar to those we find in rest-frame $5000$~\AA\ (see Figure \ref{f:hsc-doube-power-law}).

\section{Removing Progenitor Bias from QG Population} 
\label{appendix:PB-Removal}

\begin{figure*}
    \centering
	\includegraphics[width=1.95\columnwidth]{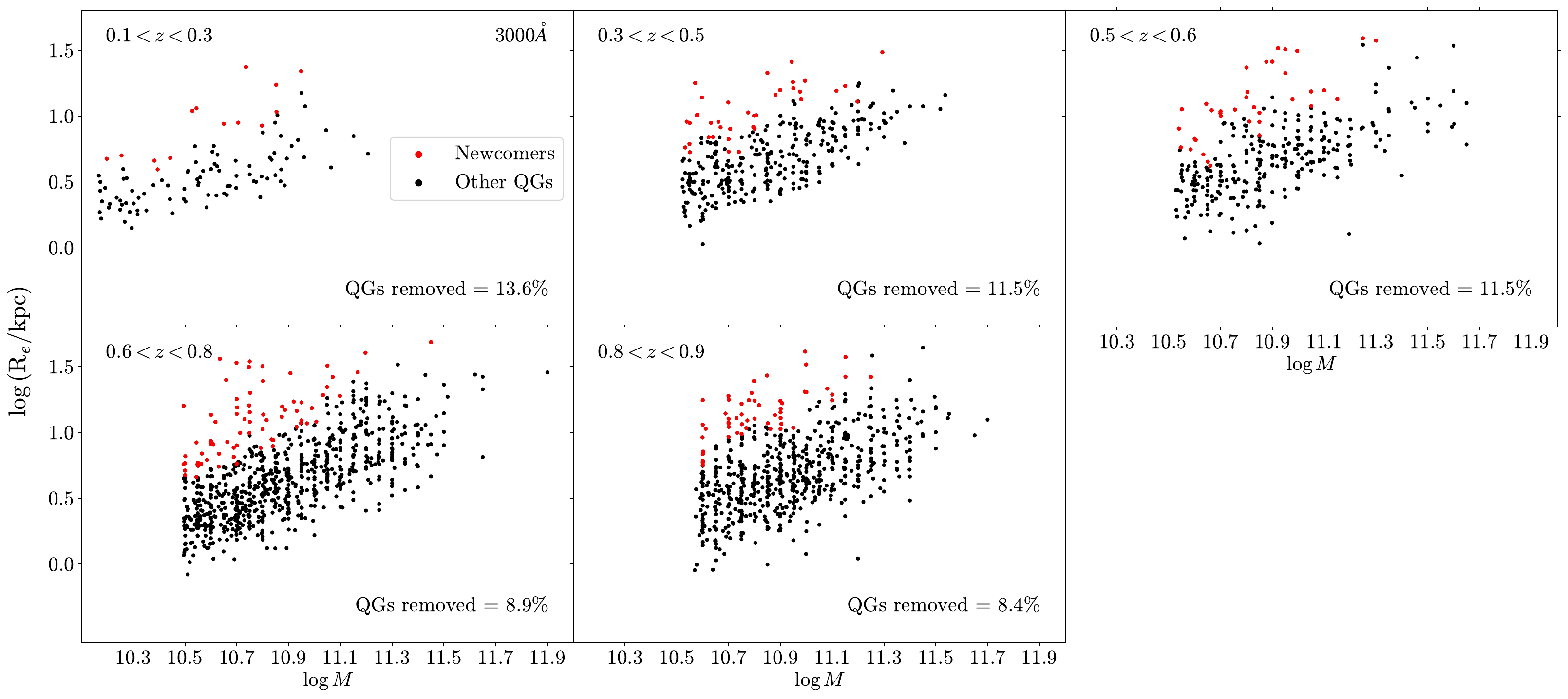}
    \caption[Removing Progenitor Bias in rest-frame $3000$~\AA]{Removing progenitor bias in rest-frame $3000$~\AA\ shown in size-mass plane. We describe the removal process in Section \ref{sec:pb-removal} and Appendix \ref{appendix:PB-Removal}. The QGs removed (newcomers) are shown in red and the remaining QGs are in black. The percentage of QGs removed are given in each panel.}
    \label{fig:pbremoval-3000}
\end{figure*}

\begin{figure*}
    \centering
	\includegraphics[width=1.95\columnwidth]{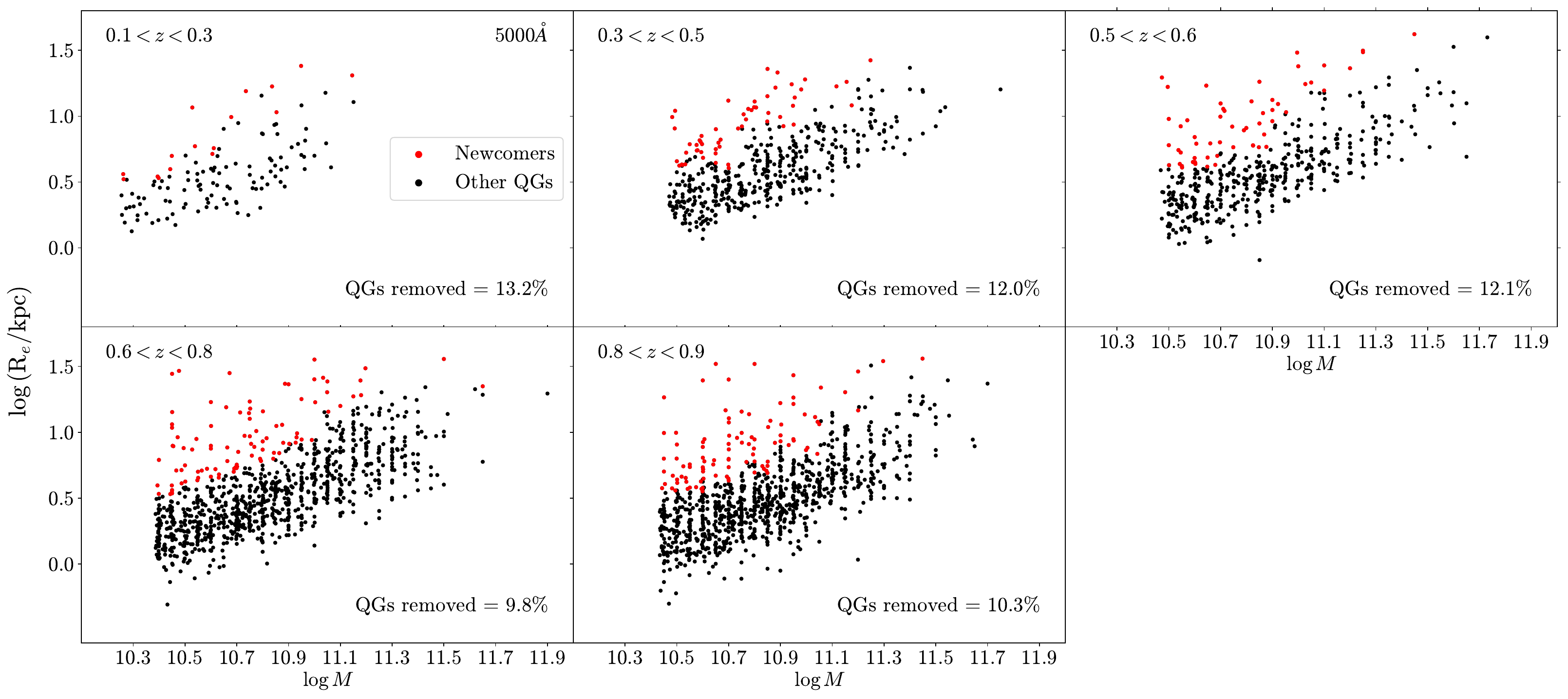} %.pdf}
    \caption[Removing Progenitor Bias in rest-frame $5000$~\AA]{Removing progenitor bias in rest-frame $5000$~\AA\ shown in size-mass plane. As in Figure \ref{fig:pbremoval-3000}, removed newcomers are in red and other QGs are in black.}
    \label{fig:pbremoval-5000}
\end{figure*}

Newly quenched galaxies introduces a bias (progenitor bias) when we analyze the size evolution of the quiescent population. Hence, we remove progenitor bias from our QG population as described in Section \ref{sec:pb-removal}. In short, we assume that the newly quenched galaxies are more extended than the galaxies those quenched earlier in a given stellar mass bin. We also assume that the quenching efficiency increases with mass and thus, the fraction of newly quenched galaxy in our QG sample varies in stellar mass bins. We estimate this fraction using Equation \ref{eq:pb-fraction}. We normalize this equation so that we remove approximately $10\%$ of QGs from the parent population in each redshift bin. 

\par 
In Figures \ref{fig:pbremoval-3000} and \ref{fig:pbremoval-5000} we show the QGs removed in the size-mass plane in rest-frames $3000$~\AA\ and $5000$~\AA\ respectively. In both wavelengths, we remove $8-13\%$ of QGs 
in each redshift bin and this percentage is given in each panel. The removed QGs (red) are more extended than the remaining QGs in their respective mass bin and their percentage increases towards higher masses by design.

\section{Central Stellar Mass Surface Density}
\label{appendix:sigma1-mass-relation}

\begin{figure*}
    \centering
	
	\includegraphics[width=1.95\columnwidth]{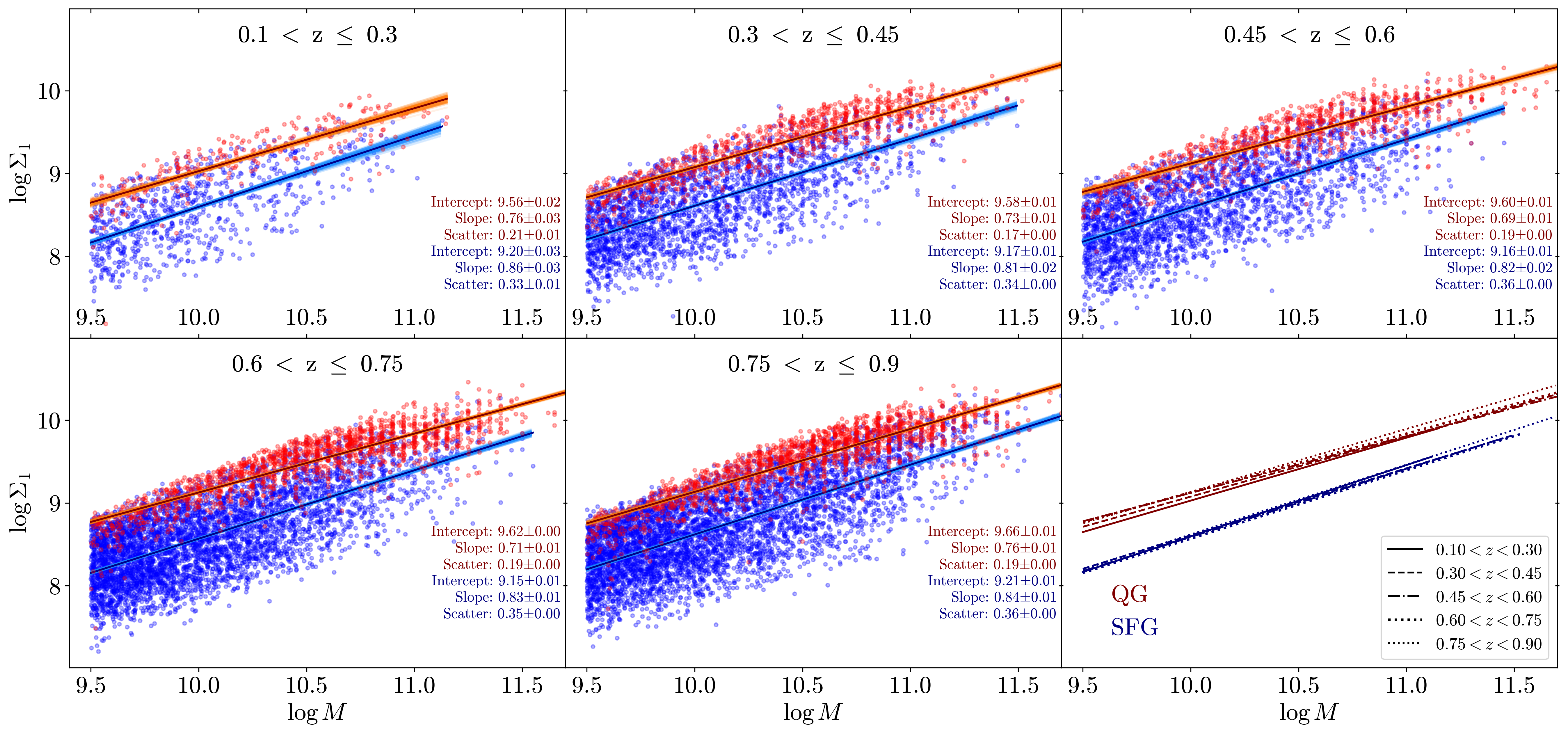} 
    \caption[Sigma1-logM relation]{
    
    The best-fitting $\log\Sigma_1 - \log M$ relation for SFGs (blue) and QGs (red) with mass $\log M>9.5$ in $5$ redshift bins. The best-fit single power-law parameters are given in each panel.
    
    }
    \label{f:sigma1-logM}
\end{figure*}

\begin{figure*}
    \centering
	\includegraphics[width=1.95\columnwidth]{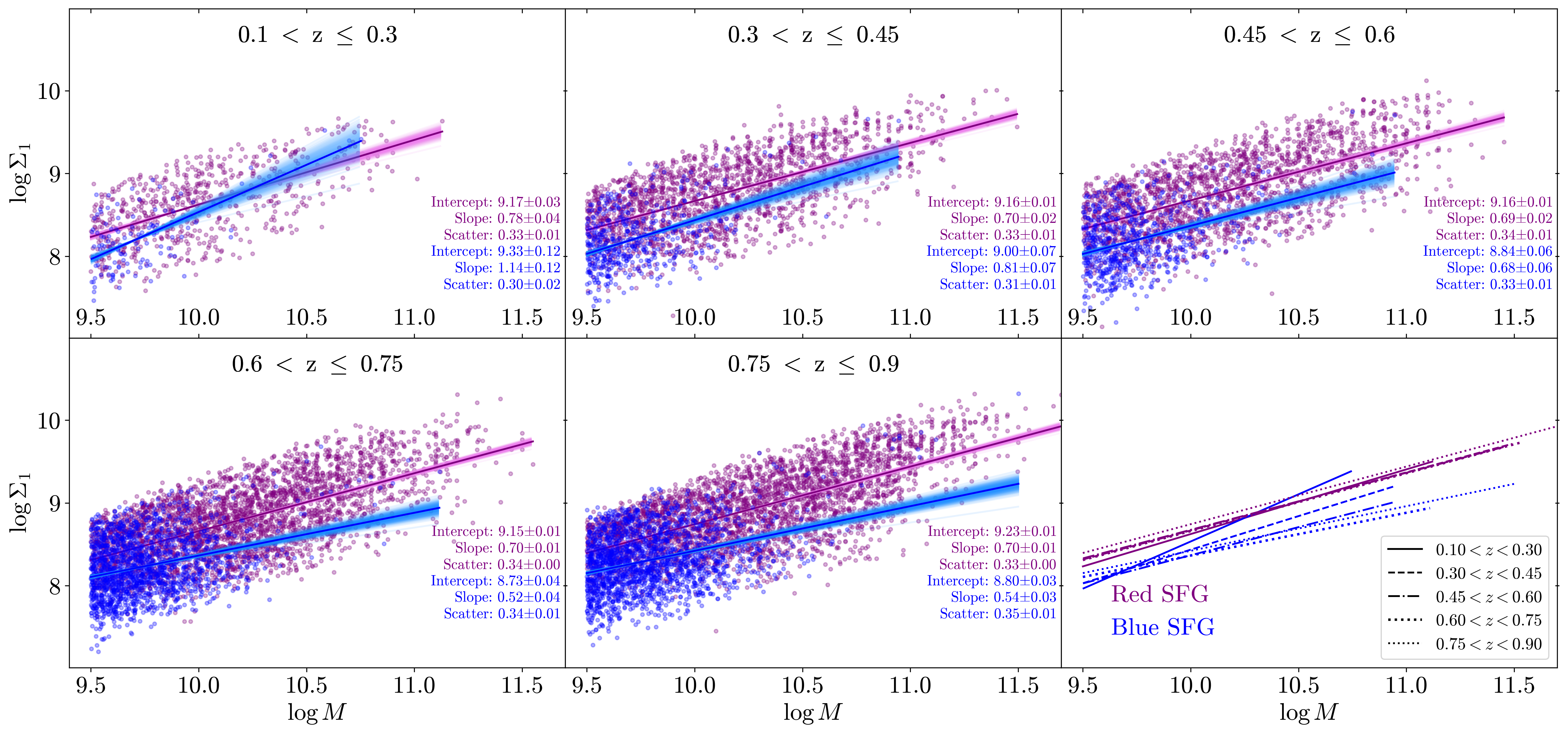} 
    \caption[Sigma1-logM relation for SFGs]{
    The best-fitting $\log\Sigma_1 - \log M$ relation for blue and red SFGs (blue and purple colours respectively) with mass $\log M>9.5$ in $5$ redshift bins. The best-fit single power-law parameters are given in each panel.
    
    }
    \label{f:sigma1-logM-SFG}
\end{figure*}

\par
Following \citet{cheungDependenceQuenchingInner2012} and \citet{jiReconstructingAssemblyMassive2022}, we calculate the projected central stellar mass surface density as
\begin{equation}
    \Sigma_1 \left[\frac{\Msun}{\text{kpc}^2}\right] = \frac{M_{1\text{kpc}}}{\pi \times 1\text{kpc}^2} = \frac{M}{\pi \times 1\text{kpc}^2} \frac{\gamma(2n,x)}{\Gamma(2n)},
    \label{eq:Sigma1}
\end{equation}
where $M_{1\text{kpc}}$ is the stellar mass within $1$~kpc radius in $\Msun$, $M$ is the total stellar mass in $\Msun$, $\gamma$ and $\Gamma$ are incomplete and complete gamma functions, and $x=b_n \left( \frac{1\text{kpc}}{R_e} \right)^{(1/n)}$. We calculate $b_n$ by solving the equation $\Gamma(2n)=2\gamma(2n,b_n) $ numerically \citep{grahamConciseReferenceProjected2005}. The caveat here is that we assume a uniform mass-to-light (M/L) ratio within each galaxy. Although this assumption is reasonable for QGs, we acknowledge that the radial variation in $M/L$ ratio is significantly higher for SFG \citep[e.g.,][]{tortoraStellarMasstolightRatio2011}. 
We attempt to minimize the effect of $M/L$ variation by analyzing the $\Sigma_1$ only at the rest-frame $5000$~\AA\ which has more contribution from bulk of the stars (old low-mass stars) than at $3000$~\AA. However, even at $5000$~\AA, radial variation of $M/L$ ratio can be significant (at least) in SFGs where the overall profile is dominated by the light from young stars in the disks. Hence, the $\Sigma_1$ that we estimate are lower limits, especially for SFGs with steep negative colour gradients. 

\par

To explore how the characteristic $\Sigma_1$ value for SFGs and QGs evolves in Section \ref{sec:central-mass-density}, we fit a single power-law to the $\Sigma_1$-stellar mass relation in log-log space for SFGs and QGs with mass $\log M>9.5$ in $5$ redshift bins (Figure \ref{f:sigma1-logM}). Across the redshift bins, we do not find any signs of significant evolution in the power-law parameters, in congruence with previous works (e.g., \citet{fangLinkStarFormation2013} at $z\sim0.1$ and \citet{barroStructuralStarformingRelations2017} at $0.5<z<3$). The only parameter that we observe changing with time is the zero point for QGs. The $\Sigma_1$ of QGs with a characteristic mass $5\times 10^{10}$ \Msun\ increases by $\sim20\%$ from the lowest to the highest redshift bin.

\par

The $\Sigma_1-$stellar mass relation is tighter for QGs than for SFGs (Figure \ref{f:sigma1-logM}). 
One of the possible reasons for this difference is the strong variation in $M/L$ ratios in SFG radial profiles compared to QG profiles \citep{tortoraStellarMasstolightRatio2011}. 
QGs in general have higher $\Sigma_1$ than SFGs in any given mass bin and this $\Sigma_1$ value is very similar for all QGs with similar stellar mass due to the very tight $\Sigma_1-$stellar mass relation. Thus, there exists a threshold $\Sigma_1$ value required for a galaxy to be quiescent and this threshold value increases with galaxy stellar mass. It means that forming a dense core is a prerequisite for quenching (or a consequence of quenching) as argued by \citet{cheungDependenceQuenchingInner2012}, \citet{fangLinkStarFormation2013}, \citet{vandokkumDenseCoresGalaxies2014}, \citet{barroStructuralStarformingRelations2017}, \citet{jiReconstructingAssemblyMassive2023} and others. However, since some SFGs also have $\Sigma_1$ close to the threshold value, having a dense core alone is insufficient for a galaxy to be quiescent.

\par 
A lack of evolution in the slopes indicates that the galaxies build up their inner core while they grow in stellar mass in a similar manner across the redshift range. However, $\Sigma_1-$stellar mass relation has $15-20\%$ steeper slope for SFGs. This difference in the slopes is at least partially due to the differences in how galaxies evolve before and after quenching. SFGs build up mass more uniformly across the galactocentric radii than QGs. However, we note that the scaling relations for QGs are also affected by recently quenched galaxies (newcomers) through the so-called progenitor bias \citep[Section \ref{sec:pb-removal}; ][]{ jiReconstructingAssemblyMassive2022, jiReconstructingAssemblyMassive2023}.    

\par

To further probe the evolution in characteristic $\Sigma_1$ for blue and red SGFs separately in Section \ref{sec:central-mass-density}, we fit the $\Sigma1-$stellar mass relations for these two SFG subpopulations in Figure \ref{f:sigma1-logM-SFG}. Because the majority of SFGs within our redshift interval is red (purple points in \ref{f:sigma1-logM-SFG}; see also Figure \ref{f:sfg-rbfrac}), the $\Sigma_1$ values of the overall SFGs reflect those of red SFGs. The blue SFGs (blue points in \ref{f:sigma1-logM-SFG}) generally have lower $\Sigma_1$ than red SFGs at a given stellar mass. This suggests that, as blue SFGs turn red, they also build up their central regions.

\par 

However, we acknowledge that the parameters of the $\Sigma_1-$mass relation for blue SFGs are quite uncertain because of the small sample that covers a narrow dynamic range in stellar masses. At $0.75<z<0.9$, around $45\%$ of SFGs are blue but this percentage drops down to below $15\%$ at $0.1<z<0.3$. Most of these blue SFGs have stellar mass in a narrow $9.5<\log M<10$ interval. Hence, the parameters of the fit for blue SFGs at $z<0.45$, where the fraction of blue galaxies in the star-forming population is below $20\%$, are not well constrained.

%%%%%%%%%%%%%%%%%%%%%%%%%%%%%%%%%%%%%%%%%%%%%%%%%%

\bsp	% typesetting comment
\label{lastpage}
\end{document}